\documentclass[aps,twocolumn,nofootinbib]{revtex4}
\usepackage{graphicx}
\usepackage{bm}

\begin{document} 

\title{\bf Large-deviation principles, stochastic effective actions,
  path entropies, and the structure and meaning of thermodynamic
  descriptions}

\author{Eric Smith}

\affiliation{Santa Fe Institute, 1399 Hyde Park Road, Santa Fe, NM
87501, USA}

\date{\today}
\begin{abstract}

The meaning of thermodynamic descriptions is found in large-deviations
scaling~\cite{Ellis:ELDSM:85,Touchette:large_dev:08} of the
probabilities for fluctuations of averaged quantities.  The central
function expressing large-deviations scaling is the entropy, which is
the basis for both fluctuation theorems and for characterizing the
thermodynamic interactions of systems.  Freidlin-Wentzell
theory~\cite{Freidlin:RPDS:98} provides a quite general formulation of
large-deviations scaling for non-equilibrium stochastic processes,
through a remarkable representation in terms of a Hamiltonian
dynamical system.  A number of related methods now exist to construct
the Freidlin-Wentzell Hamiltonian for many kinds of stochastic
processes; one method due to Doi~\cite{Doi:SecQuant:76,Doi:RDQFT:76}
and Peliti~\cite{Peliti:PIBD:85,Peliti:AAZero:86}, appropriate to
integer counting statistics, is widely used in reaction-diffusion
theory.

Using these tools together with a path-entropy method due to
Jaynes~\cite{Jaynes:caliber:80}, this review shows how to construct
entropy functions that both express large-deviations scaling of
fluctuations, and describe system-environment interactions, for
discrete stochastic processes either at or away from equilibrium.  A
collection of variational methods familiar within quantum field
theory, but less commonly applied to the Doi-Peliti construction, is
used to define a ``stochastic effective action'', which is the
large-deviations rate function for arbitrary non-equilibrium paths.

We show how common principles of entropy maximization, applied to
different ensembles of states or of histories, lead to different
entropy functions and different sets of thermodynamic state variables.
Yet the relations of among all these levels of description may be
constructed explicitly and understood in terms of information
conditions.  Although the example systems considered are limited, they
are meant to provide a self-contained introduction to methods that may
be used to systematically construct descriptions with all the features
familiar from equilibrium thermodynamics, for a much wider range of
systems describable by stochastic processes.

\end{abstract}

\maketitle

\section{Introduction}

Thermodynamics is not fundamentally a theory of energy distribution,
but a theory of statistical degeneracy~\cite{Jaynes:LOS:03}.  As such,
while most of our experience and intuition about thermodynamics is
drawn from equilibrium statistical
mechanics~\cite{Fermi:TD:56,Kittel:TP:80,Huang:SM:1987}, which
emphasizes the role of energy, we should expect that its fundamental
principles apply in much wider domains, outside equilibrium, and even
outside mechanics.

Within the last 50 years, a clear conceptual understanding of the
nature of thermodynamic
descriptions~\cite{Ellis:ELDSM:85,Touchette:large_dev:08} has combined
with new methods to analyze a wide variety of stochastic processes,
for continuous~\cite{Martin:MSR:73} and discrete
systems~\cite{Doi:SecQuant:76,Doi:RDQFT:76,Peliti:PIBD:85,Peliti:AAZero:86},
in quantum~\cite{Schwinger:MBQO:61,Keldysh::65} and classical
mechanics~\cite{Mattis:RDQFT:98,Cardy:FTNEqSM:99}, both at and away
from equilibrium, and even in other areas using information theory
such as optimal data compression and reliable
communication~\cite{Shannon:MTC:49}, and via these, robust molecular
recognition~\cite{Schneider:TMMI:91,Schneider:TMMII:91}.  These
developments confirm that thermodynamics is indeed not restricted to
equilibrium or to mechanics.  They give us insight into when
thermodynamic descriptions should exist, and they provide systematic
methods to construct such descriptions in a wide variety of
situations. 

This paper reviews the aspects of large-numbers scaling and structural
decomposition that are essential to thermodynamic descriptions, and
presents examples from which each of these may be seen in different
forms that are appropriate to equilibrium and non-equilibrium
statistical mechanics.  It also brings together construction methods
(based on generating
functions~\cite{Doi:SecQuant:76,Doi:RDQFT:76,Peliti:PIBD:85,%
  Peliti:AAZero:86}), scaling relations (based on ray
approximations~\cite{Graham:path_int:77,Freidlin:RPDS:98}), and
variational methods (based on functional Legendre
transforms~\cite{Weinberg:QTF_II:96}), which will enable the reader to
systematically construct the fluctuation theorems of thermodynamic
descriptions from their underlying stochastic processes.

\subsection{Key concepts, and source of examples}

\subsubsection{Entropy underlies large-deviations scaling and reflects
  system structure} 

The entropy, as a logarithmic measure of degeneracy, is the central
quantity in thermodynamics.  It arises as the leading term in the
log-probability for fluctuations of averaged quantities.  At the same
time, however, when sub-system components, or a system and its
environment, interact, they discover their most probable joint
configuration through fluctuations.  Therefore the competition among
entropy terms also reflects the structure of system interactions at
the macroscale.  We are reminded of the importance of this structural
role of entropies, by the fact that pure ``classical''
thermodynamics~\cite{Fermi:TD:56} is entirely devoted to the analysis
of entropy gradients.  Therefore we wish to insist on being able to
decompose entropy into its sub-system components as a criterion for
any fully-developed thermodynamic description.

The key property that defines the existence of thermodynamic limits,
and the form of their fluctuation theorems, is large-deviations
scaling~\cite{Ellis:ELDSM:85,Touchette:large_dev:08}.  It is the
precise statement of the simplification provided by the law of large
numbers, not only in the infinite limit of aggregation, but in the
asymptotic approach to infinity.  It is well-known that averaging over
ensembles of configurations for large systems removes almost all
(irrelevant) degrees of freedom, and leaves only summary statistics,
which are the \emph{state variables} of the thermodynamic
description.\footnote{In the language of Kolmogorov, the state
  variables are \emph{minimum sufficient statistics} for predicting
  the state of a system sampled from an ensemble.  No more detailed
  summary statistic provides better predictions over the whole
  ensemble.  At the same time, no further reduction makes it possible
  to write any state variable as a function of the others.}  In finite
systems, these summary statistics can still show sample fluctuation,
but in the thermodynamic limit, the fluctuation probability takes a
simple form.

For an ensemble that possesses large-deviations scaling, it is
possible to describe classes of fluctuations as having the same
structure under different degrees of aggregation.  (An example would
be fractional density fluctuations in regions of a gas, whose sample
volume may be varied).  The log-probability for any such fluctuation
then \emph{factors}, into a term that depends on overall system scale,
and a scale-invariant coefficient that depends only on the structure
of the fluctuation.  The scale-invariant coefficient is called the
\emph{rate function} of the large-deviations scaling
relation~\cite{Touchette:large_dev:08}.  (An example would be the
specific entropy of a gas.)

Large-deviations scaling presumes the existence of a \emph{separation
  of scales} -- between microscopic processes and their macroscopic
descriptions -- over which aggregation does not lead to qualitative
changes in the kinds of fluctuations that can occur.  We expect
thermodynamic limits to exist where this separation of scales is
large.  Examples of structural change that can interrupt simple
scaling under the law of large numbers include phase transitions,
which can change the space of accessible excitations.

Large-deviations scaling permits us to combine fluctuation statistics
with entropy decompositions that reflect system structure.  In
equilibrium thermodynamics, the result is the classical fluctuation
theorem for macrostates~\cite{Ellis:ELDSM:85}: the log-probability for
fluctuations (of energy, volume, particles, etc.) between sub-systems
with well-defined entropies is the difference between the sum of
entropies at the fluctuating value and the maximum value for this sum,
which is the equilibrium value.

If we wish to consider the probabilities of fluctuations with more
complex structure (whether in equilibrium thermodynamics or in more
complicated cases), we will need more flexible methods to compute
large-deviations formulae and entropy decompositions.  For this, we
introduce the notion of the \emph{stochastic effective action}, which
is defined by variational methods that generalize the familiar
Legendre transform of equilibrium thermodynamics.  The meaning of this
quantity, and the way it is used, will be most easily understood by
following its construction in the body of the paper, so we postpone
further discussion until that point.

\subsubsection{The principles in a simple progression: from
  equilibrium to non-equilibrium statistical mechanics}

We will develop examples whose purpose is to show that the definition
and properties of the entropy do not change as we extend
thermodynamics beyond equilibrium statistical mechanics.  Rather, what
changes is the state space to which we assign probabilities.

A direct example comes from comparing an equilibrium thermodynamic
system, to a non-equilibrium description constructed for the same
system.  A frequent
approach~\cite{Onsager:RRIP1:31,Onsager:RRIP2:31,DeGroot:NET:84,%
  Prigogine:MT:98} to non-equilibrium statistical mechanics (NESM)
continues to use the equilibrium state variables and equilibrium
entropy, but considers their time rates of change.  We will see that
such an approach, focused on retaining the functional form of the
equilibrium entropy, sacrifices its meaning as a large-deviations rate
function. 

Instead, we will make the transition from equilibrium to
non-equilibrium statistical mechanics by replacing an ensemble of
states (in equilibrium) by an ensemble in which entire time-dependent
trajectories -- termed \emph{histories} -- are the elementary entities
(for NESM), and then we will construct the appropriate
large-deviations limits for the ensemble of histories.  The functional
form of the entropy will necessarily change.  More importantly, the
inventory of \emph{state variables} will necessarily be enlarged, to
include not only the configuration variables of equilibrium, but also
a collection of currents that relate to the changes in configuration.
Both kinds of variables will be needed as summary statistics for an
ensemble of histories, and both will enter as arguments in the
non-equilibrium entropy.

NESM is not so far removed from equilibrium thermodynamics that it can
really do justice to the generality of large-deviations principles and
thermodynamic descriptions.  However, it allows us to begin with a
completely familiar (equilibrium) construction, then to compare it to
a construction with rather different functional forms, and finally to
derive the complete set of relations that connect the two descriptions.

\subsection{Markovian stochastic processes, the two-state model as an
  example, and the approach of the paper}

Markovian stochastic processes~\cite{Durrett:Stoch_Proc:99} provide a
general, substrate-independent framework within which to study
statistical degeneracy and large-deviations scaling.  They include
models from statistical mechanics, but they may also be used to
represent many other random processes, whose structure may have
different constraints and interpretations from those of mechanics.

This review will use the two-state random walk, in either discrete or
continuous time, as a sample system for which equilibrium and
non-equilibrium thermodynamic descriptions will be built and then
compared.  The two-state random walk is the simplest discrete
stochastic process, and most thermodynamic quantities of interest in
both ensembles can be computed for this system in closed form.
However, the constructions in the examples immediately generalize to
more complicated cases, and several generalizations and approximation
methods will be covered either in the main text or in appendices.

Sec.~\ref{sec:lg_dev_scaling} will introduce large-deviations scaling
one level ``below'' the discrete random walk, by supposing that the
discrete model emerges as a coarse-grained description from the
continuous random walk in a double-well potential.  The continuum
model illustrates the concept of \emph{concentration of measure} from
large-deviations theory, and sets the parameters (both explicit and
implicit) that define the discrete model.  It also clarifies the
nature and origin of the ``local-equilibrium''
approximation~\cite{Prigogine:MT:98} for coarse-grained descriptions
of motion on free-energy landscapes, and illustrates graphically the
dual roles that charges and currents must have in non-equilibrium
entropy principles.

Sections~\ref{sec:EQ_large_dev} and~\ref{sec:NEQ_FW_large_dev} present
the equilibrium thermodynamics of the two-state model, and its most
direct generalization through the master equation of the stochastic
process.  Sec.~\ref{sec:EQ_large_dev} uses the exact solution of the
equilibrium distribution to introduce all basic quantities of the
large-deviations theory, and derives these using generating functions
and their associated variational methods.
Sec.~\ref{sec:NEQ_FW_large_dev} then presents the same construction
for the time-dependent probability distribution, in which histories of
particle counts rather than counts at a single time are the elementary
entities.  Neither of these ensembles distinguishes particle
identities, either in states or in trajectories.
Sec.~\ref{sec:caliber} presents an alternative construction of a
thermodynamics of histories based on the entropy of
distinguishable-particle trajectories, and this second construction
naturally separates the path entropy from probability terms due to the
environment, in a form exactly analogous to the Gibbs free energy for
equilibrium.

The remainder of the introduction lists sources for the particular
methods used in later sections, and explains why they capture
different aspects of a full non-equilibrium thermodynamics.  Many
aspects of the following derivations -- the naturalness of the
generating-function representation, the role of operator algebras and
linear algebra more generally, or the information conditions and
counting statistics that relate one ensemble to another -- may be
understood in conceptual terms that are more fundamental than the
particular constructions in which they appear below.  We return to
these in the discussion, making use of examples from the text.

Numerous, diverse literatures now contribute to the understanding of
methods closely related to those used below.  A brief summary of the
history and connections among the ideas, and a broader set of
citations, are provided in App.~\ref{sec:literature}.

\subsection{Bringing together three perspectives on non-equilibrium
  thermodynamics}

The extension of large-deviations scaling to ensembles of histories is
given by {\bf Freidlin-Wentzell (F-W) theory}~\cite{Freidlin:RPDS:98}.
This approach is widely applied to the computation of escape
trajectories and first-passage
times~\cite{Maier:escape:93,Maier:non_grad:92,Maier:exit_dist:97,%
  Maier:caustics:93,Maier:scaling:96,Maier:bifurc:00,Maier:oscill:96,%
  Maier:sloshing:01,Maier:droplets:01}.  It is remarkable for the way
it reduces both problems of inference, and the description of
multiscale dynamics, to a representation which is a Hamiltonian
dynamical system~\cite{Eyink:action:96,Mattis:RDQFT:98}.

A convenient method for arriving at the Hamiltonian description of F-W
theory is the {\bf Doi-Peliti (D-P)
  construction}~\cite{Doi:SecQuant:76,Doi:RDQFT:76,Peliti:PIBD:85,%
  Peliti:AAZero:86} based on generating functionals.  The D-P
construction is only one of many closely-related methods based on
expansions in coherent states, which are reviewed in
App.~\ref{sec:literature}.  These
methods~\cite{Martin:MSR:73,Graham:path_int:77,Graham:potential:84,%
  Kamenev:DP:01} have the common feature that the field representing
sample-mean values of observables is augmented by a conjugate momentum
that generates the change in those observables.  This value/momentum
pair leads to the F-W Hamiltonian description.  The D-P method is
particular to stochastic processes with independent, discrete number
counts, but may readily be generalized to continuous or
non-independent observables~\cite{Smith:DP:08}, as well as having many
parallels in dissipative quantum
mechanics~\cite{Schwinger:MBQO:61,Keldysh::65,Lifshitz:LandL:80}.

The F-W method directly gives the scale factors and rate functions of
the large-deviations limit for histories.  However, it does not
generally decompose fluctuation probabilities into separate terms
representing sub-system components or system-environment interactions,
which we want as part of a thermodynamic description.  To produce that
decomposition, we introduce a path-entropy method due to Jaynes known
as {\bf maximum caliber}~\cite{Jaynes:caliber:80}, which has its roots
in much older analysis of the entropy rates of stochastic processes
and chaotic dynamical systems due to Kolmogorov and others.  In
addition to making the F-W representation more recognizable as a
direct counterpart to the constructions in equilibrium thermodynamics,
the maximum-caliber method will separate those events that involve
energy exchange with the environment from those that do not, allowing
us to understand the role of energy dissipation in large-deviations
formulae for paths.

\subsection{A glossary of notation}

In characterizing thermodynamic ensembles, we will need to make three
choices about the level of representation, and the basis used:

\begin{enumerate}

\item Whether we are referring to ensemble averages (hence,
  deterministic summary statistics), or quantities that fluctuate
  stochastically to represent the process of sampling;

\item Whether we are considering discrete samples that change in the
  integer basis of particle counts, or modes of collective
  fluctuation, which we describe with the continuous mean values of
  Poisson distributions;\footnote{This distinction is similar to the
    distinction between the position basis and the wavenumber basis
    exchanged by Fourier-Laplace transform.}

\item Whether we are describing absolute particle numbers distributed
  among states, or are separating total number as a scaling variable,
  from the fractional distributions of particles that have a
  scale-invariant meaning.

\end{enumerate}

Different choices will lead to objects with quite different
mathematical behavior, even if all of them represent particle numbers
in one way or another.  The notation in the following sections is
chosen to reflect the three choices above, while still permitting
readable equations.

Ensemble averages of any quantities are denoted by overbars.  This
convention is easier to incorporate in complicated equations than the
$E$ (for expectation value) commonly used in statistics.

The remaining distinctions in the notation are summarized in
Table~\ref{tab:social_makeup}.

\begin{widetext}
\begin{table}[ht]
\begin{tabular}{l|llll|l|}
  Position & 
  Domain & 
  Measure & 
  Dynamics & 
  Meaning and usage &
  Conjugate \\
  variable & & & & &
  momentum \\
\hline
  $N$        & $\left[ 0 , \infty \right]$ & Integer & Fixed & 
               Total population number & not used  \\
             &  &  &  & 
               Scale factor in large-deviations property & \\
  ${\rm n}_i$  & $\left[ 0 , N \right]$ & Integer & Stochastic &
               Values of sample points &
               $-\partial / \partial {\rm n}_i$ \\
  $n_i$        & $\left[ 0 , N \right]$ & Real    & Langevin &
               Mean of Poisson distribution & 
               ${\eta}_i$ \\
  ${\hat{\rm n}}_i \equiv {\rm n}_i / N$  & 
               $\left[ 0 , 1 \right]$ & Rational & Stochastic &
               Relative values of sample points &
               $-\left( 1/N \right) \partial / \partial {\hat{\rm n}}_i$ \\
  ${\nu}_i \equiv n_i / N$ & 
               $\left[ 0 , 1 \right]$      & Real    & Langevin &
               Distribution normalized by $N$ & 
               ${\eta}_i$ \\
             &  &  &  & 
               Structure factor in large-deviations property & \\
\hline
\end{tabular}
\parbox{\textwidth}{
\caption{
  Notations for particle number in different bases.  ${\rm n}_i$ are
  integer sample values of particle counts in state $i$.  $n_i$ are
  the corresponding continuous mean values of Poisson distributions
  used as a basis for collective fluctuations.  ${\hat{\rm n}}_i$ are
  integer sample values normalized by total number $N$ to isolate
  large-deviations scaling, and ${\nu}_i$ are the corresponding
  normalized mean values of Poisson distributions.  Each representation of
  population numbers is accompanied by a conjugate momentum or shift
  operator, shown in the last column.  Total population
  $N$ will be kept fixed in all examples, and its corresponding shift
  operator is therefore not used. 
  } 
}
\label{tab:social_makeup}
\end{table}
\end{widetext}

It will be important, in following the Doi-Peliti construction below,
to understand that math-italic $n_i$ are the mean values of Poisson
distributions that are used as a \emph{basis} in which to expand the
actual distribution that evolves under the stochastic process.  They
remain random variables that are sampled from the ensemble, but they
fluctuate with Langevin statistics rather than the Poisson statistics
of the integer particle counts ${\rm n}_i$.  The ensemble mean will be
denoted ${\bar{n}}_i$.  

The dynamics of a stochastic process may be represented in three ways
associated with these different bases, which contain equivalent
information and which will all be illustrated in the following
sections.  These are: 1) the transfer matrix of the discrete
stochastic process, which shifts probability among indices; 2) the
Liouville operator that acts on generating functions, shifting
probability among modes; and 3) the action functional of the
Doi-Peliti field-theoretic expansion in Poisson distributions, which
generates the covariance for Langevin statistics.  We will show that,
with a suitable choice of dynamical variables, one may skip over the
laborious task of interconverting these representations, and simply
copy the functional forms from one representation to another.  The
number variables and shift operators in Table~\ref{tab:social_makeup}
are those that substitute for one another under changes of
representation.

\section{Large-deviations scaling and the separation of scales:
  concentration of measure; the skeletons for equilibrium versus
  dynamics; nature of the local-equilibrium approximation; existence of
  a natural scale} 
\label{sec:lg_dev_scaling}

In later sections, the two-state random walk will be the microscopic
model, whose thermodynamic descriptions we seek.  In this section we
will take an even lower-level random walk in a continuum potential to
be the microscopic model, for which the two-state model is the
coarse-grained thermodynamic description.  Starting in the continuum
will allow us to relate the dual (charge/current) character of
non-equilibrium thermodynamic state variables to graphical features of
the underlying free-energy landscape.  The particular class of
continuum models we will consider -- landscapes with basins and
barriers -- also lead to the asymmetry between states and kinetics in
classical discrete stochastic processes, and to the nature of the
local-equilibrium approximation in such models.\footnote{This
  asymmetry is not inherent in the classical limit, however, nor is it
  a common feature of quantum statistical ensembles, where positions
  and momenta may have much more symmetric roles.  The continuum model
  therefore provides a point of departure for several, different
  macroscopic approximations.}

The main points of the section will be: 1) the way large-deviations
scaling isolates the fixed points of a stochastic dynamical system as
its control points; 2) the difference between the relevant sets of
fixed points for equilibrium versus non-equilibrium ensembles; and 3)
the way the static and kinetic properties of the underlying system
become encoded in the discrete representation.  We will also cover the
origin of the natural scale for a discrete stochastic process, which
is never expressed directly within the discrete model, but which can
be necessary to regulate and to understand divergences when its
stochastic behavior is analyzed.

\subsection{The free-energy landscape representation bridges scales,
  and applies to general stochastic processes satisfying detailed
  balance}

Entropy characterizes probabilities for systems that we describe
completely (called ``closed'' systems), while \emph{free energies}
characterize probabilities for systems coupled to
incompletely-described environments.  A review of the basic relations
of probability to entropy and free energy in classical equilibrium
thermodynamics is given in App.~\ref{sec:free_energies}.

In this and later sections, we will write probabilities of states, and
transition rates, in terms of \emph{Gibbs free
  energies}~\cite{Kittel:TP:80}.  The free energy representation is
not linked to any particular scale, and so provides the map from a
continuum potential for a stochastic particle to the parameters of its
discrete-state approximation.  This representation becomes
particularly powerful for extensions from single-particle motion, to
the conversion of several species of particles in fixed proportion,
which occurs in chemical reaction networks (considered in
App.~\ref{sec:multi_part_react}).  The origin of free energies need
not be mechanical; such a representation may be found for any
stochastic process that satisfies detailed balance in its stationary
state.

Free energies will also provide a way to interpret the generating
functions and functionals used in later sections.  A generating
function distorts the free energy landscape by changing the
one-particle free energies of stable states and transitions.  We will
see that the momentum variables that appear in the Freidlin-Wentzell
theory are precisely the distortions of free energy landscapes that
extract the \emph{past histories} most likely to lead to fluctuation
conditions imposed at any moment.

\subsection{Concentration of measure in the large-deviations limit,
  and fixed points on the free-energy landscape}
\label{sec:ctm_to_discrete}

Consider then the stochastic motion of a particle in the double-well
continuum potential illustrated in Fig.~\ref{fig:continuum_potential}.
Large-deviations theory for the
continuum~\cite{Graham:path_int:77,Graham:potential:84,%
  Freidlin:RPDS:98,Maier:escape:93} applies at low temperatures, and
characterizes three aspects of the probability of trapped motions and
escapes:

\begin{figure}[ht]
  \begin{center} 
  \includegraphics[scale=0.5]{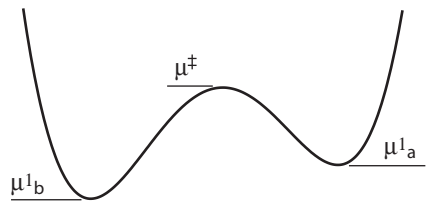}
  \caption{
  A continuum double-well potential that might underlie the two-state
  model.  The fine-grained description follows particle trajectories
  over the continuum.  The coarse-grained description is dominated by
  three chemical potentials: at the minima of the well (attracting
  fixed points) and the barrier maximum (the one-dimensional version
  of a saddle point).
    \label{fig:continuum_potential} 
  }
  \end{center}
\end{figure}

\begin{enumerate}

\item The probability for a random walker to be found away from the
  attracting fixed points (the minima of the continuum free-energy
  potential) decreases exponentially in $\beta = 1 / k_B {\tt T}$,
  where ${\tt T}$ is temperature and $k_B$ is Boltzmann's constant.
  In particular, escapes from one well to the other are exponentially
  improbable.  The leading term in the log of the escape rate is given
  by the \emph{quasipotential} of Freidlin-Wentzell
  theory~\cite{Maier:escape:93}.

\item Among escape events, the majority occur along a particular
  spatio-temporal history of thermal excitations leading uphill from
  the minima of the potential toward the interior maximum.  More
  precisely, \emph{conditional} on having observed an escape event,
  the probability that the escape trajectory deviated from this
  most-probable form decreases exponentially in $\beta = 1 / k_B {\tt
    T}$, which is a large-deviations result for histories.  The
  most-likely escape trajectory is the stationary path derived from
  the Freidlin-Wentzell dynamical system~\cite{Maier:escape:93}.  For
  one-dimensional systems, this trajectory is always the time-reverse
  of the classical diffusion trajectory from the maximum to the
  minimum.

\item The most-probable trajectory for any escape requires a specific
  finite time, which will limit the frequencies at which we can use
  coarse-grained approximations.  In one dimension, the escape time
  equals the classical diffusive relaxation time over the path whose
  time-reverse is the escape.

\end{enumerate}

The large-deviations scaling parameter in these continuum formulae is
the inverse temperature $\beta = 1 / k_B {\tt T}$.  $\beta$ will
continue to be a scaling parameter in large-deviations formulae for
the discrete approximation, but other parameters such as the number of
random walkers in the system will also enter.  

The exponential suppression of deviations from either fixed points or
stereotypical escape trajectories is the phenomenon known as
\emph{concentration of measure} for the random walk in the potential.
It is the property that leads us to make only a finite
error\footnote{The magnitude of this error decays as powers of $1 /
  k_B {\tt T}$.} even if we replace the infinitely many degrees of
freedom of a random walk on the continuum, by a (zero-dimensional!)
discrete random walk between two states, written
\begin{equation}
  b \rightleftharpoons a .  
\label{eq:min_reaction_stoich}
\end{equation}
Here $b$ and $a$ are coarse-grained labels standing for the left-hand
and right-hand wells.

\begin{figure}[ht]
  \begin{center} 
  \includegraphics[scale=0.5]{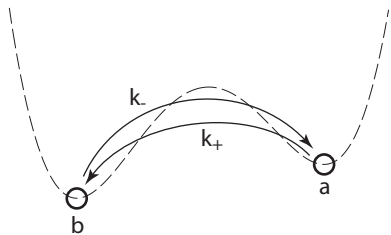}
  \caption{
  Discrete hops between well minima account for almost all structure
  in the probability distribution for the continuous random walk at
  low temperature.  The role of the barrier potential (saddle point in
  Fig.~\ref{fig:continuum_potential}) as the determiner of kinetics is
  absorbed in the transition rates $k_{\pm}$.
    \label{fig:two_state_coarse_grain} 
  }
  \end{center}
\end{figure}

The states in the discrete walk are associated approximately with the
positions of the local minima of the continuum potential, as shown in
Fig.~\ref{fig:two_state_coarse_grain}.  Each state also has a Gibbs
free energy per particle -- known in chemical thermodynamics as the
\emph{chemical potential}~\cite{Kittel:TP:80}, and written (for a
single particle) as ${\mu}^1$ -- near the value at the well minimum.

The continuum large-deviations theory for
escapes~\cite{Maier:escape:93} gives the escape rate from either well,
to leading exponential order, as a function of the chemical potentials
in the wells and at the \emph{local maximum} of the continuum
potential between the wells (indicated in
Fig.~\ref{fig:continuum_potential}), in the form
\begin{eqnarray}
  k_{+}
& = & 
  e^{
    - \beta 
    \left( {\mu}_{\ddagger} - {\mu}_a^1 \right)
  } , 
\nonumber \\
  k_{-}
& = & 
  e^{
    - \beta 
    \left( {\mu}_{\ddagger} - {\mu}_b^1 \right)
  } . 
\label{eq:rate_const_relations}
\end{eqnarray}
Note that the absolute magnitudes of probabilities which are kept to
describe transitions are generally (exponentially) smaller than
corrections to the mean properties of the fixed points, which are
omitted in the large-deviations approximation.\footnote{See
  Ref.~\cite{Coleman:AoS:85}, Ch.~7 for more on approximations of this
  kind.}  They are, however, the leading terms in the conditional
probability, and therefore the leading contribution to dynamics.

The escape rates appear in the discrete-state, continuous-time
approximation as the parameters in its \emph{master equation}, whose
form is 
\begin{eqnarray}
  \frac{\partial {\rho}_{{\rm n}_a, {\rm n}_b}}{\partial t} 
& = & 
  \left[
    k_{+} 
    \left( 
      e^{
        \partial / \partial {\rm n}_a - 
        \partial / \partial {\rm n}_b
      } - 1 
    \right)
    {\rm n}_a
  \right. 
\nonumber \\
& & 
  \mbox{} + 
  \left.
    k_{-} 
    \left( 
      e^{
        \partial / \partial {\rm n}_b - 
        \partial / \partial {\rm n}_a
      } - 1 
    \right)
    {\rm n}_b
  \right]
  {\rho}_{{\rm n}_a, {\rm n}_b} . 
\label{eq:master_equation}
\end{eqnarray}
Here, ${\rm n}_a$ and ${\rm n}_b$ are possible values for the numbers
of particles found in the right and left wells at any instant of time.
${\rho}_{{\rm n}_a, {\rm n}_b}$ is a time-dependent probability
density for an ensemble of such observations.  The master
equation~(\ref{eq:master_equation}) describes the flow of probability
among different values of the indices $\left( {\rm n}_a, {\rm n}_b
\right)$ as particles hop with rates $k_{\pm}$ per particle.  

Often the right-hand side of a master equation such as
Eq.~(\ref{eq:master_equation}) is written as a sum over all values of
$\left( {\rm n}_a, {\rm n}_b \right)$.  For stochastic processes in
which almost all change results from independent single-particle
transitions, only adjacent values $\left( {\rm n}_a \pm 1, {\rm n}_b
\mp 1 \right)$ contribute to the change in ${\rho}_{{\rm n}_a, {\rm
    n}_b}$.  Therefore we have written the shifts of indices in the
sum as the action of operators $e^{ \partial / \partial {\rm n}_b -
  \partial / \partial {\rm n}_a }$, treating the indices $\left( {\rm
  n}_a, {\rm n}_b \right)$ formally as if they lived on a continuum,
even though only values separated by integers ever appear in the time
evolution of ${\rho}_{{\rm n}_a, {\rm n}_b}$.  The operator in square
brackets in Eq.~(\ref{eq:master_equation}) is called the {\bf transfer
  matrix} of the stochastic process.  Its functional form will
re-appear throughout the subsequent analyses, in the operators that
govern the time evolution of generating functions, and as the
Hamiltonian in the Freidlin-Wentzell dynamical system.

\subsection{The discrete-state projections for equilibrium versus
  non-equilibrium systems}
\label{sec:fp_skeleton}

\begin{figure}[ht]
  \begin{center} 
  \includegraphics[scale=0.6]{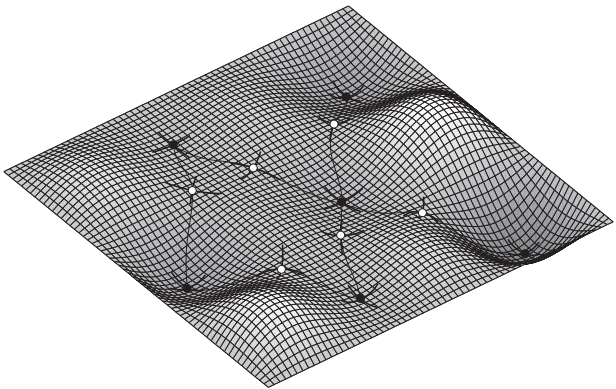}
  \caption{
  A Free energy landscape in more than one dimension, and its
  approximation by a discrete-state system.  The charge-valued state
  variables of the equilibrium ensemble live at the attracting fixed
  points (black), where the principle axes of curvature (crosses)
  point in the same direction, so the scalar curvature of the
  landscape is positive.  Current-valued state variables of the
  kinetic ensemble live on the saddle points (white), where principle
  axes of curvature point in opposite directions, and the scalar
  curvature is negative.  Trajectories mostly sit at fixed points for
  long time intervals, but when transitions happen, they are
  exponentially likely to follow specific escape paths between basins,
  indicated by narrow curved paths.
    \label{fig:fixed_saddle} 
  }
  \end{center}
\end{figure}

The net effect of concentration of measure in the continuum random
walk is to extract the parameters of its discrete approximation from
the \emph{fixed points} of the mean flow on the free-energy landscape.
The concentration toward the attracting fixed points is purely
spatial, and is easy to visualize for landscapes in any number of
dimensions.  The concentration of escape trajectories is
spatio-temporal, and is not directly illustrated in the static
potential of Fig.~\ref{fig:continuum_potential}.  Concentration of
trajectories takes on a spatial aspect for landscapes in more than one
dimension, where transitions are exponentially likely to pass through
\emph{saddle points}, as shown in Fig.~\ref{fig:fixed_saddle}.

We may now graphically characterize the difference between static and
dynamic ensembles, for random walks on landscapes with basins and
barriers.  The \emph{equilibrium} thermodynamic description for any
such system is fully specified by the chemical potentials ${\mu}^1$ at
the attracting fixed points (filled dots in
Fig.~\ref{fig:fixed_saddle}).  In this ensemble, there is no role for
barriers, and no appearance of their chemical potentials
${\mu}^{\ddagger}$.  The probabilities for state occupancy are
determined \emph{only} by their free energies, because all waiting
times for escapes are (by assumption) surpassed.

For systems away from equilibrium, transition rates become essential
to determining state occupancies, as well as transition frequencies
between pairs of states.  These rates are controlled by the saddle
points (white dots in Fig.~\ref{fig:fixed_saddle}).

With each kind of fixed point we associate a \emph{state variable} in
the thermodynamic description.  (See App.~\ref{sec:free_energies} for
discussion of the origin and role of state variables as constraints.)
The state variables of the equilibrium theory, which live on
attracting fixed points, all have the property of \emph{charges}:
their value would not change if we ran the dynamics in time-reverse.
Away from equilibrium, new state variables are introduced, which live
on the saddle points, and these have the property of \emph{currents}:
their value would change sign if we ran the dynamics in time-reverse.  

Non-equilibrium ensembles require the introduction of additional sets
of current-valued state variables~\cite{Smith:dual:05,Smith:DP:08},
which do not arise in equilibrium, and which have their origin in
properties of the saddle points of the underlying free-energy
landscape.

\subsection{The nature of the local-equilibrium approximation}

Free energy landscapes with basins and barriers lead to an extreme
asymmetry in the way charge-valued and current-valued state variables
are represented in the discrete model.  The asymmetry comes, as noted
above, from the fact that the leading probabilities for dynamics are
exponentially smaller than corrections to the properties of states
that are dropped in the large-deviations approximation.  Therefore, in
such systems, the charge-valued state variables, even in the
non-equilibrium ensemble, are nearly identical to the state variables
of equilibrium.  The one-particle chemical potentials ${\mu}^1$, and
their generalizations to concentration-dependent chemical
potentials~\cite{Kittel:TP:80}, all have the same relation to Gibbs
free energy as in equilibrium.  This property defines the
\emph{local-equilibrium approximation} for this class of models.

We note two things about the local-equilibrium approximation, to
emphasize its limits.  First, the equivalence of the non-equilibrium
charge-valued state variables with their counterparts in equilibrium
does \emph{not} extend to the entropy~\cite{Smith:dual:05}.  The
non-equilibrium entropy depends on both charges and currents, even at
a single time.  For ensembles of histories, it depends on the rate of
transitions as well as the state-occupancy statistics, as we will show
in multiple examples in Sec.~\ref{sec:NEQ_FW_large_dev} and
Sec.~\ref{sec:caliber}.  This fact will be essential to understanding
``maximum-entropy
production''~\cite{Prigogine:MT:98,Dewar:FT_MEP:03,Dewar:MEP_NESM:04,%
  Dewar:MEP_FT:05,Grinstein:MEP_error:07} as an approximation but not
a principle for non-equilibrium systems.

Second, we note that the asymmetry between states and kinetics need
not be a property of free energy landscapes if they do not have basins
and barriers.  In particular, it may not be a property of free
diffusion, and it is generally not a property of driven dissipative
quantum ensembles~\cite{Smith:dual:05}.  Therefore, if the underlying
continuum model does not have the features that create asymmetry, we
have no ground to expect that even the charge-valued state variables
in the non-equilibrium theory will closely resemble those in the
corresponding equilibrium limit.

\subsection{The implicit ``natural scale'' for a coarse-grained
  description} 
\label{sec:natural_scale}

Finally we mention an implicit limit on the use of the discrete-state
approximation.  Escapes in the continuum model are rare within the
intervals that particles spend in either basin.  However, they do
require a non-zero time, comparable to the diffusive relaxation time.
In the non-equilibrium discrete-state model, we will probe transition
probabilities with time-dependent distortions of the chemical
potentials ${\mu}^1$ and ${\mu}^{\ddagger}$.  The model will permit
these probes to have arbitrarily high-frequency time-dependence.
However, we will see when we consider path entropies in
Sec.~\ref{sec:caliber} that such sources lead to divergences in
individual entropy terms that should remain finite to be meaningful.

The solution to the problem of divergences is to recognize that the
discrete model has a \emph{natural scale}~\cite{Polchinski:RGEL:84},
which is an upper bound on the frequencies that may sensibly appear in
probes of the theory.  The natural scale is the diffusive relaxation
frequency in the continuum model.  For probes with higher frequencies,
the constants $k_{\pm}$ describing transition rates no longer retain
their meaning or values.  They were defined as lumped-parameter
representations of escape paths, in a potential which was assumed to
be fixed during the period of the escape.  Faster probes ``melt'' the
discrete approximation, and require that the description revert to the
underlying continuum.

\section{Basic quantities introduced within the equilibrium ensemble:
  generating functions and the expressions of large-deviations scaling}
\label{sec:EQ_large_dev}

We now begin the analysis of the equilibrium distribution for the
two-state system.  Since the entire equilibrium distribution may be
written down analytically (it is a binomial distribution), the purpose
of this ``analysis'' is to introduce the key quantities expressing
large-deviations scaling, along with systematic ways to compute them
using generating functions.  The methods and the asymptotics will
generalize immediately to time-dependent systems that are not exactly
solvable, or at least very inconvenient to write in closed form. 

The large-deviation result we will derive is that, for any
apportionment $\left( {\rm n}_a, {\rm n}_b \right)$ of $N$ particles
to the two states, the log-probability of this apportionment in the
equilibrium distribution is the difference of the joint entropy from
its maximizing value.  A more extensive taxonomy of large-deviation
results for equilibrium ensembles is given in
Ref.~\cite{Ellis:ELDSM:85}.  Non-maximizing values of $\left( {\rm
  n}_a, {\rm n}_b \right)$ are termed \emph{fluctuations}, and the
relation between entropies and log-probability for sample values is
therefore called a fluctuation theorem.

We will isolate this leading-exponential term in the log-probability
by using the cumulant-generating function to shift the distribution,
effectively projecting onto a sub-distribution within which $\left(
{\rm n}_a, {\rm n}_b \right)$ is the most-likely value.  The
sub-distribution, when normalized, would be the equilibrium
distribution in a two-state system with a shifted chemical potential;
the ratio between the generating function and the normalized
distribution measures the overlap of the original distribution with
the one appropriate to $\left( {\rm n}_a, {\rm n}_b \right)$.

It will be the Legendre transform of this cumulant-generating function
that gives the fluctuation probability to observe $\left( {\rm n}_a,
{\rm n}_b \right)$ in the original equilibrium distribution, and
expresses this probability as a difference of entropies.  The Legendre
transform of the cumulant-generating function is known, in some
domains of quantum field theory, as the \emph{quantum effective
  action}, and we will adapt that term here, calling it the
``stochastic effective action''.  (For more context and the relation
to literature, see App.~\ref{sec:literature}.)  Though it is only a
difference of static entropies in the equilibrium ensemble, the
stochastic effective action will become dynamical in ensembles of
histories, where it will be the strict counterpart to the quantum
effective action.

\subsection{The equilibrium distribution of the two-state stochastic
  process} 
\label{sec:genfunc_equil}

At an equilibrium steady state the solution to the master
equation~(\ref{eq:master_equation}) is the binomial distribution
\begin{eqnarray}
  {\rho}_{{\rm n}_a, {\rm n}_b}^{\mbox{\scriptsize eq}} 
& = & 
  \frac{
    k_{-}^{{\rm n}_a}
    k_{+}^{{\rm n}_b}
  }{
    {\left( k_{+} + k_{-} \right)}^N
  }
  \left(
    \begin{array}{c}
      N \\ {\rm n}_a 
    \end{array}
  \right)
\nonumber \\
& = & 
  {\bar{\nu}}_a^{{\rm n}_a}
  {\bar{\nu}}_b^{{\rm n}_b}
  \left(
    \begin{array}{c}
      N \\ {\rm n}_a 
    \end{array}
  \right) .  
\label{eq:rho_eq_binomial}
\end{eqnarray}
Total particle number $N = {\rm n}_a + {\rm n}_b$ is conserved by all
terms in the transfer matrix of Eq.~(\ref{eq:master_equation}).  The
equilibrium occupation fractions are
\begin{eqnarray}
  {\bar{\nu}}_a 
& \equiv & 
  \frac{
    k_{-}
  }{
    k_{+} + k_{-}
  } = 
  \frac{
    e^{-\beta {\mu}_a^1}
  }{
    Z_1
  }
\nonumber \\
  {\bar{\nu}}_b 
& \equiv & 
  \frac{
    k_{+}
  }{
    k_{+} + k_{-}
  } = 
  \frac{
    e^{-\beta {\mu}_b^1}
  }{
    Z_1
  } , 
\label{eq:eq_fracs_def}
\end{eqnarray}
in which 
\begin{equation}
  Z_1 \equiv 
  e^{-\beta {\mu}_a^1} + 
  e^{-\beta {\mu}_b^1} 
\label{eq:Z_1_def}
\end{equation}
is the called the \emph{partition function}~\cite{Kittel:TP:80} for a
one-particle ensemble in this two-state system.  

The Gibbs free energies for non-interacting particles scale linearly
(that is, they are ``extensive'') in particle
number~\cite{Kittel:TP:80}, so the structural terms in the
large-deviations formulae for fluctuation probabilities will be
functions of the ratios
\begin{eqnarray}
  {\hat{\rm n}}_a 
& \equiv & 
  \frac{{\rm n}_a}{N}
\nonumber \\
  {\hat{\rm n}}_b 
& \equiv & 
  \frac{{\rm n}_b}{N} .  
\label{eq:inst_fracs_def}
\end{eqnarray}
In both steady-state and time-dependent probability distributions,
Roman ${\rm n}_a$, ${\rm n}_b$, ${\hat{\rm n}}_a$, ${\hat{\rm n}}_b$
will be used for discrete particle-number indices, respectively
un-normalized or normalized by $N$.  When the description of
distributions is transferred to the generating function, the
corresponding continuous indices will be $n_a$, $n_b$ for absolute
number, and ${\nu}_a$, ${\nu}_b$ for relative numbers corresponding to
the definitions~(\ref{eq:inst_fracs_def}).

For dynamical as for static systems, it is convenient to study
open-system properties by considering an open system and its
environment to be components in a larger closed system.  Here the
closed system will be defined by $N$ as a parameter, and we introduce
the stochastic variable under the reaction ${\rm n} \equiv \left( {\rm
  n}_b - {\rm n}_a \right) / 2$, so that
\begin{eqnarray}
  {\rm n}_a 
& \equiv & 
  \frac{N}{2} - {\rm n}
\nonumber \\
  {\rm n}_b 
& \equiv & 
  \frac{N}{2} + {\rm n} . 
\label{eq:n_ns_np_def}
\end{eqnarray}
When only ${\rm n}$ is denoted explicitly, the distribution
${\rho}_{{\rm n}_a, {\rm n}_b}$ will be indexed ${\rho}_{\rm n}$.  The
relative particle number asymmetry is likewise defined as ${\hat{\rm
    n}} \equiv \left( {\hat{\rm n}}_b - {\hat{\rm n}}_a \right) / 2$.
Its counterpart in continuous variables will be denoted $\nu$.
Equilibrium values for all numbers will be indicated with overbars.

The equilibrium distribution~(\ref{eq:rho_eq_binomial}) may be cast in
a variety of instructive forms.  Using the
relation~(\ref{eq:rate_const_relations}) of the rate constants to
one-particle chemical potentials, and Stirling's formula for the
logarithms of factorials, the following expressions re equivalent: 
\begin{eqnarray}
  {\rho}_{{\rm n}_a, {\rm n}_b}^{\mbox{\scriptsize eq}} 
& \approx & 
  \frac{
    1 
  }{
    \sqrt{2\pi N {\hat{\rm n}}_a {\hat{\rm n}}_b}
  }
  e^{
    {\rm n}_a 
    \left( 
      \log {\bar{\hat{\rm n}}}_a - 
      \log {\hat{\rm n}}_a
    \right) + 
    {\rm n}_b 
    \left( 
      \log {\bar{\hat{\rm n}}}_b - 
      \log {\hat{\rm n}}_b
    \right)
  }
\nonumber \\
& \equiv &
  \frac{
    1 
  }{
    \sqrt{2\pi N {\hat{\rm n}}_a {\hat{\rm n}}_b}
  }
  e^{
    -N 
    D \left( \hat{\rm n} \parallel {\bar{\hat{\rm n}}} \right)
  } 
\nonumber \\
& = & 
  \frac{
    1 
  }{
    \sqrt{2\pi N {\hat{\rm n}}_a {\hat{\rm n}}_b}
  }
  \frac{1}{Z_1^N}
  e^{
    -\beta {\rm n}_a
    \left( {\mu}_a^1 + \tau \log {\hat{\rm n}}_a \right) - 
    \beta {\rm n}_b
    \left( {\mu}_b^1 + \tau \log {\hat{\rm n}}_b \right)
  } 
\nonumber \\
& = & 
  \frac{
    1 
  }{
    \sqrt{2\pi N {\hat{\rm n}}_a {\hat{\rm n}}_b}
  }
  e^{
    N \log \left( N / Z_1 \right)
  }
  e^{
    -\beta {\rm n}_a {\mu}_a - 
    \beta {\rm n}_b {\mu}_b
  }   
\nonumber \\
& = & 
  e^{
    N \log \left( N / Z_1 \right)
  }
  \sqrt{
    \frac{
      {\rm n}_a + {\rm n}_b
    }{
      2\pi {\rm n}_a {\rm n}_b
    }
  }
  e^{
    -\beta \left( G_a + G_b \right)
  }   
\label{eq:rho_eq_binomial_approx}
\end{eqnarray}
In the second line, $D \! \left( \hat{\rm n} \parallel {\bar{\hat{\rm
      n}}} \right)$ is the Kullback-Leibler
divergence~\cite{Cover:EIT:91}, in which $\hat{\rm n}$ and
$\bar{\hat{\rm n}}$ stand for the distributions with coefficients
$\left( {\hat{\rm n}}_a , {\hat{\rm n}}_b \right)$, $\left(
{\bar{\hat{\rm n}}}_a, {\bar{\hat{\rm n}}}_b \right)$ respectively.  
The ${\rm n}_a$- and ${\rm n}_b$-particle chemical potentials add
concentration corrections to the entropy in the one-particle
potentials, as 
\begin{eqnarray}
  {\mu}_a 
& = & 
  {\mu}_a^1 + 
  k_B {\tt T} \log {\rm n}_a 
\nonumber \\
  {\mu}_b 
& = & 
  {\mu}_b^1 + 
  k_B {\tt T} \log {\rm n}_b . 
\label{eq:n_part_chem_pots}
\end{eqnarray}
By comparing the second with the last-two lines of
Eq.~(\ref{eq:rho_eq_binomial_approx}), we see that the minimum of the
Gibbs free energies of the subsystems over ${\rm n}$ is
\begin{equation}
  \beta
  \min_{\rm n} \! 
  \left[ 
    G_a + G_b 
  \right] \equiv 
  \beta
  \min \! 
  \left[ 
    {\rm n}_a {\mu}_a + {\rm n}_b {\mu}_b
  \right] = 
  N \log \frac{N}{Z_1} . 
\label{eq:Gibbs_min_from_onepart}
\end{equation}
The one-particle minimum expressed in terms of fractional occupancies,
which is the descaled version at any $N$, gives an $N$-independent
relation between the chemical potential per particle, and the
one-particle partition function, 
\begin{eqnarray}
\lefteqn{
  \min_{\hat{\rm n}} \! 
  \left[ 
    {\hat{\rm n}}_a \left( {\mu}_a^1 + 
    k_B {\tt T} \log {\hat{\rm n}}_a \right) + 
    {\hat{\rm n}}_b \left( {\mu}_b^1 + 
    k_B {\tt T} \log {\hat{\rm n}}_b \right) 
  \right] 
} & & 
\nonumber \\
& = & 
  - k_B {\tt T} \log Z_1 . 
  \phantom{
    k_B {\tt T} \log Z_1
    k_B {\tt T} \log Z_1
    k_B {\tt T} \log Z_1
  }
\label{eq:Gibbs_min_onepart}
\end{eqnarray}
These are the standard relations for ideal gases or dilute solutions.

\subsubsection{The aggregation of state variables and the fluctuation
  probability} 

The local-equilibrium approximation for the two-state system allows us
to approximate the log-probabilities for non-equilibrium
configurations of $\left( {\rm n}_a , {\rm n}_b \right)$ by the sum of
free energies for the individual wells.  At the minimizing value of
$\left( {\rm n}_a , {\rm n}_b \right)$ for this sum, the equilibrium
free energy for the composite system is attained.  We may therefore
express the minimum joint free energy per particle, using
Eq.~(\ref{eq:Gibbs_min_from_onepart}), as a definition for the
single-particle chemical potential for the equilibrated system, in a
form equivalent to to Eq.~(\ref{eq:n_part_chem_pots}):
\begin{eqnarray}
  \frac{1}{N}
  \min \! 
  \left[ 
    G_a + G_b 
  \right] \equiv 
  {\mu}_{a \cup b} 
& = & 
  - k_B {\tt T} \log Z_1 + 
  k_B {\tt T} \log N 
\nonumber \\
& \equiv &
  {\mu}_{a \cup b}^1 + k_B {\tt T} \log N . 
\label{eq:Gibbs_min_of_N}
\end{eqnarray}
Here ${\mu}_{a \cup b}^1$ is the whole-system counterpart to the
one-particle chemical potentials ${\mu}_a^1$ and ${\mu}_b^1$ for
subsystems in the local-equilibrium approximation.

The density values in the equilibrium
distribution~(\ref{eq:rho_eq_binomial_approx}) may then be written as
exponentials of the $\left( \mbox{system} + \mbox{environment}
\right)$ entropies of Eq.~(\ref{eq:beta_Gibbs_S}), as
\begin{eqnarray}
  {\rho}_{{\rm n}_a, {\rm n}_b}^{\mbox{\scriptsize eq}} 
& \approx & 
  e^{
    \beta N {\mu}_{a \cup b}
  }
  \sqrt{
    \frac{
      {\rm n}_a + {\rm n}_b
    }{
      2\pi {\rm n}_a {\rm n}_b
    }
  }
  e^{
    -\beta \left( G_a + G_b \right)
  }   
\nonumber \\
& \approx & 
  e^{
    \beta G_{a \cup b}
  }
  \sqrt{
    \frac{
      {\rm n}_a + {\rm n}_b
    }{
      2\pi {\rm n}_a {\rm n}_b
    }
  }
  e^{
    -\beta \left( G_a + G_b \right)
  } . 
\label{eq:rho_eq_binomial_from_Gibbs}
\end{eqnarray}
As explained in App.~\ref{sec:free_energies}, all three Gibbs free
energies are functions of intensive $k_B {\tt T}$ and $p$, and of
extensive arguments ${\rm n}_a$, ${\rm n}_b$, and $N$.  Since total
energy is controlled by the environment temperature, it is the entropy
components of these $G$ values for the subsystems, as functions of
${\rm n}_a$ and ${\rm n}_b$, which control the difference of $G_a +
G_b$ from $G_{a \cup b}$.

\subsubsection{Entropies of equilibrium and residual fluctuations}

The leading-exponential approximation of large-deviations scaling
separates extensive entropies of the subsystems, which were defined
into the parameters of the two-state stochastic process from
coarse-graining the continuum model, from entropies due to chemical
fluctuation, which are sub-extensive.  To see this, following any
standard thermodynamics text~\cite{Kittel:TP:80}, we write any of the
one-particle chemical potentials ${\mu}^1 \equiv h^1 - k_B {\tt T}
s^1$, in which $h^1$ is the specific enthalpy and $s^1$ is the
specific entropy.  This decomposition yields a relation between
subsystem and whole system free energies at equilibrium, which is
\begin{eqnarray}
  h^1_{a \cup b} 
& = & 
  {\bar{\hat{\rm n}}}_a
  h^1_a + 
  {\bar{\hat{\rm n}}}_b
  h^1_b 
\nonumber \\
  s^1_{a \cup b} 
& = & 
  {\bar{\hat{\rm n}}}_a
  s^1_a + 
  {\bar{\hat{\rm n}}}_b
  s^1_b - 
  {\bar{\hat{\rm n}}}_a \log {\bar{\hat{\rm n}}}_a - 
  {\bar{\hat{\rm n}}}_b \log {\bar{\hat{\rm n}}}_b . 
\label{eq:free_energies_chainrule}
\end{eqnarray}
The free energy in the thermodynamic limit is then given by
\begin{equation}
  G_{a \cup b} = 
  N 
  \left( 
    h^1_{a \cup b} - 
    k_B {\tt T} s^1_{a \cup b} + 
    k_B {\tt T} \log N
  \right) , 
\label{eq:G_both_limit}
\end{equation}
which extensive in particle number, if we think of the $\log N$ term
as being set by pressure.  In comparison, the Shannon entropy of the
full distribution over fluctuations contains a term from the
normalization of the exponential in
Eq.~(\ref{eq:rho_eq_binomial_from_Gibbs}), due to its width, 
\begin{equation}
  - \sum_{\rm n} 
  {\rho}_{\rm n}^{\mbox{\scriptsize eq}} 
  \log {\rho}_{\rm n}^{\mbox{\scriptsize eq}} 
  \approx 
  \frac{1}{2}
  \left[ 
    1 + \log 
    \left( 
      2 \pi N {\bar{\hat{\rm n}}}_a {\bar{\hat{\rm n}}}_b 
    \right)
  \right] .  
\label{eq:resid_fluct_ents}
\end{equation}
This correction, being only logarithmic, is sub-extensive in $N$. 

\subsection{Generating functions and the stochastic effective action}
\label{sec:eq_gen_fn_SEA}

A \emph{moment-generating function} -- or ``ordinary power-series
generating function''~\cite{Wilf:gen_fun:06} -- for a distribution
indexed on the two numbers ${\rm n}_a$ and ${\rm n}_b$ has two complex
arguments, and is written
\begin{equation}
  \psi \! \left( z_a , z_b \right) \equiv 
  \sum_{{\rm n}_a, {\rm n}_b}
  z_a^{{\rm n}_a} z_b^{{\rm n}_b} 
  {\rho}_{{\rm n}_a, {\rm n}_b} . 
\label{eq:gen_func_general_def}
\end{equation}
To study the properties of ${\rho}_{\rm n}$ at fixed $N$, recognizing
that $z_a^{{\rm n}_a} z_b^{{\rm n}_a} = {\left( z_a z_b \right)}^{N/2}
{\left( z_b / z_a \right)}^{\rm n}$, we may set $z_a z_b \equiv 1$ and
denote $z_b / z_a \equiv z$.  At equilibrium we will be interested
only in the one-argument function of $z$.  However, as we pass to
dynamical descriptions, it will remain convenient in some cases to
retain both variables $z_a$ and $z_b$, even if they are applied to a
distribution with support on only one value of $N$.

The weight factors $z^{\rm n}$ have an effect similar to shift in the
subsystem chemical potentials, which will recur repeatedly in our
analysis.  Therefore we denote $z = e^{q}$, and write the one-variable
generating function as 
\begin{eqnarray}
  \psi \! \left( z \right) 
& = & 
  \sum_{\rm n} 
  z^{\rm n} {\rho}_{\rm n}^{\mbox{\scriptsize eq}}
\nonumber \\
& = & 
  \sum_{\rm n} 
  {\left( {\bar{\hat{\rm n}}}_a / \sqrt{z} \right)}^{{\rm n}_a}
  {\left( {\bar{\hat{\rm n}}}_b \sqrt{z} \right)}^{{\rm n}_b}
  \left(
    \begin{array}{c}
      N \\ {\rm n}_a 
    \end{array}
  \right)
\nonumber \\
& = & 
  {
    \left(
      \frac{Z_1^{\left( q \right)}}{Z_1}
    \right)
  }^N
   \sum_{\rm n} 
  {\tilde{\hat{\rm n}}}_a^{{\rm n}_a}
  {\tilde{\hat{\rm n}}}_b^{{\rm n}_b}
  \left(
    \begin{array}{c}
      N \\ {\rm n}_a 
    \end{array}
  \right) .  
\label{eq:gen_func_reduced_def}
\end{eqnarray}
Here new normalized fractions in the presence of $q$ -- which will be
referred to as a \emph{source} -- are defined by
\begin{eqnarray}
  {\tilde{\hat{\rm n}}}_a
& \equiv & 
  \frac{
    {\bar{\hat{\rm n}}}_a e^{-q/2}
  }{
    {\bar{\hat{\rm n}}}_a e^{-q/2} + 
    {\bar{\hat{\rm n}}}_b e^{q/2}
  } = 
  \frac{
    e^{
      -\left( \beta {\mu}_a^1 + q/2 \right)
    }
  }{
    Z_1^{\left( q \right)}
  }
\nonumber \\
  {\tilde{\hat{\rm n}}}_b
& \equiv & 
  \frac{
    {\bar{\hat{\rm n}}}_b e^{q/2}
  }{
    {\bar{\hat{\rm n}}}_a e^{-q/2} + 
    {\bar{\hat{\rm n}}}_b e^{q/2}
  } = 
  \frac{
    e^{
      -\left( \beta {\mu}_b^1 - q/2 \right)
    }
  }{
    Z_1^{\left( q \right)}
  } , 
\label{eq:normed_fracs_q}
\end{eqnarray}
and the associated one-particle partition function with source $q$ is
\begin{equation}
  Z_1^{\left( q \right)} \equiv 
  e^{-\left( \beta {\mu}_a^1 + q/2 \right)} + 
  e^{-\left( \beta {\mu}_b^1 - q/2 \right)} . 
\label{eq:Z_1_q_def}
\end{equation}

The sum in the third line of Eq.~(\ref{eq:gen_func_reduced_def})
evaluates to unity, as for any normalized binomial, but it is
instructive to use what was learned in forming
Eq.~(\ref{eq:rho_eq_binomial_from_Gibbs}) to recast the sum and
prefactor together in terms of a normalized distribution and a
``penalty'' term, as
\begin{equation}
  \psi \! \left( z \right) \approx  
  e^{
    \beta G_{a \cup b} - 
    \beta G_{a \cup b}^{\left( q \right)}
  } 
  \sum_{\rm n} 
  e^{
    \beta G_{a \cup b}^{\left( q \right)}
  }
  \sqrt{
    \frac{
      {\rm n}_a + {\rm n}_b
    }{
      2\pi {\rm n}_a {\rm n}_b
    }
  }
  e^{
    -\beta 
    \left( 
      G_a^{\left( q \right)} + G_b^{\left( q \right)} 
    \right)
  }   
\label{eq:rho_eq_binomial_source}
\end{equation}
In Eq.~(\ref{eq:rho_eq_binomial_source}) the subsystem free energies
are defined at any ${\rm n}_a$, ${\rm n}_b$ as
\begin{eqnarray}
  G_a^{\left( q \right)} 
& = & 
  G_a + {\rm n}_a k_B {\tt T} q/2 
\nonumber \\
  G_b^{\left( q \right)} 
& = & 
  G_b - {\rm n}_b k_B {\tt T} q/2 . 
\label{eq:subsys_G_with_q}
\end{eqnarray}
We wish to introduce $G_{a \cup b}^{\left( q \right)}$ as the minimum
of $G_a^{\left( q \right)} + G_b^{\left( q \right)}$ over ${\rm n}$,
as before.  However, for discrete ${\rm n}$, this gives a
discontinuous function of $q$.  The thermodynamic usage of $G_{a \cup
  b}^{\left( q \right)}$ is as a macroscopic function of its intensive
state variables, and therefore it will save intermediate steps and
notation simply to define $G_{a \cup b}^{\left( q \right)}$ as the
minimum over ${\rm n}$ treated as a continuous variable, whose
minimizing argument will then be a continuous function of $q$.

For such binomial distributions or their multinomial generalizations,
sources of the form $k_B {\tt T} q = q / \beta$ behave as shifts in
chemical potential, in this case split evenly between subsystems $a$
and $b$.  The ``penalty'' term is expressed as a function of $\left(
G_{a \cup b} - G_{a \cup b}^{\left( q \right)} \right)$: the minimum
chemical work needed to convert the system with potential difference
${\mu}_b - {\mu}_a$ and equilibrium $\bar{\hat{\rm n}}$ to one with
potential difference ${\mu}_b - {\mu}_a - q / \beta$ and an
equilibrium ${\bar{\hat{\rm n}}}^{\left( q \right)}$ determined by the
new potentials.  $\beta \left( G_{a \cup b} - G_{a \cup b}^{\left( q
  \right)} \right)$ is also the log-probability to obtain the shifted
distribution from the original through the set of weights $z^{\rm n}$.

The penalty function is referred to as the \emph{cumulant-generating
  function} and denoted $\Gamma \! \left( q \right)$.  It is defined
from the moment-generating function $\psi$ by the relation
\begin{equation}
  \psi \! \left( z \right) \equiv   
  e^{- \Gamma \left( \log z \right)} = 
  e^{- \Gamma \left( q \right)} . 
\label{eq:psi_to_Gamma}
\end{equation}
From the definition~(\ref{eq:gen_func_reduced_def}) of the
moment-generating function it follows that
\begin{equation}
  \frac{
    d \log \psi \! \left( z \right)
  }{
    d \log z 
  } = 
  - \frac{
    d \Gamma \! \left( q \right)
  }{
    dq 
  } = 
  {
    \left< {\rm n} \right>
  }^{\left( q \right)} \equiv 
  n^{\left( q \right)} , 
\label{eq:grad_Gamma_number}
\end{equation}
in which we introduce the continuous counterpart $n^{\left( q
  \right)}$ as both the gradient of $\Gamma$ and the expectation of
${\rm n}$ under the equilibrium distribution shifted by $z$.

From Eq.~(\ref{eq:rho_eq_binomial_source}), the cumulant-generating
function is just the Gibbs free energy difference 
\begin{equation}
  \Gamma \! \left( q \right) = 
  \beta G_{a \cup b}^{\left( q \right)} - 
  \beta G_{a \cup b} . 
\label{eq:Gamma_to_Gs}
\end{equation}
Recasting Eq.~(\ref{eq:grad_Gamma_number}), 
\begin{equation}
  - \frac{
    d\Gamma \! \left( q \right)
  }{
    dq 
  } = 
  \frac{
    d G_{a \cup b}^{\left( q \right)}
  }{
    d \left( - k_B {\tt T} q \right)
  } = 
  n^{\left( q \right)} , 
\label{eq:Gs_to_ns_via_mus}
\end{equation}
we recover the usual relation from equilibrium
thermodynamics~\cite{Kittel:TP:80}, that the gradient of the Gibbs
free energy with respect to the chemical potential is the particle
number.

For reference in later sections, we may compute closed forms for the
various quantities.  Define $\bar{\mu} \equiv {\mu}_b^1 - {\mu}_a^1$,
dual to the relative particle-number difference $\bar{\hat{\rm n}}$.  Then
\begin{equation}
  n^{\left( q \right)} = 
  \frac{N}{2}
  \tanh 
  \frac{1}{2}
  \left( q - \beta \bar{\mu} \right) , 
\label{eq:n_q_compute}
\end{equation}
and 
\begin{equation}
  \Gamma \! \left( q \right) = 
  N 
  \left[
    \log \cosh 
    \frac{\beta \bar{\mu}}{2} - 
    \log \cosh 
    \frac{1}{2}
    \left( q - \beta \bar{\mu} \right) 
  \right] . 
\label{eq:Gamma_q_compute}
\end{equation}

In continuum field theories with many particles and nonlinear
interactions among them, it is often necessary to approximate the
moment- and cumulant-generating functions by series expansions in the
variance of the Gaussian approximation to fluctuations.  In such an
expansion, The gradient of $\Gamma$ yields all of the \emph{connected}
graphs giving $n^{\left( q \right)}$, while the gradient of $\psi$
gives a sum over all graphs (see Ref.~\cite{Weinberg:QTF_II:96}
Ch.~16).

\subsubsection{The stochastic effective action for single-time
  fluctuations} 

App~\ref{sec:free_energies} reviews the fact that the Gibbs free
energy is a Legendre transform of the entropy.  Thus, the entropy is
the converse Legendre transform of the Gibbs free energy.  We may
therefore expect that by Legendre transforming $\Gamma \! \left( q
\right)$, we will arrive at a direct expression for the entropy
differences that govern internal fluctuations of particle number,
without reference to external temperature or pressure.  

The Legendre transform of the cumulant generating function $\Gamma$ is
known as an \emph{effective action}~\cite{Weinberg:QTF_II:96}.  When
it is computed for the single-time binomial distribution, it hardly
seems to justify its name, if we expect an action in the sense of
Hamiltonian dynamics.  However, it will become just such an action in
the time-dependent ensemble.  Here we introduce the Legendre transform
for its statistical meaning, and later we introduce the dynamical
version. 

The definition is
\begin{equation}
  S^{\mbox{\scriptsize eff}} \! \left( n \right) \equiv 
  {
    \left. 
      \left[ 
        \Gamma \! \left( q \right) + 
        n q 
      \right]
    \right| 
  }_{
    q = q^{\left( n \right)}
  } , 
\label{eq:S_eff_def_onetime}
\end{equation}
in which $q^{\left( n \right)}$ denotes the inverse function of
$n^{\left( q \right)}$ from
Eq.~(\ref{eq:Gs_to_ns_via_mus}).~\footnote{For the equilibrium
  distribution, and indeed all other distributions computed below,
  this inverse will be unique.  For problems in which it is not
  unique, such as occur along the co-existence curves of first-order
  phase transitions, the Legendre transform is replaced by the
  Legendre-Fenchel
  transform~\cite{Ellis:ELDSM:85,Touchette:large_dev:08}.}  From the
gradient relation~(\ref{eq:grad_Gamma_number}) it follows that
\begin{equation}
  \frac{
    d S^{\mbox{\scriptsize eff}} \! \left( n \right)
  }{
    dn 
  } = 
  q^{\left( n \right)} .
\label{eq:vary_S_eff_q}
\end{equation}
Therefore $S^{\mbox{\scriptsize eff}}$ must vanish at the mean value
of ${\rm n}$ in the distribution $\rho$ for which it is computed,
because no source (\emph{i.e.}, $q \! \left( n \right) = 0$) is
required to yield the equilibrium value as the mean.

Addition of the term $n^{\left( q \right)} q$ to $\Gamma$ in
Eq.~(\ref{eq:S_eff_def_onetime}) cancels only the explicit $q / \beta$
term in the shifted free energy $G_{a \cup b}^{\left( q \right)}$,
leaving the particle assignments $n_a^{\left( q \right)},n_b^{\left( q
  \right)}$ unchanged.  The terms that remain are precisely those in
the fluctuation theorem: they are the subsystem free energies
evaluated at shifted $n$s.  The effective action may be written
\begin{equation}
  S^{\mbox{\scriptsize eff}} \! \left( n \right) = 
  \beta G_a \! \left( n_a \right) + 
  \beta G_b \! \left( n_b \right) - 
  \beta G_{a \cup b} \! \left( N \right) . 
\label{eq:S_eff_G_diffs}
\end{equation}
In Eq.~(\ref{eq:S_eff_G_diffs}) the free energies $G$ are now regarded
as functions of the continuous-valued $n_a$ and $n_b$ rather than the
discrete indices ${\rm n}_a$ and ${\rm n}_b$ on which the distribution
is defined.  

Referring back to the distribution over fluctuations in the original
${\rho}_{n_s, n_p}^{\mbox{\scriptsize eq}}$ of
Eq.~(\ref{eq:rho_eq_binomial_from_Gibbs}), we see that $\exp \! \left(
- S^{\mbox{\scriptsize eff}} \right)$ directly extracts the leading
exponential dependence of the fluctuation probability.  It omits
sub-extensive normalization factors related to the width of the
distribution.  Therefore the effective action directly expresses the
large-deviations scaling for fluctuation probabilities.  The
log-probability approximated by Eq.~(\ref{eq:S_eff_G_diffs}) scales
linearly in $N$, and its structure is determined by the remaining
$N$-invariant fractions ${\nu}_a$ and ${\nu}_b$. 

For reference in comparison to later expressions for the effective
actions of fluctuating histories, the effective action for the
equilibrium distribution evaluates to
\begin{eqnarray}
  S^{\mbox{\scriptsize eff}} \! \left( n \right) 
& = & 
  N 
  \left[
    \log \cosh 
    \frac{\beta \bar{\mu}}{2} - 
    \log \cosh 
    \frac{1}{2}
    \left( q^{\left( n \right)} - \beta \bar{\mu} \right) 
  \right.
\nonumber \\
& & 
  \mbox{} + 
  \left. 
    \frac{q^{\left( n \right)}}{2}
    \tanh 
    \frac{1}{2}
    \left( q^{\left( n \right)} - \beta \bar{\mu} \right) 
  \right] , 
\label{eq:S_eff_n_compute}
\end{eqnarray}
in which $q^{\left( n \right)}$ is evaluated by inversion of
Eq.~(\ref{eq:n_q_compute}).  

\section{Ensemble of histories I: the master equation for number;
  Doi-Peliti construction of the generating functional; the
  dynamical-system representation of Freidlin and Wentzell}
\label{sec:NEQ_FW_large_dev}

We now construct a second generating function and ensemble, for the
time-dependent distribution $\rho$ evolving under the master
equation~(\ref{eq:master_equation}), rather than merely for its
equilibrium steady state.  The equilibrium fluctuation results of
Sec.~\ref{sec:EQ_large_dev} will reappear within this larger
construction, as probabilities for histories conditioned on their
behavior at a single instant but on no other information.  Here,
however, the single-time fluctuation probabilities will no longer be
computed as primitive quantities.  Instead, they will arise as
integrals of log-probability along the entire histories which are
most-likely, conditioned on achieving a particular deviation at a
single time.

In keeping with the shift from states to histories, the
cumulant-generating function and effective action in this section will
be function\emph{als}, whose arguments are sequences of occupation
numbers $\left( {\rm n}_a, {\rm n}_b \right)$.  The description
derived from the master equation will still only count particle
numbers, and will not resolve identities.  Unlike the equilibrium
constructions, the path-space large-deviations formulae will not allow
us to identify the natural path-entropy functions, for which we will
require the caliber formulation of Sec.~\ref{sec:caliber}.  However,
this section will introduce the remarkable Freidlin-Wentzell
formulation of large-deviations theory as a Hamiltonian dynamical
system, in which the stochastic effective action will become a genuine
Lagrange-Hamilton action functional.\footnote{For an introduction to
  action functionals and Hamiltonian dynamics, see
  Ref.~\cite{Goldstein:ClassMech:01}.}

\subsection{Liouville time evolution and the coherent-state expansion
  for its quadrature}
\label{sec:liouville_forms}

The reason to study generating functions at equilibrium is that the
moments of the equilibrium distribution may be more relevant and
robust descriptions of fluctuations than individual probability values
${\rho}_{{\rm n}_a, {\rm n}_b}$.  Similarly, the reason to use
generating functions to study dynamics is that the basis of
single-particle hops is often less informative than a basis in the
collective motions of many particles.  

The master equation formalism acts on probabilities in the basis of
single-particle hops.  The counterpart, which acts on the
moment-generating function, is known as a \emph{Liouville equation}.
When specified in complete detail, the Liouville equation has the same
information as its corresponding master equation.  However, low-order
approximations to the Liouville form will often contain more of the
total information than similarly reduced approximations in the basis
of single-particle hops.  Such approximations are sometimes called the
\emph{hydrodynamic limit}~\cite{Gaspard:rev_entropy:04}.  Therefore
the Liouville form is the starting point for methods of Gaussian
integration, Langevin equations, or other commonly used approximation
schemes.

\subsubsection{The master equation expressed as a function of
  mass-action rates}

The two-state master equation~(\ref{eq:master_equation}) is an
instance of a wider class of equations describing single-particle
exchanges.  A notation for the more general class is introduced here,
both because it cleanly divides the particle-hopping terms associated
with stochasticity from the probabilities which coincide with
mass-action rates, and because it immediately generalizes to more
complex reaction stoichiometries, such as those required by chemical
reaction networks.  The master equation for the two-state system is
written in this general form as
\begin{equation}
  \frac{\partial {\rho}_{\vec{{\rm n}}}}{\partial t} = 
  \sum_{i,j \in \left\{ a, b \right\}} 
  \left( 
    e^{
      \partial / \partial {\rm n}_i - 
      \partial / \partial {\rm n}_j
    } - 1 
  \right)
  r_{ji}
  {\rho}_{\vec{{\rm n}}} . 
\label{eq:general_exchange_ME}
\end{equation}
Here $\vec{\rm n}$ is a vector whose components give the number of
particles of each type $i \in 1 , \ldots , I$ (here $\vec{\rm n} =
\left( {\rm n}_a, {\rm n}_b \right)$).  For the linear reaction model,
the \emph{mass-action rates} $r_{ji}$\footnote{We write the
  mass-action rates as going from $i$ to $j$ with this index ordering,
  because they will appear again later as entries in a transfer
  matrix, which is naturally written as acting from the left on column
  vectors of occupation numbers $\left[ {\rm n}_i \right]$}
\begin{eqnarray}
  r_{ba} 
& = & 
  k_{+} {\rm n}_a 
\nonumber \\
  r_{ab} 
& = & 
  k_{-} {\rm n}_b 
\label{eq:rate_fns_linear}
\end{eqnarray}
describe the probability for particle transfer from state $i$ to state
$j$.  The form~(\ref{eq:general_exchange_ME}) extends immediately to
arbitrary many-particle exchanges with rates that may involve powers
of many different components of $\vec{\rm n}$, and to arbitrary
network connections.  The general methods derived below extend
immediately to all those cases.  Even the shortcuts we will introduce,
to pass directly from the form of the master equation to the action
functional of Freidlin-Wentzell theory, apply with no
changes.\footnote{The reader is cautioned that in cases where multiple
  reactants of the same type participate in a reaction, the assumption
  of separation of timescales may be inconsistent with using
  equilibrium state variables in the minimal way they are used here.
  An example is provided in Ref.~\cite{Nicolis:fluct_NEQ:71}, where
  the local-equilibrium approximation may still be used, but it
  requires replacing the discrete master equation with a full
  Boltzmann transport equation for continuous-valued positions and
  momenta.  An alternative solution is to add time-local corrections
  known as ``contact terms'' to represent the correct counting of
  many-particle states.  For the linear system such complications do
  not arise, so we will not digress on them further here.}

\subsubsection{Definition of the Liouville operator from the transfer
  matrix}

This section sketches the conversion from number terms and index
shifts in the master equation, to the equivalent polynomials and
differential operators in the Liouville equation.  It then converts
the representation on polynomials to the representation introduced by
Masao Doi~\cite{Doi:SecQuant:76,Doi:RDQFT:76}, in terms of raising and
lowering operators in an abstract linear vector space.  The systematic
construction of Liouville forms from master equations is now
standard~\cite{Eyink:action:96,Mattis:RDQFT:98}, and its form for the
linear mass-action kinetics of Eq.~(\ref{eq:rate_fns_linear}) is the
simplest possible.

In the vector notation, the generating function at a single time
becomes a function of a vector $\vec{z}$ whose components correspond
to those of $\vec{\rm n}$. Here the moment-generating function is 
\begin{equation}
  \psi \! \left( \vec{z} \right) = 
  \sum_{\vec{{\rm n}}}
  \prod_{i \in \left\{ a, b \right\}} 
  z_i^{{\rm n}_i}
  {\rho}_{\vec{{\rm n}}} . 
\label{eq:gen_func_any_parts}
\end{equation}
Time evolution under Eq.~(\ref{eq:general_exchange_ME}) implies time
evolution of $\psi$ according to 
\begin{eqnarray}
  \frac{\partial \psi \! \left( \vec{z} \right)}{\partial t} 
& = & 
  \sum_{\vec{{\rm n}}}
  \prod_{k \in \left\{ a, b \right\}} 
  z_k^{{\rm n}_k}
  \frac{\partial {\rho}_{\vec{{\rm n}}}}{\partial t}
\nonumber \\
& = & 
  \sum_{i,j \in \left\{ a, b \right\}} 
  \left( 
    \frac{z_j}{z_i} - 1
  \right)
  \sum_{\vec{{\rm n}}}
  \prod_{k \in \left\{ a, b \right\}} 
  z_k^{{\rm n}_k}
  r_{ji} \! \left( {\rm n}_i \right)
  {\rho}_{\vec{{\rm n}}} , 
\nonumber \\   
& = & 
  \sum_{i,j \in \left\{ a, b \right\}} 
  \left( 
    \frac{z_j}{z_i} - 1
  \right)
  r_{ji} \! 
  \left( 
    z_i \frac{\partial}{\partial z_i}
  \right)
  \sum_{\vec{{\rm n}}}
  \prod_{k \in \left\{ a, b \right\}} 
  z_k^{{\rm n}_k}
  {\rho}_{\vec{{\rm n}}} , 
\nonumber \\   
& \equiv & 
  - \mathcal{L} \! 
  \left( 
    \vec{z} , 
    \frac{\partial}{\partial \vec{z}}
  \right)
  \psi \! \left( \vec{z} \right) . 
\label{eq:transfer_to_liouville}
\end{eqnarray}
Here we have written out the functional dependence of $r_{ij}$ on the
number indices of the particle species entering half-reactions
explicitly, to show how those number indices are replaced by products
$z_i \partial / \partial z_i$ acting on the basis polynomials of the
generating function.  Successive lines of
Eq.~(\ref{eq:transfer_to_liouville}) reflect the action of successive
terms in the master equation~(\ref{eq:general_exchange_ME}).  The
extension of these steps to multi-species reactions also follows a
standard form~\cite{Mattis:RDQFT:98}, of which our example is
representative.  The differential operator $\mathcal{L} \!  \left(
\vec{z} , \partial / \partial \vec{z} \right)$ is the \emph{Liouville
  operator} for this stochastic process.

Time evolution of the density ${\rho}_{\vec{n}}$ over an interval
$\left[ 0 , T \right]$ is performed by taking the quadrature of
Eq.~(\ref{eq:transfer_to_liouville}), which we may think of as being
performed over a sequence of short time intervals of length $\delta t$: 
\begin{eqnarray}
  {\psi}_T 
& = & 
  \mathcal{T}
  e^{
    \int_0^T dt \, \mathcal{L}
  }
  {\psi}_0 
\nonumber \\
& = & 
  \mathcal{T}
  \prod_{k = 1}^{T / \delta t}
  e^{
    \delta t \, \mathcal{L}
  }
  {\psi}_0 . 
\label{eq:time_evol_quadrature}
\end{eqnarray}
$\mathcal{T}$ denotes \emph{time-ordering} in the product, and we will
use the same notation when the product of exponential terms $e^{
  \delta t \, \mathcal{L} }$ is written as the exponential of a sum or
integral.  Such expressions are known as \emph{time-ordered
  exponential integrals}.  For the Liouville operator with constant
coefficients, time-ordering serves no explicit purpose, but when
time-dependent sources are introduced below, it will become necessary.

We may efficiently approximate the
product~(\ref{eq:time_evol_quadrature}) by interposing a complete
distribution of generating functions between each of the small
increments of time evolution in the second line, which we develop in
the next sub-section.  Doing so, however, introduces a distinct set of
arguments $\vec{z}$ at each time, which are cumbersome to work with.
Therefore, at intermediate times between $0$ and $T$, it is
conventional to adopt an abstract notation for the complex vectors and
derivatives~\cite{Doi:SecQuant:76,Doi:RDQFT:76}, which more clearly
reflect the nature of the generating function as a vector in a normed
linear space (a Hilbert space).  The correspondence of the elementary
monomials,
\begin{eqnarray}
  z_j 
& \leftrightarrow & 
  a_j^{\dagger} , 
\label{eq:DP_creat_op}
\\
  \frac{\partial}{\partial z_i}
& \leftrightarrow & 
  a^i , 
\label{eq:DP_annih_op}
\end{eqnarray}
reflects the property that complex variables and their derivatives satisfy
the algebra of raising and lowering operators familiar from the
quantum harmonic oscillator,\footnote{Of course, in
  the approach of ``quantizing'' the classical Harmonic oscillator to
  a differential operator on Schr\"{o}dinger wave functions, this map
  from polynomials to operators is the origin of the raising and
  lowering operator algebra.}
\begin{equation}
  \left[ 
    \frac{\partial}{\partial z_i} , z_j
  \right] \leftrightarrow 
  \left[ 
    a^i , a_j^{\dagger}
  \right] = 
  {\delta}^i_j . 
\label{eq:commutation_alg}
\end{equation}

Polynomials built from these variables live in a function space built
on a ``vacuum'' which is simply the number 1, denoted 
\begin{equation}
  1 \leftrightarrow \left| 0 \right) . 
\label{eq:DP_right_vacuum}
\end{equation}
The conjugate operation to creating a generating function is the
evaluation of the trace -- generally with moment-sampling operators to
extract its moments -- so that the conjugate operator to the
right-hand vacuum~(\ref{eq:DP_right_vacuum}) is an integral, 
\begin{equation}
  \int d^I \! z \, 
  \delta \! \left( z \right) 
  \leftrightarrow \left( 0 \right| .  
\label{eq:DP_left_vacuum}
\end{equation}
From these definitions an entire function space with inner product can
be built up, as reviewed in
Ref's.~\cite{Mattis:RDQFT:98,Cardy:FTNEqSM:99}.\footnote{It is
  conventional to denote the inner product for classical stochastic
  processes with round braces, as in $\left| 0 \right)$, to
  distinguish them from the quantum bra-ket notation $\left| 0
  \right>$~\cite{Sakurai:ModQM:85}.  This notation reflects the fact
  that it is not in the operator algebra, but in the space of
  functions and the inner product, that the distinction between
  classical and quantum-mechanical systems lies, as discussed in the
  introduction.}

\subsubsection{The coherent-state expansion}

The transformation from discrete number-states to generating functions
is the Laplace transform.  It may be applied within the discrete time
slices of the quadrature~(\ref{eq:time_evol_quadrature}), where it
makes use of the Laplace transforms of intermediate Poisson
distributions, which are known as \emph{coherent states}.  It has
become conventional in the Doi-Peliti literature to denote the mean
values of these Poisson distributions with the field $\phi$; for a
multi-state system, we index them with a complex-valued vector field
$\vec{\phi}$.  The conjugate vectors to these Poisson distributions
are moment-sampling operators, weighted by Hermitian conjugate field
variables denoted ${\vec{\phi}}^{\dagger}$.

The ${\vec{\phi}}^{\dagger}$ will have the property that, dynamically,
they propagate information \emph{backward in time} from final data
imposed by the weights $z_a$, $z_b$.  The interpretation of this
property is that the moment-sampling operators capture
\emph{conditional distributions} at earlier times, conditioned on the
final values they produce.  The fields $\vec{\phi}$ and
${\vec{\phi}}^{\dagger}$ will turn out to be the position fields and
their conjugate-momentum fields in one version of the
Freidlin-Wentzell theory, and the Hamiltonian under which they evolve
will be the Liouville operator.  The way evolution under the dynamical
system is made consistent with the ``backward'' propagation of
information, by the conjugate momentum, is that the momentum variables
will typically have unstable evolution forward in time, while the
position variables will evolve in stable directions. 

In this section we construct the coherent-state expansion for the
quadrature~(\ref{eq:time_evol_quadrature}), which is due to
Peliti~\cite{Peliti:PIBD:85,Peliti:AAZero:86}.  

The Liouville operator is a function of raising and lowering operators
$\left\{ a^i \right\}$, $\left\{ a^{\dagger}_j \right\}$ so the
convenient expansion of Eq.~(\ref{eq:time_evol_quadrature}) will be
given by a basis of eigenstates of these operators.  The coherent
states are constructed to be such eigenstates.  The right-hand
coherent state is defined as 
\begin{eqnarray}
  \left| \vec{\phi} \right) 
& = & 
  e^{ - 
    \left( {\vec{\phi}}^{\dagger} \cdot \vec{\phi} \right) /2 
  }
  e^{
    {\vec{a}}^{\dagger} \cdot \vec{\phi}
  }
  \left| 0 \right) 
\nonumber \\
& = & 
  e^{ - 
    \left( {\vec{\phi}}^{\dagger} \cdot \vec{\phi} \right) /2
  }
  \sum_{M = 0}^{\infty}
  \frac{
    {
      \left(
        {\vec{a}}^{\dagger} \cdot \vec{\phi}
      \right)
    }^M
  }{
    M !
  }  
  \left| 0 \right) 
\nonumber \\
& = & 
  \prod_{m=1}^I
  \left(
    e^{
      - \left( {\phi}_m^{\dagger} {\phi}_m \right) /2
    }
    \sum_{n_m = 0}^{\infty}
    \frac{
      {\phi}_m^{n_m}
    }{
      n_m ! 
    }
    {
      a_m^{\dagger}
    }^{n_m}
  \right)
  \left| 0 \right) .
\label{eq:DP_open_coh_state}
\end{eqnarray}
The components of its complex parameter vector $\vec{\phi}$ correspond
to those of $\vec{z}$.  From the commutator
algebra~(\ref{eq:commutation_alg}), and the fact that all $a^i
\leftrightarrow \partial / \partial z_i$ annihilate the right
vacuum~(\ref{eq:DP_right_vacuum}), it follows that
Eq.~(\ref{eq:DP_open_coh_state}) does define an eigenstate of all the
lowering operators $a^i$ with corresponding eigenvalues ${\phi}^i$,
\begin{equation}
  a^i
  \left| \vec{\phi} \right) = 
  {\phi}^i 
  \left| \vec{\phi} \right) .
\label{eq:lower_op_right_phi}
\end{equation}

The left-hand coherent state dual to Eq.~(\ref{eq:DP_open_coh_state})
is given by 
\begin{eqnarray}
  \left( {\vec{\phi}}^{\dagger} \right| 
& = & 
  \left( 0 \right| 
  e^{
    {\vec{\phi}}^{\dagger} \cdot \vec{a}
  }
  e^{ - 
    \left( {\vec{\phi}}^{\dagger} \cdot \vec{\phi} \right) /2
  }
\nonumber \\
& = & 
  \left( 0 \right| 
  \sum_{M = 0}^{\infty}
  \frac{
    {
      \left(
        {\vec{\phi}}^{\dagger} \cdot \vec{a}
      \right)
    }^M 
  }{
    M !
  }  
  e^{ - 
    \left( {\vec{\phi}}^{\dagger} \cdot \vec{\phi} \right) /2
  } .  
\label{eq:DP_close_coh_state}
\end{eqnarray}
Eq.~(\ref{eq:DP_close_coh_state}) is likewise checked to be an
eigenstate of the raising operators, with eigenvalues
${\phi}_j^{\dagger}$,
\begin{equation}
  \left( {\vec{\phi}}^{\dagger} \right| 
  a_j^{\dagger} = 
  \left( {\vec{\phi}}^{\dagger} \right| 
  {\phi}_j^{\dagger} . 
\label{eq:raise_op_left_phi}
\end{equation}
The presence of $\vec{a}$ corresponding to $\partial / \partial
\vec{z}$ in Eq.~(\ref{eq:DP_close_coh_state}) gives the left-hand
coherent state the interpretation of a moment-sampling operator.  This
state displaces any argument $a^{\dagger}_j$ in the state that it acts
on (through the inner product), by a summand ${\phi}^{\dagger}_j$.  It
then sets the original $a^{\dagger}_j$ to zero because all left-hand
states include the trace defined by the left
vacuum~(\ref{eq:DP_left_vacuum}).

The sum of outer products of a left-hand coherent state and its dual
right-hand coherent state gives a representation of the identity
operator as an over-complete integral over states,
\begin{equation}
  \int
  \frac{
    d^I \! {\phi}^{\dagger}
    d^I \! \phi
  }{
    {\pi}^I
  }
  \left| \vec{\phi} \right) 
  \left( {\vec{\phi}}^{\dagger} \right| = 
  I . 
\label{eq:coh_state_completeness_text}
\end{equation}
The result of taking the inner product of
Eq.~(\ref{eq:coh_state_completeness_text}) with any state $\left|
\vec{\psi} \right)$ on the right is that the moment-sampling operator
$\left( {\vec{\phi}}^{\dagger} \right|$ simply transposes the weights
that it extracts from$\left| \vec{\psi} \right)$onto the left-coherent
states $\left| \vec{\phi} \right)$ in the
expansion~(\ref{eq:coh_state_completeness_text}).

If we insert a complete set of states in this form between every term
in the product~(\ref{eq:time_evol_quadrature}), we express the
moment-generating function at time $T$ as an integral over all
intermediate values of the field variables $\vec{\phi}$,
${\vec{\phi}}^{\dagger}$, of the form 
\begin{widetext}
\begin{equation}
  \psi \! \left( z_a , z_b \right) \equiv 
  e^{- {\Gamma}_T \left( \log z_a , \log z_b \right)} = 
  \int_0^T 
  \mathcal{D} {\phi}^{\dagger}_a \mathcal{D} {\phi}_a
  \mathcal{D} {\phi}^{\dagger}_b \mathcal{D} {\phi}_b
  e^{
    \left( z_a - {{\phi}^{\dagger}_a}_T \right) {{\phi}_a}_T + 
    \left( z_b - {{\phi}^{\dagger}_b}_T \right) {{\phi}_b}_T - 
    S - 
    {\Gamma}_0 
    \left( 
      \log {{\phi}^{\dagger}_a}_0 , 
      \log {{\phi}^{\dagger}_b}_0 
    \right)
  } . 
\label{eq:gen_fn_twofields_nosource}
\end{equation}
\end{widetext}
It requires some algebra, but no conceptual difficulties, to show that
the action in Eq~(\ref{eq:gen_fn_twofields_nosource}) is given by
\begin{equation}
  S = 
  \int dt
  \left[
    - \left(
      {\partial}_t {\phi}^{\dagger}_a
      {\phi}_a + 
      {\partial}_t {\phi}^{\dagger}_b
      {\phi}_b 
    \right) + 
    \mathcal{L}
  \right]  .
\label{eq:action_fields}
\end{equation}
The terms ${\partial}_t {\phi}^{\dagger}_a {\phi}_a$ come from the
inner products of states $\left( {\vec{\phi}}^{\dagger} \right|$ and
$\left| \vec{\phi} \right)$ at successive times, and the evaluation of
the Liouville operator sandwiched between these states converts
$\hat{\mathcal{L}}$ into the quantity serving as the Hamiltonian.  Its
form, 
\begin{equation}
  \mathcal{L} = 
  k_{+} 
  \left( 
    {\phi}^{\dagger}_a - 
    {\phi}^{\dagger}_b 
  \right)
  {\phi}_a + 
  k_{-} 
  \left( 
    {\phi}^{\dagger}_b - 
    {\phi}^{\dagger}_a 
  \right)
  {\phi}_b . 
\label{eq:Liouville_fields}
\end{equation}
is obtained directly from the penultimate line of
Eq.~(\ref{eq:transfer_to_liouville}) with the mass-action
rates~(\ref{eq:rate_fns_linear}).  It is a general feature, which
extends to non-linear rate equations, that factors of $z_i$ in the
rate functions on the reactant side cancel factors $1 / z_i$ from the
shift operators (the leading parenthesized term in
Eq.~(\ref{eq:transfer_to_liouville})).

\subsection{Standard computations of the Freidlin-Wentzell
  quasipotential} 
\label{sec:field_quasipotential}

Equations~(\ref{eq:gen_fn_twofields_nosource}-\ref{eq:Liouville_fields})
are the stochastic-process reformulation of the single-time generation
function~(\ref{eq:gen_func_general_def}).  The Legendre transform
$\log \psi \! \left( z_a , z_b \right)$ continues to be the
equilibrium effective action $S^{\mbox{\scriptsize eff}} \! \left( n_a
, n_b \right)$ of Eq.~(\ref{eq:S_eff_def_onetime}).  The field
integral~(\ref{eq:gen_fn_twofields_nosource}) for these quantities
will in general be dominated by the stationary paths of the
action~(\ref{eq:action_fields}).  The value of $S$ in
Eq.~(\ref{eq:action_fields}), on the stationary path, is known in
Freidlin-Wentzell theory as the \emph{quasipotential}, because it
generalizes the Gibbs free energy from equilibrium
thermodynamics.\footnote{The Gibbs free energy and other Legendre
  transforms of the entropy are conventionally known as
  \emph{thermodynamic potentials}, by yet a further analogy with the
  potential energy in mechanics.}  The quasipotential is widely used
to approximate the solutions to diffusion equations on the boundaries
of trapping
domains~\cite{Graham:path_int:77,Graham:potential:84,Freidlin:RPDS:98}. 

The field integral of the last section has converted an irreversible
stochastic process into a deterministic dynamical system perturbed by
fluctuations.  Any such Hamiltonian system offers several choices of
representation, related under the changes of (field) variables known
in Hamiltonian dynamics as \emph{canonical
  transformations}~\cite{Goldstein:ClassMech:01}.  In addition to the
canonical variables of the Hamiltonian dynamics, this action
functional also admits an unusual ``kinematic'' interpretation, in
which escape trajectories are represented as rolling in an energy
potential, even when they move opposite to the direction of classical
diffusion.\footnote{A thorough treatment of this kinematic potential
  is provided in Ref.~\cite{Smith:evo_games:}.}  Finally, the leading
expansion about the mean behavior in the
integral~(\ref{eq:gen_fn_twofields_nosource}) comes from Gaussian
fluctuations, which are readily converted to standard approximation
methods such as the Langevin formulation.  For the bilinear action,
the Langevin approximation in the original fields $\vec{\phi}$,
${\vec{\phi}}^{\dagger}$ is exact, though in other variables or for
more complex reactions it will not be.  We will consider each of these
aspects of the Doi-Peliti field integral, and the Freidlin-Wentzell
quasipotential, in turn in this section.

All introductions to Freidlin-Wentzell theory that I have
seen in the reaction-diffusion literature compute the quasipotential
by studying the stationary points in the field variables $\phi$,
${\phi}^{\dagger}$~\cite{Mattis:RDQFT:98,Cardy:FTNEqSM:99}, or in the
equivalent operator expectations~\cite{Eyink:action:96}.  For the
two-state system, this choice is technically the easiest, because the
bilinear functional integral~(\ref{eq:gen_fn_twofields_nosource}) is
manifestly the ``simple harmonic oscillator'' of stochastic processes.
However, field variables are the least intuitive choice, because
individually neither $\phi$ nor ${\phi}^{\dagger}$ represents an
observable quantity.  (Recall that the function of $\phi$ as the
expectation value of a Poisson distribution is only propagated through
time by the moment-sampling weight given by ${\phi}^{\dagger}$.)  An
elementary canonical transformation to \emph{action-angle
  variables}~\cite{Goldstein:ClassMech:01} will produce fields which,
unlike $\phi$ and ${\phi}^{\dagger}$, correspond directly to observed
average number occupancies, and to a conjugate momentum that has the
physical interpretation of a chemical potential.

\subsubsection{Diagonalization and descaling in field variables}

The quasipotential provides the leading log-probability for
fluctuations.  When large-deviations scaling holds, the value of the
quasipotential on its stationary path should factor into an overall
scale factor, and a scale-independent rate function.  This scaling is
particularly easy to achieve for the the Gaussian functional
integral~(\ref{eq:gen_fn_twofields_nosource}) of the two-state model.
The field variable $\vec{\phi}$ is rescaled to remove a factor $N$,
which may be moved outside the entire action (\ref{eq:action_fields}),
leaving a bilinear functional of normalized fields that do not depend
on system scale, at any field values.  In standard treatments using
coherent-state variables~\cite{Eyink:action:96,Mattis:RDQFT:98}, all
scale factors come from the Poisson field $\vec{\phi}$, and their
Hamiltonian-conjugate momenta ${\vec{\phi}}^{\dagger}$ are unchanged.

Field rescaling is one of several changes of variable that may be made
in the bilinear field basis given by $\vec{\phi}$, and
${\vec{\phi}}^{\dagger}$.  Another is a rotation of components from
$\left( {\phi}_a , {\phi}_b \right)$ to a basis in which the
independent dynamical quantity $n \equiv \left( n_b - n_a \right) / 2$
may be separated from the non-dynamical conserved quantity $N$.
However, this rotation must be used with care, as certain terms needed
to define correlation functions in terms of the equilibrium
distribution disappear from the na\"{\i}ve continuous-time limit, and
must be reconstructed from the explicit discrete-time
forms~\cite{Kamenev:DP:01}.  We will also see that correct calculation
of the cumulant-generating functional can rely on contributions from
the apparently non-dynamical number $N$, depending on how total
derivatives are handled in the action~(\ref{eq:action_fields}).  

In order to work out the correct treatment of such technical issues,
we will begin in this section with the generating function $\psi \!
\left( z_a , z_b \right)$ of Eq.~(\ref{eq:gen_func_general_def}), with
its apparently superfluous second complex argument.  Once the
two-variable generating function has been understood, we will perform
the more complicated, but more direct, evaluation of the
single-argument form $\psi \!  \left( z \right)$ from
Eq.~(\ref{eq:gen_func_reduced_def}).

We will see in this section that the coherent-state variables
$\vec{\phi}$, and ${\vec{\phi}}^{\dagger}$ provide a powerful and
simple route to scaling and a variety of exact solutions in the
Gaussian model.  However, they are not directly related to observable
quantities, and their simplicity in the case of non-interacting
particles can quickly give way to quite complicated correlation
functions if particle interactions must be considered.  They also
produce a continuum limit which, although simple in form, does not
explicitly represent all quantities needed to compute correlation
functions.  Therefore we will abandon the field basis of $\vec{\phi}$,
and ${\vec{\phi}}^{\dagger}$ as soon as the structure of its
generating function has been understood, and move to a set of
transformed variables that lead to slightly more complex algebra, but
do not suffer from any of these shortcomings.

The field components ${\phi}_a$, ${\phi}_b$ correspond to expectation
values $n_a$, $n_b$ defined as in
Eq.~(\ref{eq:grad_Gamma_number}).\footnote{For the case
  ${\vec{\phi}}^{\dagger} \equiv 1$ representing thermal equilibrium
  and all classical diffusion solutions, ${\phi}_a = n_a$ and
  ${\phi}_b = n_b$ as an expectation in the functional integral, but
  not as a general insertion in higher-order correlation functions.}
Therefore, as for the equilibrium distribution, we may define a
rotated basis corresponding in a similar manner to $n$ and $N$, by the
orthogonal transformation
\begin{eqnarray}
  \phi 
& \equiv & 
  \left( {\phi}_b - {\phi}_a \right) / 2 
\nonumber \\
  {\phi}^{\dagger} 
& \equiv & 
  {\phi}^{\dagger}_b -  
  {\phi}^{\dagger}_a 
\nonumber \\
  \Phi 
& \equiv & 
  {\phi}_b + 
  {\phi}_a 
\nonumber \\
  {\Phi}^{\dagger} 
& \equiv & 
  \left( {\phi}^{\dagger}_b + {\phi}^{\dagger}_a \right) / 2 . 
\label{eq:fields_rotate_basis}
\end{eqnarray}
A scale factor proportional to total particle number may then be
removed from the Poisson fields $\phi$, $\Phi$ only, defining
normalized fields 
\begin{eqnarray}
  \hat{\phi}
& \equiv &
  \phi / N 
\nonumber \\ 
  \hat{\Phi}
& \equiv &
  \Phi / N , 
\label{eq:descale_fields}
\end{eqnarray}
Here the relation of absolute to relative number fields repeats the
notation used to relate absolute to relative number indices ${\rm n}$
and $\hat{\rm n} \equiv {\rm n} / N$.  Even though the relation
between variables $\phi$, ${\phi}^{\dagger}$ and the expected number
indices is not simple, in general it is the $\phi$ variables that
carry the scaling with total system size.

A final coordinate transformation reflects the fact that the
coordinate timescale $dt$ is not characteristic of the dynamics for
either average behavior or fluctuations.  We therefore introduce a
rescaled time variable $\tau$, with Jacobean 
\begin{equation}
  \frac{d\tau}{dt} \equiv 
  k_{+} + k_{-} = 
  \frac{
    e^{
      -\beta
      \left( 
        {\mu}_{\ddagger} - {\mu}_{a \cup b}^1
      \right)
    }
  }{
    {\bar{\nu}}_a 
    {\bar{\nu}}_b
  } = 
  \frac{
    e^{- \beta {\mu}_{\ddagger}}
  }{
    Z_1 
    {\bar{\nu}}_a 
    {\bar{\nu}}_b
  } , 
\label{eq:time_rescale_def}
\end{equation}
as the physically relevant time variable.  Note that this natural
timescale for rare events relates the transition-state chemical
potential to that for the equilibrated joint system, defined from
Eq.~(\ref{eq:Gibbs_min_of_N}).  

In these new variables the action~(\ref{eq:action_fields}) becomes 
\begin{eqnarray}
  S 
& = & 
  N \int d\tau 
  \left[ 
    - {\partial}_{\tau}
    {\Phi}^{\dagger}
    \hat{\Phi} - 
    {\partial}_{\tau}
    {\phi}^{\dagger}
    \hat{\phi} + 
    {\phi}^{\dagger}
    \left( 
      \hat{\phi} - 
      \bar{\nu} \hat{\Phi}
    \right)
  \right]
\nonumber \\
& \equiv & 
  N \int d\tau 
  \left( 
    - {\partial}_{\tau}
    {\Phi}^{\dagger}
    \hat{\Phi} - 
    {\partial}_{\tau}
    {\phi}^{\dagger}
    \hat{\phi} + 
    \hat{\mathcal{L}}
  \right) . 
\label{eq:action_twofields_descaled}
\end{eqnarray}

\subsubsection{Calculation of the two-argument cumulant-generating
  function} 

The leading large-$N$ exponential dependence of the functional
integral~(\ref{eq:gen_fn_twofields_nosource}) comes from the
stationary point of the action and the boundary terms at times $0$ and
$T$.  The stationary-point conditions are vanishing of the first
variational derivatives of the
action~(\ref{eq:action_twofields_descaled}), which take the form 
\begin{eqnarray}
  {\partial}_{\tau} {\Phi}^{\dagger}
& = & 
  \frac{\partial \hat{\mathcal{L}}}{\partial \hat{\Phi}} = 
  - \bar{\nu} {\phi}^{\dagger}
\nonumber \\
  {\partial}_{\tau} \hat{\Phi} 
& = & 
  - \frac{\partial \hat{\mathcal{L}}}{\partial {\Phi}^{\dagger}} = 
  0 
\nonumber \\
  {\partial}_{\tau} {\phi}^{\dagger}
& = & 
  \frac{\partial \hat{\mathcal{L}}}{\partial \hat{\phi}} = 
  {\phi}^{\dagger}
\nonumber \\
  {\partial}_{\tau} \hat{\phi} 
& = & 
  - \frac{\partial \hat{\mathcal{L}}}{\partial {\phi}^{\dagger}} = 
  - \left( 
    \hat{\phi} - 
    \bar{\nu} \hat{\Phi} 
  \right)
\label{eq:Hamilton_EOM_twofields}
\end{eqnarray}
Eq.~(\ref{eq:Hamilton_EOM_twofields}) is the promised conversion of
the first-moment dynamics of the stochastic
process~(\ref{eq:general_exchange_ME}) into a deterministic dynamical
system with Hamiltonian $\hat{\cal L}$.  It follows immediately from
these equations and from the lack of explicit time-dependence in
$\hat{\mathcal{L}}$, that this Liouville-Hamiltonian is a constant of
the motion along the stationary path,
\begin{equation}
  \frac{
    d\hat{\mathcal{L}}
  }{
    d\tau 
  } \equiv 
  0 . 
\label{eq:conservation_Liouville}
\end{equation}

The boundary terms at time $\hat{T} \equiv T \left( k_{+} + k_{-}
\right)$ in the variation of the functional
integral~(\ref{eq:gen_fn_twofields_nosource}) vanish at
${{\phi}^{\dagger}_a}_{\hat{T}} = z_a$,
${{\phi}^{\dagger}_b}_{\hat{T}} = z_b$.  The intermediate
time-dependence of ${\phi}^{\dagger}$ is thus 
\begin{equation}
  {\phi}^{\dagger}_{\tau} = 
  \left( z_b - z_a \right)
  e^{\tau - \hat{T}} , 
\label{eq:phi_dag_soln_genfunc}
\end{equation}
and the corresponding solution for ${\Phi}^{\dagger}$ is 
\begin{equation}
  {\Phi}^{\dagger}_{\tau} = 
  \frac{1}{2}
  \left( z_b + z_a \right) + 
  \bar{\nu}
  \left( z_b - z_b \right) 
  \left( 
    1 - e^{\tau - T}
  \right) . 
\label{eq:Phi_dag_soln_gen_func}
\end{equation}

The field $\hat{\Phi}$ is non-dynamical, and a steady-state solution
for the relative number field $\hat{\phi}$ in terms of $\hat{\Phi}$ is
therefore give by 
\begin{equation}
  \hat{\phi} = \bar{\nu} \hat{\Phi} . 
\label{eq:phi_diag_const}
\end{equation}
The product ${\phi}^{\dagger}_a {\phi}_a + {\phi}^{\dagger}_b
{\phi}_b$ is the expectation of the total number operator, and
therefore must equal $N$.  In the rotated and descaled basis, this
equality takes the form 
\begin{eqnarray}
  1 
& = & 
  {\Phi}^{\dagger} \hat{\Phi} + 
  {\phi}^{\dagger} \hat{\phi} 
\nonumber \\
& = & 
  \hat{\Phi}
  \left[ 
    \frac{1}{2}
    \left( z_b + z_a \right) + 
    \bar{\nu}
    \left( z_b - z_a \right) 
  \right] , 
\label{eq:Phi_from_N_solve}
\end{eqnarray}
which may therefore be used to assign a value to $\hat{\Phi}$ in
terms of the arguments $z_a$, $z_b$ of the generating function.  

The equivalent expression for the relative number-operator difference
$\left( {\phi}^{\dagger}_b {\phi}_b - {\phi}^{\dagger}_a {\phi}_a
\right) / N$ is then 
\begin{equation}
  2 \nu = 
  \frac{1}{2}
  {\phi}^{\dagger} \hat{\Phi} + 
  2 {\Phi}^{\dagger} \hat{\phi} .
\label{eq:nu_from_Phi_solve}
\end{equation}
Eq.~(\ref{eq:nu_from_Phi_solve}) is immediately solved from the
boundary
values~(\ref{eq:phi_dag_soln_genfunc},\ref{eq:Phi_dag_soln_gen_func})
at time $\hat{T}$ to recover Eq.~(\ref{eq:n_q_compute}) as 
\begin{equation}
  2 {\nu}_T = 
  \tanh 
  \frac{1}{2}
  \left( 
    \log z - \beta \bar{\mu}
  \right) .
\label{eq:init_solve}
\end{equation}
In terms of this final value for the function $\nu$, the functional
dependence at earlier times is given by 
\begin{equation}
  {\nu}_{\tau} - \bar{\nu} = 
  \left( {\nu}_T - \bar{\nu} \right) 
  e^{\tau - \hat{T}} . 
\label{eq:nu_tau_time_genfunc}
\end{equation}

The algebra of the preceding solution is simple and linear, and yet it
is profoundly obscure as a representation of the physical
distributions of random walkers in the two-state model.  This
obscurity is the reason we will soon abandon coherent-state variables.
Note that the field $\hat{\phi}$ -- na\"{\i}vely corresponding to the
observable $n$ -- is actually constant by
Eq.~(\ref{eq:phi_diag_const}).  All particle number-dynamics in this
solution comes from the fields ${\phi}^{\dagger}$ and
${\Phi}^{\dagger}$.  These fields, known as \emph{response fields} in
the Doi-Peliti literature~\cite{Kamenev:DP:01}, are acting to
propagate information from the boundary condition at the final time
$\hat{T}$ backward into the interior of the functional integral, by
selecting moments of the intermediate states with time-dependent
weights.  

The stationary-path solution~(\ref{eq:nu_tau_time_genfunc}) is the
least-improbable sequence of fluctuations to have led to the final
value ${\nu}_{\tau}$.  The single-time probability of ${\nu}_{\tau}$
in turn results from the accumulation of probabilities for successive
accumulating fluctuations in the
integral~(\ref{eq:action_twofields_descaled}) for $S$.  However, to
identify the particle distribution in this solution requires
incorporating the response fields, including the (strangely dynamical)
response field ${\Phi}^{\dagger}$ associated with the (unchanging)
total particle number $N$.

To show that, despite the difficulties of interpretation, the
quasipotential recovers the single-time fluctuation theorems, we
evaluate the cumulant-generating function ${\Gamma}_{\hat{T}}$.  The
calculation is easiest if we take $\hat{T} \gg 1$ so that the initial
condition ${\Gamma}_0$ may be evaluated on the equilibrium
distribution.\footnote{Any other initial condition would decay toward
  the equilibrium distribution for sufficiently large $\hat{T}$.}  For
the exact bilinear action~(\ref{eq:action_fields}) a conservation law
greatly simplifies the calculations, although it again renders the
correct answer in a most cryptic fashion.

We begin with the conservation law~(\ref{eq:conservation_Liouville}).
To obtain a reference value for $\hat{\mathcal{L}}$ we note that in
the equilibrium distribution where the mass-action equations balance,
$\hat{\mathcal{L}} = 0$.  The
action~(\ref{eq:action_twofields_descaled})  is bilinear in the
fields $\phi$ and ${\phi}^{\dagger}$, so that it has a simple relation
to either of its gradients, 
\begin{equation}
  \hat{\mathcal{L}} = 
  \sum_{i = a,b}
  {\phi}^{\dagger}_i
  \frac{
    \partial \hat{\mathcal{L}}
  }{
    \partial {\phi}^{\dagger}_i
  } = 
  \sum_{i = a,b}
  \frac{
    \partial \hat{\mathcal{L}}
  }{
    \partial {\hat{\phi}}_i
  } 
  {\hat{\phi}}_i .
\label{eq:bilinear_hat_L}
\end{equation}
Therefore, the stationary point conditions, together with
$\hat{\mathcal{L}} \equiv 0$, imply $S \equiv 0$ when evaluated on
\emph{any} stationary path.  Thus the only contribution to ${\Gamma}_T
\! \left( \log z_a , \log z_b \right)$ must come from the boundary
term ${\Gamma}_0 \!  \left( \log {{\phi}^{\dagger}_a}_0 , \log
{{\phi}^{\dagger}_b}_0 \right)$ in
Eq.~(\ref{eq:gen_fn_twofields_nosource}).  Since the initial
distribution has support only on a single value $n_a + n_b = N$, from
the definition~(\ref{eq:gen_func_general_def}) we may write
\begin{equation}
  e^{
    - {\Gamma}_0 
    \left( 
      \log {{\phi}^{\dagger}_a}_0 , 
    \log {{\phi}^{\dagger}_b}_0 , 
    \right) 
  } = 
  e^{
    \frac{N}{2}
    \log 
    \left( 
      {{\phi}^{\dagger}_a}_0 {{\phi}^{\dagger}_b}_0 
    \right) 
  }
  e^{ 
    - {\Gamma}_0 
    \left( 
      \log 
      \left( 
        {{\phi}^{\dagger}_b}_0 / 
        {{\phi}^{\dagger}_a}_0 
      \right) 
    \right) 
  } , 
\label{eq:gen_func_twofields}
\end{equation}
in which the single-argument cumulant-generating function corresponds
to the logarithm of $\psi \! \left( z \right)$ in
Eq.~(\ref{eq:gen_func_reduced_def}).  From
Eq.~(\ref{eq:phi_dag_soln_genfunc}) at $\tau \rightarrow 0$ and large
$\hat{T}$, ${\phi}^{\dagger}_b - {\phi}^{\dagger}_a \rightarrow 0$,
and as we will see ${\phi}^{\dagger}_b$ and ${\phi}^{\dagger}_a$
remain nonzero, so $\log \left( {{\phi}^{\dagger}_b}_0 /
{{\phi}^{\dagger}_a}_0 \right) \rightarrow 0$.  Therefore, for the
equilibrium distribution ${\Gamma}_0 \! \left( \log \left(
{{\phi}^{\dagger}_b}_0 / {{\phi}^{\dagger}_a}_0 \right) \right)
\rightarrow 0$ as well.  The only contribution to the generating
function at time $\hat{T}$, even with $z_b z_a \equiv 1$, is the term
$\frac{N}{2} \log \! \left( {{\phi}^{\dagger}_a}_0
{{\phi}^{\dagger}_b}_0 \right)$ corresponding to the total number,
which is not even dynamical!

In the same limit $e^{-\hat{T}} \rightarrow 0$, setting $z_a z_b = 1$,
we have
\begin{equation}
  {{\phi}^{\dagger}_a}_0 \rightarrow 
  {{\phi}^{\dagger}_b}_0 \rightarrow 
  {\Phi}^{\dagger}_0 \rightarrow
  \frac{
    \cosh \frac{1}{2}
    \left( 
      \log z - \beta \bar{\mu}
    \right)
  }{
    \cosh \frac{1}{2}
    \beta \bar{\mu}
  } . 
\label{eq:lims_Phi_dagger}
\end{equation}
The resulting evaluation for the cumulant-generating function is then
\begin{eqnarray}
\lefteqn{
  {\Gamma}_{\hat{T}} \! 
  \left( 
    \log z_a, \log z_b 
  \right) = 
  {\Gamma}_0 \! 
  \left( 
    \log {{\phi}^{\dagger}_a}_0 , 
  \log {{\phi}^{\dagger}_b}_0 
  \right) 
} & & 
\nonumber \\
& = & 
  - \frac{N}{2}
  \log \! 
  \left( 
    {{\phi}^{\dagger}_a}_0 {{\phi}^{\dagger}_b}_0 
  \right) 
\nonumber \\
& \rightarrow & 
  N 
  \left[ 
    \log \cosh \frac{1}{2}
    \beta \bar{\mu} - 
    \log \cosh \frac{1}{2}
    \left( 
      \log z - \beta \bar{\mu}
    \right) 
  \right] , 
\nonumber \\
\label{eq:gen_func_fields_Phis}
\end{eqnarray}
recovering Eq.~(\ref{eq:Gamma_q_compute}).  We obtain the correct
answer from a collection of surface terms, while the single-argument
cumulant-generating functional that should have controlled dynamics
dropped out.  How shall we understand this?  

\subsubsection{Cancellation of surface terms, and the quasipotential
  as an integral of time-local log likelihoods}

The puzzle of the Gaussian evaluation of the cumulant-generating
function is solved by taking care with the freedom we have to include
total derivatives in the field action.  The particular choice that
moves all fluctuation probabilities into the expected, single-variable
cumulant-generating function $\Gamma \! \left( \log z \right)$ also
sets up the action-angle change of variables in the next section,
which will isolate only the dynamical observables.

The property $S \equiv 0$, together with the general feature
$\hat{\mathcal{L}} \equiv 0$, implies \emph{for a bilinear action}
that the sum of time-derivative terms equals zero independently.  In
particular, we may decompose these as
\begin{eqnarray}
  0 
& = & 
  {\partial}_t {\phi}^{\dagger}_a
  {\hat{\phi}}_a + 
  {\partial}_t {\phi}^{\dagger}_b
  {\hat{\phi}}_b 
\nonumber \\
& = & 
  \frac{1}{2}
  {\partial}_t 
  \log \! 
  \left( 
    {\phi}^{\dagger}_b
    {\phi}^{\dagger}_a
  \right) 
  \left( 
    {\phi}^{\dagger}_b {\hat{\phi}}_b + 
    {\phi}^{\dagger}_a {\hat{\phi}}_a
  \right) 
\nonumber \\
& & 
  \mbox{} + 
  {\partial}_t 
  \log \! 
  \left( 
    \frac{
      {\phi}^{\dagger}_b
    }{
      {\phi}^{\dagger}_a
    } 
  \right) 
  \frac{1}{2}
  \left( 
    {\phi}^{\dagger}_b {\hat{\phi}}_b -  
    {\phi}^{\dagger}_a {\hat{\phi}}_a
  \right) . 
\label{eq:time_der_zero_transform}
\end{eqnarray}
The combination $\left( {\phi}^{\dagger}_b {\hat{\phi}}_b +
{\phi}^{\dagger}_a {\hat{\phi}}_a \right)$ is nothing more than the
conserved number 1, while the combination $\left( {\phi}^{\dagger}_b
{\hat{\phi}}_b - {\phi}^{\dagger}_a {\hat{\phi}}_a \right) / 2$ is
precisely the observable expected number asymmetry $\nu$.  The total
derivative $-\left( N/2 \right) {\partial}_t \log \!  \left(
{\phi}^{\dagger}_b {\phi}^{\dagger}_a \right)$ could have been removed
from the action to cancel the term in the final generating function,
leaving only the argument ${\Gamma}_0 \!  \left( \log \left(
{{\phi}^{\dagger}_b}_0 / {{\phi}^{\dagger}_a}_0 \right) \right)$ and
causing the magnitude of ${\Gamma}_{\hat{T}} \! \left( \log z \right)$
to originate from the remaining term in the action rather than from a
boundary term.

Let us summarize what has been accomplished so far: We have recovered
the value of the single-time generating function and the mean value of
its associated distribution, but we have also derived a new inference
about the most probable path of previous observations
\emph{conditioned} on those values at a time $\hat{T}$.  The fact that
the stationary value ${\nu}_{\tau} \neq \bar{\nu}$ for $\tau <
\hat{T}$ does not reflect the influence of sources, or in fact
\emph{any} causal influence at times $\tau < \hat{T}$, but rather the
conditional probability structure generated by the stationary points
of the field functional integral.  This very powerful feature makes
such functional integrals extremely useful, but it is also the feature
that requires us to go through the full exercise of defining the
generating functional of continuous-time sources, and then computing
the Legendre transform to a Stochastic Effective Action, to identify
the history-dependent observations for which we are actually computing
probabilities.

\subsubsection{Action-angle variables}

We next compute the same generating function using a set of
transformed variables in which the expected number indices $\vec{n}$
are given by elementary fields rather than by bilinear forms.  In
these variables the response field has the physically simple
interpretation of a chemical potential.  The transformed variables
have advantages and disadvantages relative to the coherent-state field
variables.  The Liouville/Hamiltonian operator in the transformed
variables will no longer be bilinear, making solutions to the
equations of motion less obvious, even though they may still be found
exactly.  On the other hand, the kinematic nature of the two-field
model as a Hamiltonian system will be clearer.  Also -- although we
have not considered fluctuations yet and have only talked about mean
values -- the transformed Hamiltonian will show explicitly all terms
needed to compute fluctuations, including those that vanished in the
continuous-time limit in coherent-state variables $\vec{\phi}$ and
${\vec{\phi}}^{\dagger}$.  In the transformed variables it is no
longer necessary to appeal to the underlying discrete-time form to
compute correlation functions.

The canonical transformation, from coherent-state fields to
action-angle fields (including the now-familiar factoring out of the
scale factor $N$), is defined by
\begin{eqnarray}
  {\phi}^{\dagger}_i 
& \equiv & 
  e^{{\eta}_i}
\nonumber \\
  {\phi}_i
& \equiv & 
  e^{- {\eta}_i}
  n_i \equiv 
  e^{- {\eta}_i}
  N {\nu}_i , 
\label{eq:action_angle_transform}
\end{eqnarray}
for $i \in \left\{ a, b \right\}$.  The fields $n_i$ now correspond to
expectation values from Eq.~(\ref{eq:grad_Gamma_number}), not only as
single insertions, but in general products of fields in the Doi-Peliti
functional integral~(\ref{eq:gen_fn_twofields_nosource}).  In the
transformed variables, this integral becomes 
\begin{widetext}
\begin{equation}
  \psi \! \left( z_a , z_b \right) \equiv 
  e^{- {\Gamma}_T \left( \log z_a , \log z_b \right)} = 
  \int_0^T 
  \mathcal{D} {\eta}_a \mathcal{D} n_a
  \mathcal{D} {\eta}_b \mathcal{D} n_b
  e^{
    \left( z_a e^{- {{\eta}_a}_T} - 1 \right) {n_a}_T + 
    \left( z_b e^{- {{\eta}_b}_T} - 1 \right) {n_b}_T - 
    S - 
    {\Gamma}_0 
    \left( 
      {{\eta}_a}_0 , 
      {{\eta}_b}_0 
    \right)
  } . 
\label{eq:gen_fn_AA_nosource}
\end{equation}
\end{widetext}
The action in transformed variables becomes 
\begin{eqnarray}
  S 
& = & 
  \int dt
  \left[
    - \left(
      {\partial}_t {\eta}_a
      n_a + 
      {\partial}_t {\eta}_b
      n_b 
    \right) + 
    \mathcal{L}
  \right] 
\nonumber \\
& = & 
  N \int d\tau 
  \left[
    - \left(
      {\partial}_{\tau} {\eta}_a
      {\nu}_a + 
      {\partial}_{\tau} {\eta}_b
      {\nu}_b 
    \right) + 
    \hat{\mathcal{L}}
  \right] , 
\label{eq:action_actionangle}
\end{eqnarray}
in which the Liouville/Hamiltonian has the form 
\begin{eqnarray}
  \mathcal{L} 
& = & 
  k_{+} 
  \left( 
    1 - 
    e^{
      {\eta}_b - 
      {\eta}_a
    } 
  \right)
  n_a + 
  k_{-} 
  \left( 
    1 - 
    e^{
      {\eta}_a - 
      {\eta}_b
    }
  \right)
  n_b , \mbox{ or }
\nonumber \\
  \hat{\mathcal{L}}
& = & 
  {\bar{\nu}}_b
  \left( 
    1 - 
    e^{
      {\eta}_b - 
      {\eta}_a
    } 
  \right)
  {\nu}_a + 
  {\bar{\nu}}_a
  \left( 
    1 - 
    e^{
      {\eta}_a - 
      {\eta}_b
    }
  \right)
  {\nu}_b . 
\label{eq:Liouville_actionangle}
\end{eqnarray}

We now observe an important and general relation between the field
action~(\ref{eq:Liouville_actionangle}) and the transfer-matrix terms
in the original master equation~(\ref{eq:master_equation}).  (Recall
that the Liouville operator is defined with a minus sign relative to
the transfer matrix.)  The original index shifting operators $\partial
/ \partial {\rm n}_i$ have been replaced by momentum fields $-
{\eta}_i$, and the discrete indices ${\rm n}_i$ have been replaced by
field variables $n_i$, whose expectation values in the functional
integral are those of the indices ${\rm n}_i$ in the time-dependent
density ${\rho}_{{\rm n}_a, {\rm n}_b}$.  Thus we see that the tedious
task of expressing the operator algebra and coherent-state expansion
may be bypassed by a simple notational replacement, to arrive at the
functional integral directly.  The glossary indicating which number
terms are substituted in passing between the master equation and
Liouville operator is given in Table.~\ref{tab:social_makeup}.  The
same forms hold for the master equation~(\ref{eq:general_exchange_ME})
with more general non-linear reaction rates, and the forms for
large-deviation rates in the more general case are provided in
App.~\ref{sec:multi_part_react}.

The basis rotation in action-angle variables is now identical to that
performed in Sec.~\ref{sec:genfunc_equil} for the particle numbers
\emph{and the chemical potentials}, with $\eta \equiv {\eta}_b -
     {\eta}_a$ transforming as the chemical potential dual to $n$.
     The resulting action becomes
\begin{eqnarray}
\lefteqn{
  S = 
  {
    \left. 
      - \frac{N}{2}
      \left( {\eta}_b + {\eta}_a \right) 
    \right|
  }_0^{\hat T}
} & & 
\nonumber \\
& & 
  \mbox{} + 
  N \int d\tau 
  \left\{ 
    - {\partial}_{\tau}
    \eta \, 
    \nu + 
    {\bar{\nu}}_b {\nu}_a
    \left(
      1 - 
      e^{\eta}
    \right) + 
    {\bar{\nu}}_a {\nu}_b
    \left(
      1 - 
      e^{-\eta}
    \right) 
  \right\} . 
\nonumber \\
\label{eq:S_ass_diss_AA_red}
\end{eqnarray}
The first term, arising from the total derivative discussed at the end
of the preceding subsection, is precisely the one needed to cancel a
factor of $\left( N / 2 \right) \left( {\eta}_b + {\eta}_a \right)$ in
the initial generating function ${\Gamma}_0 \!  \left( {{\eta}_a}_0 ,
{{\eta}_b}_0 \right)$ in Eq.~(\ref{eq:gen_fn_AA_nosource}), leaving
only the single-argument function ${\Gamma}_0 \!  \left( {\eta}_0
\right)$ for the response field conjugate to $\nu$.  The net effect of
the boundary terms at time $\hat{T}$ is to set ${{\eta}_a}_{\hat{T}} =
\log z_a$ and ${{\eta}_b}_{\hat{T}} = \log z_b$ with the result that
${\Gamma}_{\hat{T}} \! \left( \log z_a , \log z_b \right) =
     {\Gamma}_{\hat{T}} \!  \left( \log z \right) \equiv
     {\Gamma}_{\hat{T}} \!  \left( q \right)$ from
     Eq.~(\ref{eq:psi_to_Gamma}).  We therefore proceed to solve for
     the classical stationary paths as in the preceding section, and
     dispense with further consideration of these extra, canceling
     terms.

The stationary-path equations, which are the \emph{equations of
  motion} with respect to the dynamical system, in action-angle
variables are
\begin{eqnarray}
  {\partial}_{\tau} \eta 
& = & 
  \frac{\partial \hat{\mathcal{L}}}{\partial \nu}
\nonumber \\
  {\partial}_{\tau} \nu 
& = & 
  - \frac{\partial \hat{\mathcal{L}}}{\partial \eta} . 
\label{eq:Hamilton_EOM}
\end{eqnarray}
Again $\hat{\mathcal{L}} \equiv 0$, but now a non-trivial relation
exists between the generating function and the stationary-path action,
which may be written 
\begin{equation}
  {\Gamma}_{\hat{T}} \! \left( \log z \right) = 
  N \int_0^{\hat{T}} d\tau 
  \left( 
    -\frac{\partial \hat{\mathcal{L}}}{\partial \nu} \nu + 
    \hat{\mathcal{L}} 
  \right) + 
  {\Gamma}_0 \! \left( 0 \right) .  
\label{eq:gamma_simple_form}
\end{equation}
Here $\nu$ and (the implicitly present) $\eta$ are evaluated over a
stationary path.  Of course the second factor $\hat{\mathcal{L}} \equiv 0$
need not have been written, but it serves to emphasize the
Legendre-dual relation between the Liouville/Hamiltonian operator and
its stochastic ``Lagrangian'' $-{\partial}_{\tau} \eta \nu +
\hat{\mathcal{L}}$, which is similar to the duality between the
generating functional and the effective action.

\subsubsection{The canonical versus the kinematic description, and its
  consequences for time-reversal in one-dimensional systems.}
\label{sec:kin_description}

The cumulant-generating function, evaluated as a stationary-path
integral in Eq.~(\ref{eq:gamma_simple_form}), is the
\emph{quasipotential} of Freidlin-Wentzell theory.  For single-time
fluctuations, it is a difference of Gibbs free energies, by
Eq.~(\ref{eq:Gamma_to_Gs}).  With respect to the underlying dynamical
system, however, the quasipotential is an \emph{action}, with the
Liouville operator acting as its conserved Hamiltonian.  The Liouville
operator, however, still masks the deep simplicity of one-dimensional
stochastic processes, which is provided by yet another level of
description, in terms of a \emph{kinematic} potential.  The kinematic
potential is the counterpart to the familiar potential energy in
Hamiltonian dynamics, and it provides the most intuitive connection to
finite-temperature instanton methods~\cite{Coleman:AoS:85}.  Here we
show how the kinematic description is extracted, and use it to prove
that, for one-dimensional systems, the most-likely path to arrive at
any fluctuation is the time-reverse of the classical
diffusive-relaxation path from that fluctuation.  This extension of
Onsager's near-equilibrium
results~\cite{Onsager:RRIP1:31,Onsager:RRIP2:31} is independent of the
magnitude of the fluctuation or the form of the potential.  The
ability to prove it so easily in the general case is an example of the
power of Freidlin-Wentzell methods in some circumstances.

In dynamical-systems terms, $\nu$ is a field with canonical momentum
$\eta$ with respect to the Hamiltonian $\hat{\mathcal{L}}$.  While
$\eta$ is a \emph{canonical} momentum, however, it does not play the
role of a \emph{kinematic} momentum in the stationary-path solutions.
To express the kinematic variables, we recall that ${\bar{\nu}}_a /
{\bar{\nu}}_b = e^{\beta \bar{\mu}}$ from
Sec.~\ref{sec:genfunc_equil}, or equivalently $2 \bar{\nu} = - \tanh
\left( \beta \bar{\mu} / 2 \right)$.  Correspondingly, as a purely
notational device, for any instantaneous value of $\nu$, we may denote
\begin{equation}
  \frac{{\nu}_a}{{\nu}_b} \equiv 
  e^{\beta \mu} ,  
\label{eq:nus_zeta_def}
\end{equation}
so that $\mu$ is the chemical potential for which that value of $\nu$
would result at an equilibrium.  With this notation we may recast
Eq.~(\ref{eq:S_ass_diss_AA_red}) -- as promised, now ignoring the
surface term -- as
\begin{widetext}
\begin{eqnarray}
  S 
& = & 
  N 
  \int d\tau 
  \left\{ 
    - {\partial}_{\tau}
    \eta \, 
    \nu + 
    2 \sqrt{
      {\bar{\nu}}_a {\bar{\nu}}_b
      {\nu}_a {\nu}_b
    }
    \left[
      \cosh 
      \left( 
        \frac{\beta}{2}
        \left( \mu - \bar{\mu} \right)
      \right) - 
      \cosh 
      \left( 
        \eta + 
        \frac{\beta}{2}
        \left( \mu - \bar{\mu} \right)
      \right) 
    \right]
  \right\} 
\nonumber \\
& = & 
  N 
  \int d\tau 
  \left\{ 
    - {\partial}_{\tau}
    \eta \, 
    \nu + 
    \frac{
      \sinh \left( \eta / 2 \right)
    }{
      \cosh \left( \beta \bar{\mu} / 2 \right)
    }
    \left[
      \sinh 
      \left( 
        \frac{\beta \bar{\mu} - \eta}{2}
      \right) + 
      2 \nu 
      \cosh 
      \left( 
        \frac{\beta \bar{\mu} - \eta}{2}
      \right)
    \right]
  \right\} . 
\label{eq:S_ass_diss_AA_simplfd}
\end{eqnarray}
\end{widetext}

The first line of Eq.~(\ref{eq:S_ass_diss_AA_simplfd}) shows us that
$\eta - \beta \left( \bar{\mu} - \mu \right) / 2$ is, to leading
quadratic order about zero, the term that creates the ``kinetic
energy'' in the Hamiltonian $\hat{\cal L}$.  Its vanishing results in
the stationary solution ${\partial}_{\tau} \nu = 0$ for its conjugate
number field.  The $\eta$-independent part of the Hamiltonian, $- 2
\sqrt{ {\bar{\nu}}_a {\bar{\nu}}_b {\nu}_a {\nu}_b } \left\{ \cosh
\left[ \beta \left( \bar{\mu} - \mu \right) / 2 \right] - 1 \right\}$,
defines the kinematic potential for the dynamical system.  The fields
$\nu$, $\eta - \beta \left( \bar{\mu} - \mu \right) / 2$ follow a
familiar Hamiltonian phase-space dynamics in this potential, with the
kinematic momentum vanishing at the turning points $\hat{\cal L} = 0$,
and the potential vanishing only where the mass-action rates satisfy
conditions of detailed balance.  These total stationary points are all
unstable, just as for problems of barrier penetration or escape in
equilibrium Hamiltonian statistical mechanics~\cite{Coleman:AoS:85}.
A more detailed treatment, for problems with multiple basins of
attraction and an interesting instanton structure as a result, is
provided in Ref.~\cite{Smith:evo_games:}.

The very strong consequence of the form in the first line, in one
dimension, is that the conservation law $\hat{\mathcal{L}} \equiv 0$
has only two solutions at any value of $\nu$: $\eta = 0$ and $\eta =
\beta \left( \bar{\mu} - \mu \right)$, by symmetry of the second
$\cosh$.  Then, by antisymmetry of the gradient of the same $\cosh$,
${\partial}_{\tau} \nu$ is \emph{equal in magnitude and opposite in
  sign} at these two solutions.  Thus, the stationary-path solution at
nonzero $\eta$ is the time-reverse in $\nu$ of the classical diffusion
solution.  Moreover, we have $\beta \mu = \beta \bar{\mu} - \eta$ at
this solution, which together with the form in the second line of
Eq.~(\ref{eq:S_ass_diss_AA_simplfd}) immediately gives
\begin{equation}
  \nu = 
  \frac{1}{2}
  \tanh 
  \frac{1}{2}
  \left( \eta - \beta \bar{\mu} \right) . 
\label{eq:time_reverse_nu}
\end{equation}

For these solutions we may recover the value of the single-time
cumulant-generating function from the action-angle equations of motion.
Vanishing $\hat{\mathcal{L}}$ gives the action as an integral over its
kinetic terms along the stationary path, which no longer vanish, 
\begin{eqnarray}
  {\Gamma}_T \! \left( \log z \right) - 
  {\Gamma}_0 \! \left( 0 \right) 
& = & 
  - N \int_0^{\hat T} d\tau \, 
  {\partial}_{\tau} \eta \, 
  \nu
\nonumber \\
& = & 
  - N \int_0^{\log z} d\eta \, \nu . 
\label{eq:shift_Gamma_formal}
\end{eqnarray}
Using the expression~(\ref{eq:time_reverse_nu}) in the last line of
Eq.~(\ref{eq:shift_Gamma_formal}), we then obtain
\begin{equation}
  {\Gamma}_T \! \left( \log z \right) - 
  {\Gamma}_0 \! \left( 0 \right) = 
  {
    \left. 
      - N \log \cosh 
      \frac{1}{2}
      \left( \eta - \beta \bar{\mu} \right) 
    \right| 
  }_0^{\log z} , 
\label{eq:shift_Gamma_solve}
\end{equation}
which again reproduces Eq.~(\ref{eq:Gamma_q_compute}).

\subsubsection{Langevin approximation and the magnitude of
  fluctuations about time-dependent solutions}
\label{sec:Langevin_approx}

It is important that the equilibrium generating
function~(\ref{eq:gen_func_reduced_def}) produces not only an offset
equilibrium value, but an entire distribution corresponding to an
effective shift in the chemical potential between the $a$ and $b$
states.  The width of this distribution is identical to the
fluctuation variance in $\nu$ produced by the Gaussian integral with
action~(\ref{eq:S_ass_diss_AA_simplfd}), as we now show.  The easiest
demonstration is by means of the Langevin approximation, which is also
useful to show how the classical Langevin equation is generated from
the more complete Doi-Peliti functional integral.  This construction
is particularly elegant in action-angle variables, which lead to a
Langevin equation directly for the particle numbers.  The
coherent-state variables in which the Langevin equation is usually
constructed~\cite{Krishnamurthy:Signaling:07} only offer this
interpretation for fluctuations about classical diffusion paths, where
${\phi}^{\dagger} \equiv 1$, and they can become quite complicated to
compute about other backgrounds.  In action-angle variables, the same
interpretation is valid in for all paths, including those that
propagate conditions about single-time fluctuation backward in time. 

The fact that the Gaussian integral delivers the fluctuation magnitude
(and more generally, the correct correlation structure) even for
time-dependent solutions is an important improvement over an approach
often taken with Langevin equations, which is simply to ``guess'' a
fluctuation magnitude as an independent input to
models~\cite{Ghosh:caliber:06,Krishnamurthy:Signaling:07}.  The
variance produced is also the unique value for which the terms in the
path entropies of Sec.~\ref{sec:caliber} will cancel to produce the
continuous-time hydrodynamic limit of
Ref.~\cite{Gaspard:rev_entropy:04}.

The Langevin stochastic differential equation encodes exactly the same
approximations as the Gaussian approximation to fluctuations in the
functional integral.  About any solutions ${\eta}^{\mbox{\scriptsize
    cl}}$, ${\nu}^{\mbox{\scriptsize cl}}$, to the stationary-point
equations, we write general $\eta = {\eta}^{\mbox{\scriptsize cl}} +
{\eta}^{\prime}$, $\nu = {\nu}^{\mbox{\scriptsize cl}} +
{\nu}^{\prime}$, and expand $S$ from Eq.~(\ref{eq:S_ass_diss_AA_red})
to second order in primes.
\begin{widetext}
\begin{equation}
  S = 
  S^{\mbox{\scriptsize cl}} + 
  N \int d\tau
  \left\{ 
    {\eta}^{\prime}
    \left[ 
      {\partial}_{\tau} + 
      \left(
        {\bar{\nu}}_b 
        e^{
          {\eta}^{\mbox{\scriptsize cl}}
        } + 
        {\bar{\nu}}_a 
        e^{
          - {\eta}^{\mbox{\scriptsize cl}}
        } 
      \right)
    \right]
    {\nu}^{\prime} - 
    \frac{1}{2}
    \left(
      {\bar{\nu}}_b {\nu}_a^{\mbox{\scriptsize cl}} 
      e^{
        {\eta}^{\mbox{\scriptsize cl}}
      } + 
      {\bar{\nu}}_a {\nu}_b^{\mbox{\scriptsize cl}}
      e^{
        - {\eta}^{\mbox{\scriptsize cl}}
      } 
    \right)
    {{\eta}^{\prime}}^2
  \right\}
\label{eq:S_2nd_ord_expand}
\end{equation}
In Eq.~(\ref{eq:S_2nd_ord_expand}) we denote by $S^{\mbox{\scriptsize
cl}} = N \int d\tau \left[ -
{\partial}_{\tau}{\eta}^{\mbox{\scriptsize cl}}
{\nu}^{\mbox{\scriptsize cl}} + \hat{\cal L} \! \left(
{\eta}^{\mbox{\scriptsize cl}}, {\nu}^{\mbox{\scriptsize cl}} \right)
\right] $ the action of the classical stationary path.  Terms linear
in ${\eta}^{\prime}$ and ${\nu}^{\prime}$ vanish as the condition for
the stationary-point solutions to hold.

We could complete the square in ${\eta}^{\prime}$ in
Eq.~(\ref{eq:S_2nd_ord_expand}), leading to the construction of
Onsager and Machlup~\cite{Onsager:Machlup:53}, and we will do this in
a later section.  An alternative approach, pursued here, is to perform
the \emph{Hubbard-Stratonovich}
transformation~\cite{Weinberg:QTF_I:95,Weinberg:QTF_II:96} by
introducing an auxiliary field $\lambda$, with a functional integral
${\left( \mbox{Det} \!  \left[ {\bar{\nu}}_b
    {\nu}_a^{\mbox{\scriptsize cl}} + {\bar{\nu}}_a
    {\nu}_b^{\mbox{\scriptsize cl}} \right] \right)}^{1/2} \int {\cal
  D}\lambda \exp S^{\mbox{\scriptsize Aux}}$ which is just a
representation of unity, in which the auxiliary field action is given
by
\begin{equation}
  S^{\mbox{\scriptsize Aux}} \equiv 
  \frac{N}{2} 
  \int d\tau 
  \frac{
    1
  }{
    \left(
      {\bar{\nu}}_b {\nu}_a^{\mbox{\scriptsize cl}} 
      e^{
        {\eta}^{\mbox{\scriptsize cl}}
      } + 
      {\bar{\nu}}_a {\nu}_b^{\mbox{\scriptsize cl}}
      e^{
        - {\eta}^{\mbox{\scriptsize cl}}
      } 
    \right)
  }
  {
    \left[ 
      \lambda - 
      \left(
        {\bar{\nu}}_b {\nu}_a^{\mbox{\scriptsize cl}} 
        e^{
          {\eta}^{\mbox{\scriptsize cl}}
        } + 
        {\bar{\nu}}_a {\nu}_b^{\mbox{\scriptsize cl}}
        e^{
          - {\eta}^{\mbox{\scriptsize cl}}
        } 
      \right)
      {\eta}^{\prime}
    \right]
  }^2
\label{eq:aux_action_def}
\end{equation}
The sum of actions for the original and auxiliary fields becomes 
\begin{equation}
  S + 
  S^{\mbox{\scriptsize Aux}} = 
  \frac{N}{2} 
  \int d\tau 
  \frac{
    {\lambda}^2 
  }{
    \left(
      {\bar{\nu}}_b {\nu}_a^{\mbox{\scriptsize cl}} 
      e^{
        {\eta}^{\mbox{\scriptsize cl}}
      } + 
      {\bar{\nu}}_a {\nu}_b^{\mbox{\scriptsize cl}}
      e^{
        - {\eta}^{\mbox{\scriptsize cl}}
      } 
    \right)
  } + 
  2 {\eta}^{\prime}
  \left\{
    \left[ 
      {\partial}_{\tau} + 
      \left(
        {\bar{\nu}}_b 
        e^{
          {\eta}^{\mbox{\scriptsize cl}}
        } + 
        {\bar{\nu}}_a 
        e^{
          - {\eta}^{\mbox{\scriptsize cl}}
        } 
      \right)
    \right]
    {\nu}^{\prime} - 
    \lambda 
  \right\}
\label{eq:sum_actions_HS}
\end{equation}
The integral $\int {\cal D} {\eta}^{\prime}$ in the original
functional integral, if rotated to an imaginary contour of
integration,\footnote{Note that ${\eta}^{\prime}$ is always integrated
  along an imaginary contour, as the condition for stability of the
  Gaussian integral.  This is true in the Onsager-Machlup
  construction, and it is also the imaginary part of ${\eta}^{\prime}$
  that defines the quantity behaving as a momentum in the kinematic
  description.} simply produces the functional $\delta$-function
\begin{equation}
  \int {\cal D} {\eta}^{\prime}
  e^{- 
    \left( 
      S + 
      S^{\mbox{\scriptsize Aux}}
    \right) 
  } = 
  \delta \! 
  \left[ 
    \left[ 
      {\partial}_{\tau} + 
      \left(
        {\bar{\nu}}_b 
        e^{
          {\eta}^{\mbox{\scriptsize cl}}
        } + 
        {\bar{\nu}}_a 
        e^{
          - {\eta}^{\mbox{\scriptsize cl}}
        } 
      \right)
    \right]
    {\nu}^{\prime} - 
    \lambda 
  \right]
  \exp
  \left\{ 
    - \frac{N}{2} 
    \int d\tau
    \frac{
      {\lambda}^2 
    }{
      \left(
        {\bar{\nu}}_b {\nu}_a^{\mbox{\scriptsize cl}} 
        e^{
          {\eta}^{\mbox{\scriptsize cl}}
        } + 
        {\bar{\nu}}_a {\nu}_b^{\mbox{\scriptsize cl}}
        e^{
          - {\eta}^{\mbox{\scriptsize cl}}
        } 
      \right)
    }
  \right\} . 
\label{eq:int_deta_pr_delta}
\end{equation}
\end{widetext}
The last of the original functional integrals, over ${\nu}^{\prime}$
admits only those solutions which satisfy the \emph{Langevin equation}
\begin{equation}
  \left[ 
    {\partial}_{\tau} + 
    \left(
      {\bar{\nu}}_b 
      e^{
        {\eta}^{\mbox{\scriptsize cl}}
      } + 
      {\bar{\nu}}_a 
      e^{
        - {\eta}^{\mbox{\scriptsize cl}}
      } 
    \right)
  \right]
  {\nu}^{\prime} = 
  \lambda . 
\label{eq:Langevin}
\end{equation}
$\lambda$, known as the \emph{Langevin field}, has the correlation
function at any two times
\begin{equation}
  \left<
    {\lambda}_{\tau}
    {\lambda}_{{\tau}^{\prime}}
  \right> = 
  \frac{
    {\bar{\nu}}_b {\nu}_a^{\mbox{\scriptsize cl}} 
    e^{
      {\eta}^{\mbox{\scriptsize cl}}
    } + 
    {\bar{\nu}}_a {\nu}_b^{\mbox{\scriptsize cl}}
    e^{
      - {\eta}^{\mbox{\scriptsize cl}}
    } 
  }{
    N 
  }
  \delta \! 
  \left( 
    \tau - {\tau}^{\prime}
  \right) , 
\label{eq:lambda_corr_fn}
\end{equation}
as a result of the Gaussian kernel of integration. 

A case of special interest, for its relation to the single-time
distribution -- at or away from equilibrium -- is the fluctuation
variance at equal times $\left< { \left( {\nu}^{\prime}_{\tau} \right)
}^2 \right>$ about a constant or long-time persistent background.  We
will not provide details about the inversion of
Eq.~(\ref{eq:Langevin}) to express ${\nu}^{\prime}$ in terms of
$\lambda$ and a retarded Green's function, which may be found in
Ref.~\cite{Kamenev:DP:01}.  However, the general result is that for
backgrounds that persist much longer than the decay time $ \left(
{\bar{\nu}}_b e^{ {\eta}^{\mbox{\scriptsize cl}} } + {\bar{\nu}}_a e^{
- {\eta}^{\mbox{\scriptsize cl}} } \right)$, the single-time variance
is given by
\begin{equation}
  \left<
    { 
      \left( {\nu}^{\prime}_{\tau} \right)
    }^2
  \right> = 
  \frac{
    {\bar{\nu}}_b {\nu}_a^{\mbox{\scriptsize cl}} 
    e^{
      {\eta}^{\mbox{\scriptsize cl}}
    } + 
    {\bar{\nu}}_a {\nu}_b^{\mbox{\scriptsize cl}}
    e^{
      - {\eta}^{\mbox{\scriptsize cl}}
    } 
  }{
    2 N 
    \left(
      {\bar{\nu}}_b 
      e^{
        {\eta}^{\mbox{\scriptsize cl}}
      } + 
      {\bar{\nu}}_a 
      e^{
        - {\eta}^{\mbox{\scriptsize cl}}
      } 
    \right)
  } . 
\label{eq:nu_pr_corr_fn}
\end{equation}
We will return in Sec.~\ref{sec:fixed_diseq} to check that this result
agrees with the variance of the single-time distribution given by the
equilibrium generating function~(\ref{eq:rho_eq_binomial_source}).

We have given a thorough treatment of the single-time generating
function to provide multiple points of reference for terms that cause
${\phi}^{\dagger}$ fields to deviate from unity, or $\eta$ to deviate
from zero, and to interpret their effect on stationary points.  It
will now be straightforward to place all such terms entirely within
the functional integral rather than in boundary terms.  We will do so
first for a discrete source with an identical effect to the
single-time generating function, and examine the future as well as
past properties of the stationary paths.  We will then define exact
solutions for the more general case of continuous sources, and finally
study the analytic structure of the small-fluctuation limit, which is
simply a linear expansion in point sources.

\subsection{Generating functionals and arbitrary time-dependent
  sources} 
\label{sec:DP_gen_fn_srcs}

When sources interior to functional integrals are used to compute the
probabilities of complicated histories, the construction is usually
done in two disconnected stages.  First, the final-time arguments
$z_i$ are set to one, so that the moment-generating function $\psi \!
\left( z_a, z_b \right)$ becomes simply the trace of the probability
distribution at time $T$.  That is, the elaborate field
integrals~(\ref{eq:gen_fn_twofields_nosource},\ref{eq:gen_fn_AA_nosource})
are simply complicated expressions for the number 1.  New terms
involving the integration variables and external fields are then
simply inserted into the action, with the understanding that these are
``sources'' that perturb the particle motion.

Here we will introduce sources systematically as part of the
construction of the functional integral, to make clear their
connection to the original construction of the generating function.
The construction involves modifying the map by which we have, up to
now, simply re-interpreted intermediate variables $z$ and $\partial /
\partial z$ as operators $a^{\dagger}$ and $a$.  

A generating \emph{functional} is the time-evolved product of a
sequence of \emph{generating functions} produced by adding small
continuous sources to all previous distributions.  That is, we replace
the identification~(\ref{eq:DP_annih_op}) with one in which the
abstract operators differ from variables $z$ by the addition of
sources.  At any time $\tau$, we identify
\begin{eqnarray}
  z_j 
& \leftrightarrow & 
  a_j^{\dagger} 
  e^{q_{\tau}},  \mbox{ keeping} 
\nonumber \\
  \frac{\partial}{\partial z_i}
& \leftrightarrow & 
  a^i . 
\label{eq:DP_annih_op_j}
\end{eqnarray}
Moreover, to separate the discretization time $\delta \tau$ from
characteristic timescales associated with the sources, we will write
\begin{equation}
  q_{\tau} = 
  j \! \left( \tau \right) \delta \tau , 
\label{eq:q_to_js}
\end{equation}
and take $j \! \left( \tau \right) \equiv j_{\tau}$, called a
\emph{current}, to be a smooth function of $\tau$ in the limit $\delta
\tau \rightarrow 0$.  Because both relaxation and fluctuation effects
are governed by timescales in $\Delta \tau \sim 1$, we may readily
consider sources $j_{\tau}$ that have large magnitude over some range
in $\tau$ that is $\ll 1$.  Currents of this form may be constructed
to approximate the point source $j_{\tau} \approx q \delta \!  \left(
\tau - {\tau}_C \right)$ for some particular time ${\tau}_C$.

Finally we set values $z_a = 1$, $z_b = 1$ at time $\hat{T}$, keeping
only sources from nonzero $j \! \left( \tau \right)$ in the range $0 <
\tau < \hat{T}$.  The field integral at $\tau = T$ therefore simply
computes a trace, and all correlations are studied internally in the
functional integral.  The notation $\psi \!  \left( z_a , z_b \right)$
is no longer needed, and we simply refer explicitly to the
cumulant-generating functional $\Gamma \! \left[ j \right]$ whose
argument is the function $j$.

Both coherent-state field variables and action-angle variables will be
of interest, and so we provide both forms here.
Eq.~(\ref{eq:gen_fn_twofields_nosource}) is replaced by 
\begin{widetext}
\begin{equation}
  e^{- \Gamma \left[ j \right]} = 
  \int_0^T 
  \mathcal{D} {\phi}^{\dagger}_a \mathcal{D} {\phi}_a
  \mathcal{D} {\phi}^{\dagger}_b \mathcal{D} {\phi}_b
  e^{
    \left( 1 - {{\phi}^{\dagger}_a}_T \right) {{\phi}_a}_T + 
    \left( 1 - {{\phi}^{\dagger}_b}_T \right) {{\phi}_b}_T - 
    S_j - 
    {\Gamma}_0 
    \left( 
      \log {{\phi}^{\dagger}_a}_0 , 
      \log {{\phi}^{\dagger}_b}_0 
    \right)
  } , 
\label{eq:gen_fn_twofields_source}
\end{equation}
in which 
\begin{equation}
  S_j \equiv 
  N \int d\tau 
  \left[ 
    - \left( 
      {\partial}_{\tau}
      {\phi}^{\dagger}_a {\hat{\phi}}_a + 
      {\partial}_{\tau}
      {\phi}^{\dagger}_b {\hat{\phi}}_b 
    \right) - 
    \frac{j}{2} 
    \left( 
      {\phi}^{\dagger}_b {\hat{\phi}}_b - 
      {\phi}^{\dagger}_a {\hat{\phi}}_a 
    \right) + 
    \hat{\mathcal{L}}
  \right] . 
\label{eq:action_twofields_descaled_j}
\end{equation}
\end{widetext}
Likewise, Eq.~(\ref{eq:gen_fn_AA_nosource}) -- removing the un-needed
integration variable ${\eta}_a + {\eta}_b$ conjugate to $N$, and
writing the single-argument generating function alone -- becomes 
\begin{equation}
  e^{- \Gamma \left[ j \right]} = 
  \int_0^T 
  \mathcal{D} \eta \mathcal{D} n
  e^{
    \left( e^{- {\eta}_T} - 1 \right) n_T - 
    S_j - 
    {\Gamma}_0 
    \left( 
      {\eta}_0 
    \right)
  } , 
\label{eq:gen_fn_AA_source}
\end{equation}
in which 
\begin{equation}
  S_j = 
  N \int d\tau 
  \left[  
    - {\partial}_{\tau}
    \eta \, 
    \nu - 
    j \nu + 
    \hat{\mathcal{L}} 
  \right] 
\label{eq:action_actionangle_j}
\end{equation}
with Liouville operator
\begin{equation}
  \hat{\mathcal{L}} \equiv 
  {\bar{\nu}}_b {\nu}_a
  \left(
    1 - 
    e^{\eta}
  \right) +
  {\bar{\nu}}_a {\nu}_b
  \left(
    1 - 
    e^{-\eta}
  \right) . 
\label{eq:hatcalL_AA_def}
\end{equation}

The upper time limit $T$ may now be taken to infinity and dropped from
the notation.  If we consider sources $j_{\tau} \rightarrow 0$ at both
$\tau \rightarrow 0$ and $\tau \rightarrow \infty$ and which are
smooth on the scale of the discrete sum (not a very restrictive
condition) it follows from the stationary-path conditions in the
presence of the source $j$ that
\begin{eqnarray}
  {\hat{\mathcal{L}}}_j 
& \equiv & 
  \hat{\mathcal{L}} - j 
  \left( 
    {\phi}^{\dagger}_b {\hat{\phi}}_b - 
    {\phi}^{\dagger}_a {\hat{\phi}}_a 
  \right) / 2 
\nonumber \\
& = & 
  \hat{\mathcal{L}} - j \nu
\label{eq:hatcal_L_j_def}
\end{eqnarray}
is the Hamiltonian of the dynamical system in the presence of sources.
Because ${\hat{\mathcal{L}}}_j$ now depends explicitly on time through
$j$, Eq.~(\ref{eq:conservation_Liouville}) is replaced by
\begin{equation}
  \frac{
    d {\hat{\mathcal{L}}}
  }{
    d\tau 
  } \equiv 
  - \nu 
  \frac{
    d j 
  }{
    d\tau
  } . 
\label{eq:non_cons_L_j}
\end{equation}

We also have as boundary conditions that ${\eta}_{\infty} = 0$ (more
generally ${{\eta}_a}_{\infty} = 0$ and ${{\eta}_b}_{\infty} = 0$
independently if we had kept both sets of integration variables), or
equivalently ${{\phi}^{\dagger}_a}_{\infty} = 1$ and
${{\phi}^{\dagger}_b}_{\infty} = 1$.  To understand how these
conditions lead to stationary-point solutions, which involves the
stability structure of the functional integrals, it is easiest to
solve some examples.  The general exact solution, in coherent-state
fields, is given in App.~\ref{sec:exact_gen_fun}.  

\subsection{General solutions}

In all the following cases, we will suppose that the initial
distribution is the equilibrium
distribution~(\ref{eq:rho_eq_binomial}).  Other initial conditions can
easily be considered, but all converge to the equilibrium
exponentially in $\tau$, and so may be decoupled to any desired degree
from the influence of $j$ at later times.  We begin with a point
source which recovers the single-time generating function, and then
consider sources extended in time.

\subsubsection{Point sources}
\label{sec:point_sources}

Suppose that $j_{\tau} \rightarrow q \delta \!  \left( \tau - {\tau}_C
\right)$ for some time ${\tau}_C$ in the sense of convergence of
smooth distributions with compact support and fixed area $q$.  The
exact solution in this case is most simply expressed in action-angle
variables, which can then be converted to coherent-state field
variables if desired.

From the action form~(\ref{eq:action_actionangle_j}) with this source,
stationary solutions are just those of the unperturbed action except
at ${\tau}_C$, where $\eta$ has the discontinuity 
\begin{equation}
  {\eta}_{{\tau}_C + \epsilon} = 
  {\eta}_{{\tau}_C - \epsilon} -
  q 
\label{eq:eta_disc_q}
\end{equation}
as $\epsilon \rightarrow 0$.  To understand what this implies, we
return to the kinematic form for~(\ref{eq:S_ass_diss_AA_simplfd}) for
the action, using ${\eta}^{\mbox{\scriptsize cl}} + {\eta}^{\prime}$
as in Sec.~\ref{sec:Langevin_approx}.  $\beta \mu$ is a function of
$\nu$ from Eq.~(\ref{eq:nus_zeta_def}), which we indicate on the
stationary path by writing ${\mu}^{\mbox{\scriptsize cl}}$.  For free
solutions we already know that $\hat{\mathcal L} = 0$ requires either
${\eta}^{\mbox{\scriptsize cl}} = 0$ or ${\eta}^{\mbox{\scriptsize
    cl}} = \beta \left( \bar{\mu} - {\mu}^{\mbox{\scriptsize cl}}
\right)$.  Therefore we may write
\begin{eqnarray}
\lefteqn{
  \cosh 
  \left( 
    \frac{\beta}{2}
    \left( 
      {\mu}^{\mbox{\scriptsize cl}} - \bar{\mu} 
    \right)
  \right) - 
  \cosh 
  \left( 
    \eta + 
    \frac{\beta}{2}
    \left( 
      {\mu}^{\mbox{\scriptsize cl}} - \bar{\mu} 
    \right)
  \right) = 
} & & 
\nonumber \\
& & 
  \cosh 
  \left( 
    \frac{\beta}{2}
    \left( 
      {\mu}^{\mbox{\scriptsize cl}} - \bar{\mu} 
    \right)
  \right) 
  \left( 
    1 - \cosh {\eta}^{\prime}
  \right) 
\nonumber \\
& & 
  \mbox{} + 
  \sinh 
  \left( 
    {\eta}^{\mbox{\scriptsize cl}} + 
    \frac{\beta}{2}
    \left( 
      {\mu}^{\mbox{\scriptsize cl}} - \bar{\mu} 
    \right)
  \right) 
  \sinh {\eta}^{\prime} . 
\label{eq:flucts_eta_AA}
\end{eqnarray}
Our concern is with the second line of Eq.~(\ref{eq:flucts_eta_AA}).
This quantity appears with positive sign in $\hat{\mathcal{L}}$ and so
describes \emph{divergent} fluctuations for ${\eta}^{\prime}$
integrated along a real contour.  As in the case of the
Hubbard-Stratonovich transformation, the convergent contour of
integration for ${\eta}^{\prime}$ is imaginary, where it behaves like
an ordinary momentum variable in finite-temperature field
theory~\cite{Mahan:MPP:90}.

\begin{figure}[ht]
  \begin{center} 
  \includegraphics[scale=0.5]{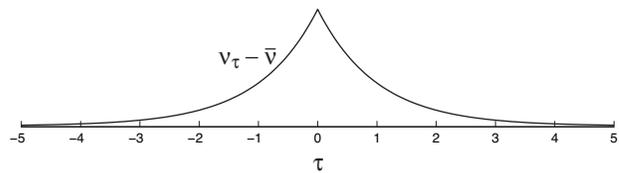}
  \caption{
  The time-symmetric, exponentially decaying stationary path
  associated with single-time $\delta$-function sources.  This
  solution illustrates the embedding of the single-time equilibrium
  large-deviations theory within the richer time-dependent
  large-deviations theory of the stochastic process.  In the forward
  direction, classical diffusion is responsible for the decay, and the
  response field $\eta \equiv 0$.  In the reverse direction, $\eta$
  back-propagates the constraint that a fluctuation of magnitude
  $\left( {\nu}_{{\tau}_C} - \bar{\nu} \right)$ is observed at time
  ${\tau}_C$, and the dynamical-system representation generates the
  least-improbable trajectory of fluctuations that achieve that
  constraint. 
    \label{fig:sym_exp_decay_soln} 
  }
  \end{center}
\end{figure}

The instability that governs the integration contour for fluctuations
also implies that time evolution in the forward direction for
real-valued $\eta$ is \emph{unstable} in general.  This should not be
surprising, as we have already seen that the evolution equations for
$\eta$ are stable going backward in time, as $\eta$ propagates context
from observables to their most-likely prior causes.  The implication
for our stationary path is that if ${\eta}^{\mbox{\scriptsize
    cl}}_{\infty} = 0$ and $j_{\tau} \equiv 0$ for $\tau > {\tau}_C +
\epsilon$, we must have ${\eta}^{\mbox{\scriptsize cl}}_{\tau} \equiv
0$ for all $\tau > {\tau}_C + \epsilon$.  Hence
${\eta}^{\mbox{\scriptsize cl}}_{{\tau}_C - \epsilon} = q$ and for
$\tau < {\tau}_C + \epsilon$ we are back to the stationary path
evaluated in Sec.~\ref{sec:kin_description}.  The consequence for the
number field is that its magnitude at $\tau = {\tau}_C$ is just that
of Eq.~(\ref{eq:n_q_compute}), and that for times either before or
after ${\tau}_C$, ${\nu}^{\mbox{\scriptsize cl}}$ satisfies the
generalization of Eq.~(\ref{eq:nu_tau_time_genfunc}) to
\begin{equation}
  {\nu}_{\tau}^{\mbox{\scriptsize cl}} - 
  \bar{\nu} = 
  \left( 
    {\nu}^{\mbox{\scriptsize cl}}_{{\tau}_C} - \bar{\nu} 
  \right) 
  e^{
    - \left| \tau - {\tau}_C \right|
  } . 
\label{eq:nu_tau_time_genfunctal}
\end{equation}
The time-dependence of ${\nu}^{\mbox{\scriptsize cl}}$ is shown in
Fig.~\ref{fig:sym_exp_decay_soln}.  

The generating functional for the point source has the value already
computed, 
\begin{eqnarray}
  \Gamma \! \left[ j \right] 
& = & 
  {\Gamma}_0 \! \left( 0 \right) - 
  N \int_0^{\infty} d\tau \, 
  {\partial}_{\tau} {\eta}^{\mbox{\scriptsize cl}} \, 
  {\nu}^{\mbox{\scriptsize cl}}
\nonumber \\
& = & 
  {
    \left. 
      - N \log \cosh 
      \frac{1}{2}
      \left( 
        {\eta}^{\mbox{\scriptsize cl}} - 
        \beta \bar{\mu} 
      \right) 
    \right| 
  }_0^q , 
\label{eq:shift_Gamma_AA_pt_src}
\end{eqnarray}
again recovering Eq.~(\ref{eq:Gamma_q_compute}).  Note that although
${\nu}^{\mbox{\scriptsize cl}} \neq 0$ at all times, the integral
receives contributions only from $\tau < {\tau}_C$ where $\eta \neq 0$.

The effective action is numerically just that given in
Eq.~(\ref{eq:S_eff_n_compute}), because it differs from the generating
functional by subtraction of the point source $q {\nu}_{{\tau}_C}$ as
before.  However, as a Legendre transform on a space of
\emph{histories}, we now regard it as the functional
\begin{equation}
  S^{\mbox{\scriptsize eff}} \! 
  \left[ n \right] = 
  S^{\mbox{\scriptsize eff}} \! 
  \left[ 
    N 
    {\nu}^{\mbox{\scriptsize cl}}
  \right] , 
\label{eq:S_eff_pt_src}
\end{equation}
whose argument ${\nu}^{\mbox{\scriptsize cl}}$ is the whole
exponential history given by Eq.~(\ref{eq:nu_tau_time_genfunctal})
with ${\nu}_{{\tau}_C}$ fixed by Eq.~(\ref{eq:n_q_compute}).  The
entropy difference of Eq.~(\ref{eq:S_eff_G_diffs}) is now expressed
not only as a probability of a fluctuation of given magnitude, but of
the most-likely sequence of previous and subsequent configurations
consistent with that observed fluctuation but with no other
information given.

To understand the meaning of the stochastic effective action as a
probability on histories, it is necessary to appreciate the non-local
relation between histories and the structure of the moment-sampling
currents $j$ which contain the minimal information to specify them.
To extend the formulae for the effective action to more general
histories, a variety of methods are known which produce both non-local
and exact, or local but approximate, solutions. 

\subsubsection{Continuous sources; linear-response approximation and
  analytic structure}

We now return to the second-order expansion of
Eq.~(\ref{eq:action_actionangle_j}) in fluctuations, from
Sec.~\ref{sec:Langevin_approx}, but this time we complete the square
directly rather than through a Hubbard-Stratonovich transformation.
Following
Onsager~\cite{Onsager:RRIP1:31,Onsager:RRIP2:31,Onsager:Machlup:53},
we will expand about the equilibrium background where
${\nu}^{\mbox{\scriptsize cl}} \equiv \bar{\nu}$ and
${\eta}^{\mbox{\scriptsize cl}} \equiv 0$, and will simply write
${\eta}^{\prime} = \eta$.  (The more general problem of expanding
about non-classical backgrounds conditioned on sources could of course
be considered as well.)  The quadratic expansion of the action takes
the simple form
\begin{eqnarray}
  S_j 
& = & 
  - N \bar{\nu} 
  \int_0^{\infty} d\tau
  j_{\tau} 
\nonumber \\
& & 
  \mbox{} + 
  \int_0^{\infty} d\tau  
  \left\{
    \left( \nu - \bar{\nu} \right) 
    \left[ 
      \left( - {\partial}_{\tau} + 1 \right) 
      \eta - 
      j 
    \right] - 
    {\bar{\nu}}_q
    {\bar{\nu}}_b
    {\eta}^2
  \right\} . 
\nonumber \\
\label{eq:S_j_Gauss_AA}
\end{eqnarray}

The stationary-point equations are now linear-response equations, 
\begin{eqnarray}
  \left( - {\partial}_{\tau} + 1 \right) 
  \eta 
& = & 
  j 
\nonumber \\
  \left( {\partial}_{\tau} + 1 \right) 
  \left( \nu - \bar{\nu} \right) 
& = & 
  2 {\bar{\nu}}_a
  {\bar{\nu}}_b
  \eta . 
\label{eq:eta_EOM_lin_gen}
\end{eqnarray}
$\eta$ is the response to $j$ through the advanced Green's function: 
\begin{equation}
  {\eta}_{\tau} = 
  \int_{\tau}^{\infty} d{\tau}^{\prime} 
  e^{
    \tau - {\tau}^{\prime}
  }
  j_{{\tau}^{\prime}} , 
\label{eq:eta_from_j}
\end{equation}
while $\nu - \bar{\nu}$ responds through the symmetric Green's
function: 
\begin{equation}
  {\nu}_{\tau} - 
  \bar{\nu} = 
  {\bar{\nu}}_a
  {\bar{\nu}}_b
  \int_0^{\infty} d{\tau}^{\prime} 
  e^{
    - \left| \tau - {\tau}^{\prime} \right| 
  }
  j_{{\tau}^{\prime}} . 
\label{eq:nu_m_barnu_from_j}
\end{equation}
Eq.~(\ref{eq:nu_m_barnu_from_j}) reproduces the exact time dependence
of Eq.~(\ref{eq:nu_tau_time_genfunctal}) for a point source, and gives
the linear small-$q$ approximation to the
magnitude~(\ref{eq:n_q_compute}). 

We now have a closed-form \emph{local} expression for the generating
functional in terms of the response field $\eta$ and the source $j$,
because in these variables $\Gamma \! \left[ j \right]$ is just $S_j$
evaluated on the stationary solution.  From
Eq.~(\ref{eq:eta_EOM_lin_gen}), this is 
\begin{equation}
  \Gamma \! \left[ j \right] = 
    - N \bar{\nu} 
  \int_0^{\hat T} d\tau \, 
  j_{\tau} - 
  \frac{N}{2}
  \int_0^{\hat T} d\tau 
  \left( 
    2 {\bar{\nu}}_a
    {\bar{\nu}}_b
  \right) 
  {\eta}^2 , 
\label{eq:Gamma_lin_AA}
\end{equation}
with $\eta$ given by Eq.~(\ref{eq:eta_from_j}).  We recognize from
Eq.~(\ref{eq:rho_eq_binomial_approx}), the quantity $N {\bar{\nu}}_a
{\bar{\nu}}_b$ as the variance of the equilibrium distribution and
hence the susceptibility to perturbations in chemical potential.

From the definition~(\ref{eq:S_eff_exact_fields}) and use of the
equations of motion~(\ref{eq:eta_EOM_lin_gen}), the effective action
is similarly expressed as a local functional, 
\begin{eqnarray}
  S^{\mbox{\scriptsize eff}} \! 
  \left[ n \right] 
& = & 
  \frac{N}{2} 
  \int_0^{\infty} d\tau
  \frac{
    {
      \left[ 
        \left( {\partial}_{\tau} + 1 \right)
        \left( \nu - \bar{\nu} \right)
      \right]
    }^2 
  }{
    2 {\bar{\nu}}_a
    {\bar{\nu}}_b
  } 
\nonumber \\
& = & 
  \frac{1}{2} 
  \int_0^{\infty} d\tau
  \frac{
    {
      \left[ 
        \left( {\partial}_{\tau} + 1 \right)
        \left( n - \bar{n} \right)
      \right]
    }^2 
  }{
    2 N
    {\bar{\nu}}_a
    {\bar{\nu}}_b
  } . 
\label{eq:S_eff_Onsager_Machlup}
\end{eqnarray}
Eq.~(\ref{eq:S_eff_Onsager_Machlup}) is the expression first due to
Onsager and Machlup~\cite{Onsager:Machlup:53}.  The first line, as
usual, makes explicit the separation of large-deviations scaling in
$N$ from the integral that is the rate function, which is a function
only of fractional displacements $\nu - \bar{\nu}$.  

\subsection{The relation to Onsager's ``minimum entropy production''
  property}

Eq.~(\ref{eq:S_eff_Onsager_Machlup}) is a relation between an offset
$n - \bar{n}$ and a transport current ${\partial}_{\tau} n$, which we
might call $v$, a \emph{reaction velocity} in appropriate units of
time.  Recalling Eq.~(\ref{eq:S_eff_G_diffs}) for the log-probability
of single-time fluctuations, we may recognize the combination ${\left(
  n - \bar{n} \right)}^2 / \left( 2 N {\bar{\nu}}_a {\bar{\nu}}_b
\right)$ as none other than $S^{\mbox{\scriptsize eff}} \! \left( n
\right)$, the entropy deficit from the equilibrium large-deviations
principle.  Immediately we would then recognize $\left( n - \bar{n}
\right) / \left( N {\bar{\nu}}_a {\bar{\nu}}_b \right)$ as $\partial
S^{\mbox{\scriptsize eff}} \! \left( n \right) / \partial n$, and
$\left( n - \bar{n} \right) v / \left( N {\bar{\nu}}_a {\bar{\nu}}_b
\right)$ as the rate of change in the equilibrium entropy of
quasi-equilibrated subsystems.  From these associations we could
rewrite Eq.~(\ref{eq:S_eff_Onsager_Machlup}) in the form
\begin{eqnarray}
  S^{\mbox{\scriptsize eff}} \! 
  \left[ n \right] 
& = & 
  \frac{1}{2} 
  \int_0^{\infty} d\tau
  \left\{
    S^{\mbox{\scriptsize eff}} \! 
    \left( n \right) + 
    \frac{
      \partial 
      S^{\mbox{\scriptsize eff}} \! 
      \left( n \right) 
    }{
      \partial n
    } 
    v + 
    \frac{1}{2}
    \frac{
      {\partial}^2 
      S^{\mbox{\scriptsize eff}} \! 
      \left( n \right) 
    }{
      {\partial n}^2
    } 
    v^2 
  \right\} , 
\nonumber \\ 
\label{eq:S_eff_Onsager_Machlup_from_eq}
\end{eqnarray}
in which the term ${\partial}^2 S^{\mbox{\scriptsize eff}} \!  \left(
n \right) / {\partial n}^2$ gives the linear-response coefficients $1
/ \left( N {\bar{\nu}}_a {\bar{\nu}}_b \right)$. 

If we chose to regard minimization of $S^{\mbox{\scriptsize eff}} \!
\left[ n \right]$ as an integral over time of two separate criteria --
the first being the probability of a fluctuation to $n$ from
$S^{\mbox{\scriptsize eff}} \!  \left( n \right)$ and the second being
a minimization over $v$ -- the minimized function of $v$ given $n$
would be the ``entropy production'' relative to a bilinear form $v^2
{\partial}^2 S^{\mbox{\scriptsize eff}} \!  \left( n \right) /
{\partial n}^2$ parametrized by the near-equilibrium response
coefficients.  In two papers in
1931~\cite{Onsager:RRIP1:31,Onsager:RRIP2:31}, this was the result
derived by Onsager as a consequence of microscopic reversibility and
referred to as a ``minimum entropy production'' property.  Onsager
referred to the bilinear form of currents $v^2
{\partial}^2 S^{\mbox{\scriptsize eff}} \!  \left( n \right) /
{\partial n}^2$ as the ``dissipation function''.  

Note three things: 1) It is not the ``entropy production'' per se that
is minimized, but only its value relative to the dissipation function,
which from the form of the system entropy may be arbitrary; 2) there
is no obvious reason to interpret the dissipation function as a
``constraint'' on the entropy production, and thus entropy production
is not ``minimized subject to constraints'', as an analogy to the way
the entropy \emph{is} maximized subject to constraints in equilibrium;
and 3) the presence of the response coefficients in this form results
from the near-equilibrium expansion, as is well-known.

We see, then, how we may speak precisely about the range of
assumptions needed to make the ``production'' of the equilibrium
entropy a quantity that is informative about dynamics.  We have
required the particular form of the local-equilibrium approximation
that comes from a basins-and-barriers model, so that the dependence of
the stochastic effective action $S^{\mbox{\scriptsize eff}} \!  \left[
  n \right]$ on its instantaneous configuration variables reduces at
leading order to the equilibrium quantity $S^{\mbox{\scriptsize eff}}
\!  \left( n \right)$.  We have assumed small fluctuations in order to
expand the linear velocity dependence in terms of $\partial
S^{\mbox{\scriptsize eff}} \!  \left( n \right) / \partial n$, and to
approximate the linear response coefficients by their equilibrium
values.  Had all of these conditions not been satisfied, the
expansion~(\ref{eq:S_eff_Onsager_Machlup_from_eq}) need not have been
valid.  Yet, within the more general framework of the stochastic
process, we could readily have derived the correct alternative form
from the exact solution, which is derived as
Eq.~(\ref{eq:S_eff_exact_fields}) in App.~\ref{sec:exact_gen_fun}.
In the more general regime of non-linear response and large
perturbations from the equilibrium distribution, the relation between
a source $j \! \left( \tau \right)$ and the path $\nu \! \left( \tau
\right)$ that it induces will generally be non-local in $\tau$.  

\subsection{A second example: fixed dis-equilibria and an entropy rate
  for a large-deviation rate function}
\label{sec:fixed_diseq}

We now consider a second problem, aimed at relating dynamics to
reference static distributions.  Suppose that, rather than assuming
free relaxation immediately after a fluctuation of
magnitude~(\ref{eq:n_q_compute}), we ask for the probability of a
history which \emph{remains} at that value for a fixed time, and then
freely decays.  The difference of the effective action for this
fixed-disequilibrium trajectory from that for a single-time
fluctuation will give the \emph{entropy rate} difference, between the
stochastic process about the non-equilibrium steady state and the same
process acting on the equilibrium distribution.

\begin{figure}[ht]
  \begin{center} 
  \includegraphics[scale=0.5]{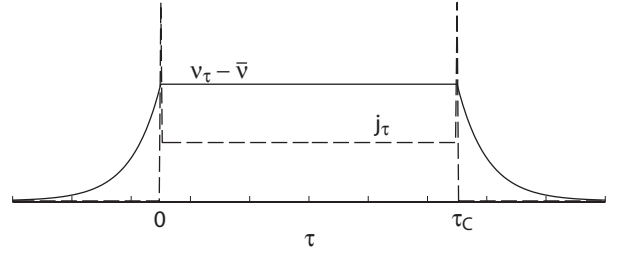}
  \caption{
  The source current (dashed) and resulting stationary solution
  (solid), for a source that raises $n_{\tau}$ to $n^{\left( q
    \right)}$ at $\tau = 0$, holds it at this value until $\tau =
  {\tau}_C$, and then releases it to free diffusive decay.
    \label{fig:ext_dis_eq_bw} 
  }
  \end{center}
\end{figure}

For this example, it is convenient to put the initial distribution at
time $\tau \rightarrow -\infty$, to put the initial fluctuation at
$\tau = 0$ and hold it until ${\tau}_C$, and then to permit free decay
after ${\tau}_C$ as before.  From the constructions above, it is easy
to define a source protocol that will produce this history.  The
source current takes the form 
\begin{eqnarray}
  j_{\tau} 
& = & 
  \Theta \! \left( {\tau}_C - \tau \right)
  \Theta \! \left( \tau - 0 \right)
  \left[ 
    \sinh \frac{q}{2} + 
    2 \bar{\nu}
    \left( \cosh \frac{q}{2} - 1 \right)
  \right]
\nonumber \\
& & 
  \mbox{} + 
  \frac{q}{2}
  \left[
    \delta \! \left( \tau - {\tau}_C \right) + 
    \delta \! \left( \tau - 0 \right)
  \right] . 
\label{eq:holder_current}
\end{eqnarray}
The stationary-point background evolves ${\eta}^{\mbox{\scriptsize
    cl}}$ unstably from its initial asymptotic value of zero at $\tau
\rightarrow -\infty$ to value $q$ at $\tau = 0 - \epsilon$.  The
$\delta$-function in the source then lowers ${\eta}^{\mbox{\scriptsize
    cl}}$ to value $q/2$ at $\tau = 0 + \epsilon$, which solves the
steady-state condition
\begin{equation}
  \frac{
    \partial \hat{\mathcal{L}}
  }{
    \partial \nu
  } = 
  j , 
\label{eq:steady_NEQ_sp}
\end{equation}
and holds it there until ${\tau}_C - \epsilon$, at which point the
second $\delta$-function term takes ${\eta}^{\mbox{\scriptsize cl}}
\rightarrow 0$ at $\tau = {\tau}_C + \epsilon$.  These results are
easy to check from the exact equations of motion in action-angle
coordinates, and the time-dependence of both $j_{\tau}$ and the
associated ${\nu}^{\mbox{\scriptsize cl}}$ are shown in
Fig.~\ref{fig:ext_dis_eq_bw}.

Because of the time-dependence of this source protocol $j$, neither
$\hat{\mathcal{L}}$ nor ${\hat{\mathcal{L}}}_j$ vanishes.  Indeed,
because ${\partial}_{\tau} \eta = 0$ except at the support of the
$\delta$-functions of $j_{\tau}$, and because these two terms cancel
in the action, the nonzero values of ${\hat{\mathcal{L}}}_j$ and
$\hat{\mathcal{L}}$ determine $\Gamma$ and $S^{\mbox{\scriptsize
    eff}}$.  We can check that these potentials evaluate to
\begin{eqnarray}
  \Gamma \! \left[ j \right] 
& = & 
  N \int d\tau
  \left( 
    - {\partial}_{\tau} \eta \nu + 
    {\hat{\mathcal{L}}}_j 
  \right)
\nonumber \\
& = & 
  N 
  \left[
    \log \cosh 
    \frac{\beta \bar{\mu}}{2} - 
    \log \cosh 
    \frac{1}{2}
    \left( q - \beta \bar{\mu} \right) 
  \right] 
\nonumber \\
& & 
  \mbox{} - 
  {\tau}_C
  \frac{N}{2}
  \left[
    \left( \cosh \frac{q}{2} - 1 \right) + 
    2 \bar{\nu} \sinh \frac{q}{2}
  \right] 
\label{eq:Gamma_holder_eval}
\end{eqnarray}
and 
\begin{eqnarray}
  S^{\mbox{\scriptsize eff}} \! 
  \left[ n \right] 
& = & 
  N \int d\tau
  \left(
    - {\partial}_{\tau} \eta \nu + 
    \hat{\mathcal{L}} 
  \right)
\nonumber \\
& = & 
  S^{\mbox{\scriptsize eff}} \! 
  \left( n_{\tau = \left\{ 0 , {\tau}_C \right\} } \right) + 
  {\tau}_C
  \frac{N}{2}
  \frac{
    \cosh \frac{q}{2} - 1
  }{
    \cosh \frac{\beta \bar{\mu}}{2}
    \cosh \frac{q  - \beta \bar{\mu}}{2}
  } 
\nonumber \\
& = & 
  S^{\mbox{\scriptsize eff}} \! 
  \left( n_{\tau = \left\{ 0 , {\tau}_C \right\} } \right) + 
  {\tau}_C
  N 
  {
    \left( 
      \sqrt{{\bar{\nu}}_b {\nu}_a} - 
      \sqrt{{\bar{\nu}}_a {\nu}_b}
    \right)
  }^2 . 
\nonumber \\
\label{eq:S_eff_holder_eval}
\end{eqnarray}
${\hat{\mathcal{L}}}_j$ may be of either sign, while
$\hat{\mathcal{L}} \ge 0$ always.

We may evaluate the $q \ll 1$ limit of
Eq.~(\ref{eq:S_eff_holder_eval}) for comparison to both the
equilibrium effective action and the Onsager-Machlup linear
approximation.  Noting that in the linear-response regime $\left( \nu
- \bar{\nu} \right) \rightarrow {\bar{\nu}}_a {\bar{\nu}}_b q$, the
small-$q$ expansion becomes 
\begin{equation}
  S^{\mbox{\scriptsize eff}} \! 
  \left[ n \right] \rightarrow
  \frac{
    N
    {\bar{\nu}}_a
    {\bar{\nu}}_b 
    q^2 
  }{
    2
  } +  
  {\tau}_C
  \frac{
    N
    {\bar{\nu}}_a
    {\bar{\nu}}_b 
    q^2 
  }{
    4
  } . 
\label{eq:S_eff_smallq_holder}
\end{equation}

The probability of an initial fluctuation to non-equilibrium number
occupancies is given by the equilibrium effective action, while the
probability of \emph{persistence} at that value is governed by a new
term with the structure of an \emph{entropy rate}~\cite{Cover:EIT:91}.
We will see in the Sec.~\ref{sec:caliber} that this part of the
effective action is a free-energy-rate difference, just as the
equilibrium $S^{\mbox{\scriptsize eff}} \!  \left( n \right)$ of
Eq.~(\ref{eq:S_eff_G_diffs}) is a free-energy difference.  The
entropy-rate difference for persistent dis-equilibrium may be
extracted as
\begin{eqnarray}
  \frac{
    \partial 
  }{
    \partial {\tau}_C
  } 
  S^{\mbox{\scriptsize eff}} \! 
  \left[ n \right] 
& = & 
  N 
  {
    \hat{\mathcal{L}}
  }_{
    \tau \in \left[ 0_{+} , {\tau}_C \right]
  }
\nonumber \\
& = & 
  \frac{N}{2}
  \frac{
    \cosh \frac{q}{2} - 1
  }{
    \cosh \frac{\beta \bar{\mu}}{2}
    \cosh \frac{q  - \beta \bar{\mu}}{2}
  } 
\nonumber \\
& = & 
  N 
  {
    \left( 
      \sqrt{{\bar{\nu}}_b {\nu}_a} - 
      \sqrt{{\bar{\nu}}_a {\nu}_b}
    \right)
  }^2 .   
\label{eq:holding_ent_rate}
\end{eqnarray}
Eq.~(\ref{eq:holding_ent_rate}) is a new result.  The combination of
square-root dependence on quantities that equal the expected escape
rates -- ${\bar{\nu}}_b {\nu}_a$ and ${\bar{\nu}}_a {\nu}_b$ -- and
    the squared difference of these in the exact entropy rate, extend
    to general multiparticle chemical reactions as shown in
    App.~\ref{sec:multi_part_react}.

To make contact with the Langevin treatment in
Sec.~\ref{sec:Langevin_approx}, we check that in the persistent
non-equilibrium domain where $\eta = q/2$, the fluctuation of the
Langevin field in Eq.~(\ref{eq:lambda_corr_fn}) is given by 
\begin{equation}
  {\bar{\nu}}_b {\nu}_a^{\mbox{\scriptsize cl}} 
  e^{
    {\eta}^{\mbox{\scriptsize cl}}
  } + 
  {\bar{\nu}}_a {\nu}_b^{\mbox{\scriptsize cl}}
  e^{
    - {\eta}^{\mbox{\scriptsize cl}}
  } = 
  2 
  \sqrt{
    {\bar{\nu}}_a
    {\bar{\nu}}_b
    {\nu}_a^{\mbox{\scriptsize cl}} 
    {\nu}_b^{\mbox{\scriptsize cl}} 
  } . 
\label{eq:lambda_variance_eval}
\end{equation}
The decay rate in the Langevin equation~(\ref{eq:Langevin}) evaluates
to
\begin{equation}
  {\bar{\nu}}_b 
  e^{
    {\eta}^{\mbox{\scriptsize cl}}
  } + 
  {\bar{\nu}}_a 
  e^{
    - {\eta}^{\mbox{\scriptsize cl}}
  } = 
  \sqrt{
    \frac{
      {\bar{\nu}}_a
      {\bar{\nu}}_b
    }{
      {\nu}_a^{\mbox{\scriptsize cl}} 
      {\nu}_b^{\mbox{\scriptsize cl}} 
    }
  } . 
\label{eq:rate_decay_eval}
\end{equation}
Hence the single-time fluctuation expression~(\ref{eq:nu_pr_corr_fn})
evaluates to 
\begin{equation}
  \left<
    { 
      \left( {\nu}^{\prime}_{\tau} \right)
    }^2
  \right> = 
  \frac{
    {\nu}_a^{\mbox{\scriptsize cl}} 
    {\nu}_b^{\mbox{\scriptsize cl}} 
  }{
    N 
  } , 
\label{eq:nu_variance_eval}
\end{equation}
in agreement with the variance obtained from
Eq.~(\ref{eq:nu_pr_corr_fn}) for the distribution produced by the
equilibrium generating function. 

\subsubsection{The non-equilibrium entropy rate in relation to
  gradients of the equilibrium entropy}

How does the entropy-rate difference defined from the dynamical
stochastic effective action relate to changes in the equilibrium
entropy of subsystems under free decay?  We may recast
Eq.~(\ref{eq:S_eff_G_diffs}) as
\begin{eqnarray}
  S^{\mbox{\scriptsize eff}} \! \left( n \right) 
& = & 
  N D \! 
  \left( \nu \parallel \bar{\nu} \right) 
\nonumber \\
& = & 
  \frac{N}{2} 
  \left[ 
    \log
    \left(
      \frac{
        {\nu}_b {\nu}_a 
      }{
        {\bar{\nu}}_b {\bar{\nu}}_a
      } 
    \right) + 
    q
    \left( {\nu}_b - {\nu}_a \right)
  \right] . 
\label{eq:Eq_ent_KL}
\end{eqnarray}
For the period of free decay after ${\tau}_C$, the change in
equilibrium entropy with $n$ is
\begin{equation}
  \frac{\partial}{\partial n}
  S^{\mbox{\scriptsize eff}} \! \left( n \right) = 
  q . 
\label{eq:Eq_ent_KL_grad}
\end{equation}

The time-dependence of the equilibrium-form entropy follows from the
time-dependence of $n$.  Its gradient at the moment after the source
perturbation is turned off is given by 
\begin{eqnarray}
  {\partial}_{\tau} n 
& = & 
  - N \left( \nu - \bar{\nu} \right) 
\nonumber \\
& = & 
  - \frac{N}{2}
  \frac{
    \sinh \frac{q}{2}
  }{
    \cosh \frac{\beta \bar{\mu}}{2}
    \cosh \frac{q  - \beta \bar{\mu}}{2}
  } .   
\label{eq:n_free_release}
\end{eqnarray}
Therefore, under free diffusion from a chemical-potential perturbation
by $q$, the initial rate of subsystem entropy change is given by
\begin{eqnarray}
  {\partial}_{\tau} 
  S^{\mbox{\scriptsize eff}} \! \left( n \right)
& = & 
  \frac{
    \partial
    S^{\mbox{\scriptsize eff}} \! \left( n \right)
  }{
    \partial n
  }
  {\partial}_{\tau} n 
\nonumber \\
& = & 
  - \frac{N}{2}
  \frac{
    q \sinh \frac{q}{2}
  }{
    \cosh \frac{\beta \bar{\mu}}{2}
    \cosh \frac{q  - \beta \bar{\mu}}{2}
  } .   
\label{eq:S_free_release}
\end{eqnarray}
At small $q$, $- {\partial}_{\tau} S^{\mbox{\scriptsize eff}} \!
\left( n \right) \rightarrow 4 \partial S^{\mbox{\scriptsize eff}} \!
\left[ n \right] / \partial {\tau}_C$ from
Eq.~(\ref{eq:holding_ent_rate}).  Thus the stochastic-process entropy
rates, and the rate of change of equilibrium entropies, are not equal
even in this limit; more generally they are distinct functions
altogether.  We can check from Eq.~(\ref{eq:holding_ent_rate}) that in
the interval $\left[ 0 , {\tau}_C \right]$, $\hat{\mathcal{L}}$ has
the limits $\hat{\mathcal{L}} \rightarrow {\bar{\nu}}_s$ as $\tanh
\left( q/2 \right) \rightarrow 1$ and $\hat{\mathcal{L}} \rightarrow
     {\bar{\nu}}_p$ as $\tanh \left( q/2 \right) \rightarrow -1$.
     Thus the entropy rate of the stochastic process is bounded by
     leaving-rates from the two respective states, while the
     corresponding rate of entropy change in
     Eq.~(\ref{eq:S_free_release}) is unbounded.

We may summarize this section as follows: It should not be surprising
that the stochastic effective action for histories includes
entropy-rate terms that have no simple relation to rates of change in
the equilibrium entropy of states.  The former measures uncertainty
about rates of transition, while the latter measures uncertainty about
the occupation frequencies for states.  It may be that, in some cases,
dynamics is near enough to equilibrium that the state occupancy at
successive moments of time places tight constraints on the possible
entropy of transitions.  However, this is not a general result, and it
is not even implied by the local-equilibrium approximation of the form
produced by the double-well potential.  To say more about the origin
of the rate term in the stochastic effective
action~(\ref{eq:S_eff_holder_eval}), however, requires a separation
between system and environment terms that the continuous-time
two-state model does not readily provide.  For that separation we turn
to the method of maximum caliber.

\section{Ensemble of histories II: resolved single-particle histories;
  path entropies and maximum caliber; structural decomposition into
  system and environment; connection to the stochastic-process entropy
  rate}
\label{sec:caliber}

The construction of equilibrium Gibbs free energies from the entropy,
reviewed in App.~\ref{sec:free_energies}, begins with an explicit
division between entropy terms for the system and its environment.
The Freidlin-Wentzell construction of the preceding section does not
naturally provide such a decomposition for paths, because the
contribution of the ``environment'' comes from the form of the
transition-rate terms $k_{\pm}$, which are embedded in the master
equation.  To separate these two contributions in the ensemble over
histories, we require an explicit combinatorial formula for properties
of paths, which is defined independently of the probabilities given to
such paths by the transition rates set by the environment.  The method
of maximum caliber, as formulated in
Ref's~\cite{Jaynes:caliber:80,Ghosh:caliber:06,Stock:caliber:08,%
  Wu:caliber:09}, provides such a decomposition.

\begin{figure}[ht]
  \begin{center} 
  \includegraphics[scale=0.5]{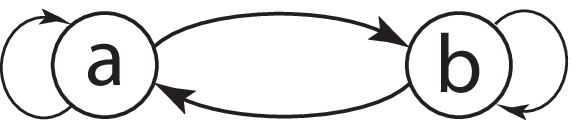}
  \caption{
  Representation of the coarse-grained two-state system as a discrete
  finite-state model with forced transitions on every interval of
  length $\Delta t$.  
    \label{fig:two_states} 
  }
  \end{center}
\end{figure}

This section will introduce two changes of representation.  One is the
consideration of histories of an individual random walker, as
explained in the introduction.  The other, which will be more
important to the ability to isolate system-environment interactions,
is the replacement of the continuous-time stochastic process of the
master equation~(\ref{eq:master_equation}), by a discrete-time
two-state Markov process, shown in Fig.~\ref{fig:two_states}.  The
continuous-time and discrete-time models, in appropriate limits,
represent the same stochastic process.  However, the discrete-time
model has the feature that, in every time interval of length $\Delta
t$, some transition of the particle's state must occur, even if it is
a transition back to the same state.  

It may come as a surprise when we find, below, that the entropy rate
in the effective action~(\ref{eq:S_eff_holder_eval}) draws its form
entirely from the probabilities of \emph{no change}.  When we separate
the large-deviations formula into contributions from the path entropy
and the environmental probabilities, other terms associated with
changes will also appear.  However, these cancel exactly in the
hydrodynamic limit represented by Eq.~(\ref{eq:S_eff_holder_eval}).
Thus we will see that, among the many terms that could have been used
to probe the generating functional of the previous section, the
particular coupling to the field $\nu$ that we studied was the form
that produces the hydrodynamic limit.  We return at the end of this
section to what such a decomposition implies about the role of
``energy dissipation'' as an explanation for large-deviations
probabilities of histories. 

In the next two sections, we will solve for the path ensemble first in
its stationary distribution, and then perturbed with sources.  With
appropriate choices for the transition parameters in the discrete
model, the path ensemble will be constructed from the transfer matrix
for the single-particle continuous-time model.  With a particular
choice of perturbing sources in the generating functional for this
distribution, we will be able to show that the continuous-time
Langevin equation gives the Gaussian approximation to the fluctuating
single-particle trajectory. 

\subsection{Variational approach for a path ensemble in its stationary
  distribution, and connection to the continuous-time transfer matrix}
\label{sec:var_approach}

The implementation of maximum-caliber in
Ref's.~\cite{Stock:caliber:08,Wu:caliber:09} considers only steady
trajectories displaced from the equilibrium.  It does not consider
time-dependent displacements, for which the Doi-Peliti formalism may
provide easier constructions.  However, the steady-state
non-equilibrium problem does make contact directly with the
constant-$j$ solutions of the preceding section.

For the stochastic process of Fig.~\ref{fig:two_states} and a finite
time interval $T$, consider four observables $M^{\alpha \beta}$, for
$\alpha , \beta \in \left\{ a , b \right\}$.  These are the number of
transitions from state $\beta$ to state $\alpha$ between $0$ and $T$.
If we specify no more than these as constraints on a path entropy,
they will tend toward uniform distribution on a long interval $T$,
irrespective of the distribution we use to initialize paths.

The idea is to label each path with an index $j$, and to compute a
path entropy 
\begin{equation}
  S^{\mbox{\scriptsize path}} \equiv 
  - \sum_j 
  p_j \log p_j , 
\label{eq:path_ent_genform}
\end{equation}
subject to some set of appropriate constraints on properties of the
paths, such as its $M^{\alpha \beta}$ values, which will be denoted
$M^{\alpha \beta}_j$.  It will be convenient here to divide the path
specification into the state in which the path starts, labeled with
index $\sigma$, and the remainder of the path conditioned on that
starting state, indexed $j \mid \sigma$.  We could think, thus, of $j$
as a string denoting the ordered states through which the trajectory
passed, and $j \mid \sigma$ as the string excluding its first letter.
In this indexing $p_j = p_{j \mid \sigma} p_{\sigma}$.

It simplifies the treatment, without omitting any results that matter
here, to exclude the distribution over over initial states from the
variational problem, and to consider only variations of $p_{j \mid
  \sigma}$.  For any path $j$, the numbers $M^{\alpha \beta}_j$ count
the state-transitions from state $\beta$ to state $\alpha$ along that
path.  Denote by ${\bar{M}}^{\alpha \beta}$ the constraint values on
maximum entropy for this path ensemble.  Then the Lagrangian for the
path-entropy maximization problem becomes
\begin{widetext}
\begin{equation}
  {\mathcal{L}}^{\mbox{\scriptsize path}} \equiv 
  - \sum_{\sigma} \sum_{j \mid \sigma}
  p_{j \mid \sigma} p_{\sigma}
  \log p_{j \mid \sigma} p_{\sigma} - 
  \sum_{\alpha \beta}
  {\lambda}^{\alpha \beta}
  \left[ 
    \sum_{\sigma} \sum_{j \mid \sigma}
    p_{j \mid \sigma} p_{\sigma}
    M^{\alpha \beta}_j - 
    {\bar{M}}^{\alpha \beta}
  \right] - 
  \sum_{\sigma}
  {\eta}_{\sigma}
  \left[ 
    \sum_{j \mid \sigma}
    p_{j \mid \sigma} - 
    1 
  \right] . 
\label{eq:path_ent_Lagrangian}
\end{equation}
\end{widetext}

Lagrangians of the form~(\ref{eq:path_ent_Lagrangian}) produce Gibbs
(exponential) distributions as their maximizing distributions.
Therefore, in terms of the Lagrange multipliers $\left\{
{\lambda}^{\alpha \beta} \right\}$, the maximum-entropy probabilities
have the form:
\begin{equation}
  p_j = 
  \frac{1}{Q}
  {\left( {\gamma}^{aa} \right)}^{M^{aa}_j}
  {\left( {\gamma}^{ab} \right)}^{M^{ab}_j}
  {\left( {\gamma}^{ba} \right)}^{M^{ba}_j}
  {\left( {\gamma}^{bb} \right)}^{M^{bb}_j}
  p_{\sigma} . 
\label{eq:path_probs_Gibbs_form}
\end{equation}
The entropy on this distribution was given the name \emph{caliber} by
Jaynes~\cite{Jaynes:caliber:80}, to reflect the fact that it is a
logarithmic measure of the width of a ``tube'' of microhistories which
are typical within the distribution that produces macrohistories
characterized by the $\left\{ {\bar{M}}^{\alpha \beta} \right\}$.

Following Ref's.~\cite{Stock:caliber:08,Wu:caliber:09} (apart from a
sign convention for the $\left\{ {\lambda}^{\alpha \beta} \right\}$)
we define
\begin{equation}
  {\gamma}^{\alpha \beta} \equiv 
  e^{
    - {\lambda}^{\alpha \beta} 
  } , 
\label{eq:gammas_from_lambdas}
\end{equation}
and choose $p_{\sigma}$ to be the starting probability appropriate
to whichever is the first state in history $j$.  For this set of
observables, the order in which transitions occur does not affect the
weight function, though it does restrict the set of possible paths.

$Q$ is the partition function on histories, defined as the sum over
$j$ of the other terms in Eq.~(\ref{eq:path_probs_Gibbs_form}).
Because histories are in 1-1 correspondence with monomials that arise
from the $M$th power of a $2 \times 2$ matrix, the partition function
is immediately expressed as the usual trace of a transfer matrix, just
as we could have computed working directly from
Eq.~(\ref{eq:general_exchange_ME}) above:
\begin{equation}
  Q = 
  \begin{array}{c}
    \left[ 
      \begin{array}{cc}
        1 & 1 
      \end{array}
    \right] \\
    \phantom{
      \left[ 
        \begin{array}{cc}
          1 & 1 
        \end{array}
      \right]
    }
  \end{array}
  {
    \left( 
      \left[
        \begin{array}{cc}
          {\gamma}^{aa} & {\gamma}^{ab} \\
          {\gamma}^{ba} & {\gamma}^{bb}
        \end{array}
      \right]
    \right) 
  }^M
  \left[
    \begin{array}{c}
      p_a \\
      p_b 
    \end{array}
  \right] . 
\label{eq:Q_trace_eval}
\end{equation}

Now we may make some simplifying assumptions without loss of
generality, to bring the expression~(\ref{eq:Q_trace_eval}) into
direct correspondence with the transfer matrix of the continuous-time
model.  The matrix $\gamma$ does not actually have four independent
values $\left\{ {\gamma}^{\alpha \beta} \right\}$, because the
original Lagrangian~(\ref{eq:path_ent_Lagrangian}) did not have four
independent constraints $\left\{ {\bar{M}}^{\alpha \beta} \right\}$.
The sum $\sum_{\alpha \beta} {\bar{M}}^{\alpha \beta} \equiv M$,
because for each possible path also $\sum_{\alpha \beta} M^{\alpha
  \beta}_j \equiv M$.  Moreover, for large $M$, it is not possible to
set ${\bar{M}}^{ab} \neq {\bar{M}}^{ba}$ by more than $\pm 1$, so
effectively these must be chosen equal.\footnote{Note that if we were
  imposing more finely resolved time-dependent constraints, the same
  might not be true, but for this finite set of four observables
  covering indefinite $M$, we have no other choice consistent with
  only two states.}  Therefore two normalizations of the $\left\{
{\gamma}^{\alpha \beta} \right\}$ may be chosen arbitrarily.  These
may be chosen so that ${\gamma}^{aa} + {\gamma}^{ba} = 1$,
${\gamma}^{ab} + {\gamma}^{bb} = 1$, effectively choosing unit
normalization for the partition function $Q$.  The matrix $\left[
  {\gamma}^{\alpha \beta} \right]$ is then called a \emph{stochastic
  matrix}.

Next, in keeping with the omission of the initial state from the
variational problem, suppose $p_{\sigma}$ is chosen to be the
stationary distribution of largest eigenvalue preserved by whatever
values of ${\gamma}^{\alpha \beta}$ solve the variational problem.
Then it follows immediately -- by collapse of the matrix $\left[
  {\gamma}^{\alpha \beta} \right]$ with $\left[ \begin{array}{cc} 1 &
    1 \end{array} \right]$ on the left, and with
${\left[ \begin{array}{cc} p_a & p_b \end{array} \right]}^T$ on the
right -- that
\begin{equation}
  \frac{
    \partial \log Q
  }{
    \partial \log {\gamma}^{\alpha \beta}
  } = 
  \left< 
    M^{\alpha \beta}_j
  \right> = 
  M 
  {\gamma}^{\alpha \beta}
  p_{\beta} .
\label{eq:vary_lgamma_get_M}
\end{equation}
We solve for the $\left\{ {\gamma}^{\alpha \beta} \right\}$ by setting
Eq.~(\ref{eq:vary_lgamma_get_M}) equal to ${\bar{M}}^{\alpha
  \beta}$.\footnote{If we had not supposed that $\left\{ p_{\alpha}
  \right\}$ was the stationary distribution, the same result would be
  obtained asymptotically in large $M$, through the approximation of
  $\log Q$ by its largest log-eigenvalue, denoted $\log
  {\lambda}_{+}$.  This is the approach taken in
  Ref's.~\cite{Stock:caliber:08,Wu:caliber:09}.} It follows from our
choice to normalize $\left[ {\gamma}^{\alpha \beta} \right]$ to be a
stochastic matrix that
\begin{equation}
  \sum_{\alpha}
  \frac{
    {\bar{M}}^{\alpha \beta}
  }{
    M 
  } = 
  p_{\beta} . 
\label{eq:M_alphabeta_sum_p}
\end{equation}

The relation of the $\left\{ {\gamma}^{\alpha \beta} \right\}$ to the
$\left\{ p_{\beta} \right\}$ may now be made explicit.  The assumption
behind the use of a discrete stochastic-process model is that $\Delta
t$ is a scaling variable that may be changed in the discrete model
while leaving important physical observables unchanged, though
rescaling $\Delta t$ may introducing overall renormalization constants
associated with the discretization.  In general the $\left\{
{\gamma}^{\alpha \beta} \right\}$ will then depend on the $\Delta t$
but the stationary distribution of the stochastic process should not.
Denote the stationary distribution that is self-consistent with the
$\left\{ {\gamma}^{\alpha \beta} \right\}$ appropriate to the given
$\left\{ {\bar{M}}^{\alpha \beta} \right\}$ by $\left[ {\pi}_{\beta}
  \right]$.\footnote{The distribution $\left[ {\pi}_{\beta} \right]$
  clearly corresponds to the number fraction $\left[ {\nu}_{\beta}
    \right]$ of the preceding sections.}  Then the most general
stochastic matrix $\left[ {\gamma}^{\alpha \beta} \right]$ with
$\left[ {\pi}_{\beta} \right]$ as the dominant eigenvector is
\begin{equation}
  \left[
    \begin{array}{cc}
      {\gamma}^{aa} & {\gamma}^{ab} \\
      {\gamma}^{ba} & {\gamma}^{bb}
    \end{array}
  \right] = 
  \left\{ 
    \left[
      \begin{array}{cc}
        1 &   \\
          & 1
      \end{array}
    \right] + 
    \left[
      \begin{array}{cc}
        -{\pi}_b &  {\pi}_a \\
        {\pi}_b  & -{\pi}_a
      \end{array}
    \right] 
    \left( 
      1 - r^{-R}
    \right)
  \right\} . 
\label{eq:general_gamma_mat}
\end{equation}
The constant $R$ sets the frequency of transition events $M^{ab} =
M^{ba}$.  To model ensure that the discrete stochastic process has the
same continuum limit as the continuous-time model of the previous
sections, we require that for sufficiently short $\Delta t$, $R$ must
converge to
\begin{equation}
  R = 
  \left( k_{+} + k_{-} \right)
  \Delta t . 
\label{eq:R_time_consist}
\end{equation}
The form~(\ref{eq:general_gamma_mat}) is chosen so that under
refinement or coarsening of the time increment, the
definition~(\ref{eq:R_time_consist}) remains consistent. 

The interpretation of $k_{\pm}$ as rate constants is completed if we
set their ratio from the fractions of time intervals in which the
system starts, respectively, in states $a$ or $b$: 
\begin{eqnarray}
  {\pi}_a
& = & 
  \frac{k_{-}}{k_{+} + k_{-}}
\nonumber \\
  {\pi}_b
& = & 
  \frac{k_{+}}{k_{+} + k_{-}} , 
\label{eq:pis_ks_relate}
\end{eqnarray}
corresponding to Eq.~(\ref{eq:eq_fracs_def}).  Values of $k_{\pm}$ are
then chosen to satisfy the constraints~(\ref{eq:M_alphabeta_sum_p}).
With these choices the matrix $\left[ {\gamma}^{\alpha \beta} \right]$
becomes precisely the transfer matrix for the continuous-time
one-particle problem.

Note that the relation between the rate constants, equilibrium
frequencies, and transition numbers may be written
\begin{equation}
  \frac{{\bar{M}}^{ba}}{T} = 
  \frac{{\bar{M}}^{ab}}{T} = 
  \sqrt{{\pi}_a k_{+}}
  \sqrt{{\pi}_b k_{-}} . 
\label{eq:transition_forms_eq}
\end{equation}
This form -- a variant expression for either ${\pi}_a k_{+}$ or
${\pi}_b k_{-}$ -- which characterizes the equilibrium distribution,
will arise again when we use the generating function to probe steady
non-equilibrium states.

From the variational property~(\ref{eq:vary_lgamma_get_M}) and the
original Lagrangian formulation of the problem, we recognize that
$\log Q$ will be the negative of a Gibbs free energy.  From the fact
that $\log Q \equiv 0$ when the $\left\{ {\gamma}^{\alpha \beta}
\right\}$ are evaluated at the above conditions, we know that the free
energy may be offset by a constant at the entropy-maximizing,
equilibrium distribution.  We may derive the exact relation by
calculating the entropy of paths (the caliber) on the
solution~(\ref{eq:path_probs_Gibbs_form}),
\begin{eqnarray}
\lefteqn{
  S^{\mbox{\scriptsize path}} \! 
  \left[ 
    \left\{ 
      {\bar{M}}^{\alpha \beta}
    \right\}
  \right] = 
  - \sum_{\alpha}
  {\pi}_{\alpha}
  \log {\pi}_{\alpha} + 
  \sum_{\alpha \beta}
  \left< M^{\alpha \beta} \right> 
  {\lambda}^{\alpha \beta}
} & & 
\nonumber \\
& = & 
  - \sum_{\alpha}
  {\pi}_{\alpha}
  \log {\pi}_{\alpha} - 
  \sum_{\alpha \beta}
  {\bar{M}}^{\alpha \beta} 
  \log 
  \frac{
    {\bar{M}}^{\alpha \beta} 
  }{
    M {\pi}_{\beta}
  }
\nonumber \\
& = & 
  - \sum_{\alpha}
  {\pi}_{\alpha}
  \log {\pi}_{\alpha} - 
  M \sum_{\beta}
  {\pi}_{\beta}
  \sum_{\alpha}
  {\gamma}^{\alpha \beta} 
  \log 
  {\gamma}^{\alpha \beta} . 
\nonumber \\
\label{eq:S_path_stat_eval}
\end{eqnarray}
The first line of Eq.~(\ref{eq:S_path_stat_eval}) is simply the
evaluation of the definition~(\ref{eq:path_ent_genform}), recalling
that $\log Q = 0$ in the normalization of $p_j$.  The second line is
the proper extensive-form dependence of the entropy on the constraint
variables which are its macroscopic arguments.  The third line,
interpreting the $\left\{ {\gamma}^{\alpha \beta} \right\}$ as
transition probabilities in the state diagram of
Fig.~(\ref{fig:two_states}), expresses the path entropy as a sum of
the entropy of the initial distribution, with the time-integral of the
\emph{entropy rate} of the stochastic process in that
distribution~\cite{Cover:EIT:91}.  This expression is the first in the
paper, in which the entropy rate is given its familiar form from the
stochastic process of a finite-state system.

The combinatorial interpretation of the second line in
Eq.~(\ref{eq:S_path_stat_eval}) is immediate.  Using $S^0 \! \left(
\left\{ {\pi}^{\alpha} \right\} \right)$ to denote the single-time
entropy of the initial distribution, we may express
\begin{equation}
  S^{\mbox{\scriptsize path}} \! 
  \left[ 
    \left\{ 
      {\bar{M}}^{\alpha \beta}
    \right\}
  \right] = 
  S^0 \! 
  \left( 
    \left\{ {\pi}^{\alpha} \right\}
  \right) + 
  \log 
  \left(
    \begin{array}{c}
      M {\pi}_a \\
      {\bar{M}}^{ba} 
    \end{array}
  \right) + 
  \log 
  \left(
    \begin{array}{c}
      M {\pi}_b \\
      {\bar{M}}^{ab} 
    \end{array}
  \right) .
\label{eq:path_ent_multinomials}
\end{equation}
The combinatorial entropy is the Stirling approximation for the logs
of the two independent binomial factors for distributing
${\bar{M}}^{ba}$ transitions $a \rightarrow b$ among $M {\pi}_a =
{\bar{M}}^{aa} + {\bar{M}}^{ba}$ total exits from state $a$, and
similarly for distributing ${\bar{M}}^{ab}$ exits $b \rightarrow a$
among $M {\pi}_b$ total exits from $b$.  Note that the entropy rate
term in Eq.~(\ref{eq:S_path_stat_eval}) will diverge logarithmically
in $\Delta t$ if a finite number of transitions ${\bar{M}}^{ba} =
{\bar{M}}^{ab}$ is fixed in a growing number of opportunities $M$.

Finally, from the explicit expression for the entropy either in terms
of $\left\{ {\bar{M}}^{\alpha \beta} \right\}$ or the associated
Lagrange multipliers $\left\{ {\lambda}^{\alpha \beta} = - \log
{\gamma}^{\alpha \beta} \right\}$, we may identify the relation
between the partition function $Q$ and a free energy that leads to the
relations~(\ref{eq:vary_lgamma_get_M})
\begin{equation}
  - \log Q = 
  \sum_{\alpha \beta}
  {\lambda}^{\alpha \beta}
  {\bar{M}}^{\alpha \beta}_{\lambda} - 
    S^{\mbox{\scriptsize path}} \! 
  \left[ 
    \left\{ 
      {\bar{M}}^{\alpha \beta}_{\lambda}
    \right\}
  \right] + 
  S^0 \! 
  \left( 
    \left\{ {\pi}^{\alpha} \right\}
  \right) . 
\label{eq:loq_Q_Gibbs_reln}
\end{equation}
Here each ${\bar{M}}^{\alpha \beta}_{\lambda}$ denotes the $\alpha
\beta$-component of the set of $\bar{M}$ jointly satisfying 
\begin{equation}
  \frac{
    \partial 
    S^{\mbox{\scriptsize path}} \! 
    \left[ 
      \left\{ 
        {\bar{M}}^{\alpha \beta}
      \right\}
    \right]
  }{
    \partial {\bar{M}}^{\alpha \beta}
  } = 
  {\lambda}^{\alpha \beta} . 
\label{eq:ent_max_bar_Ms}
\end{equation}

From the developments of the single-time equilibrium, we may
anticipate that the effective action for other configurations $\left\{
{\bar{M}}^{\alpha \beta} \right\}$ than the most-probable ones will
have the same form as Eq.~(\ref{eq:loq_Q_Gibbs_reln}), with $\left\{
{\lambda}^{\alpha \beta} \right\}$ fixed and $\left\{
{\bar{M}}^{\alpha \beta} \right\}$ permitted to vary independently of
them.  We will return to that and to its physical and combinatorial
evaluation in a moment, but first we do the construction methodically
through a generating function for the path-probability distribution. 

\subsection{Generating function for steady non-equilibrium states, and
  connection to the continuous-time Langevin equation}

A generating function for the discrete-time path ensemble may be
constructed as it was in the Doi-Peliti method.  Because we will
consider only steady non-equilibrium distributions, we may do this in
a direct and simple way, as a modification of the underlying transfer
matrix by shifts in the chemical potential.  There are many
perturbations that could be chosen for the discrete-time model, but
only the one equivalent to the coupling to the number field in
Sec.~\ref{sec:DP_gen_fn_srcs} produces the same hydrodynamic limit as
$\Delta t \rightarrow 0$.

Because the discrete model has four parameters $\left\{
{\gamma}^{\alpha \beta} \right\}$ for transition rates, we introduce
four complex variables 
\begin{equation}
  z^{\alpha \beta} \equiv 
  e^{
    {\zeta}^{\alpha \beta}
  } 
\label{eq:z_paths_intro}
\end{equation}
as arguments of the moment-generating function.  Suppose that, like
the ${\lambda}^{\alpha \beta}$, all ${\zeta}^{\alpha \beta}$ scale
$\sim \left( k_{+} + k_{-} \right) \Delta t \ll 1$.  The terms
${\zeta}^{ba}$ and ${\zeta}^{ab}$ modify the transition currents,
while terms ${\zeta}^{aa}$ and ${\zeta}^{bb}$ modify the persistence
probabilities.  The counterpart to the current $j$ of
Sec.~\ref{sec:DP_gen_fn_srcs}, which effectively modifies the chemical
potentials of the states, is the combination ${\zeta}^{aa} -
{\zeta}^{bb}$.  Therefore, to compare the path entropy to this
perturbation in the Doi-Peliti formulation, we set ${\zeta}^{ba} =
{\zeta}^{ab} \equiv 0$, and consider only the remaining two
perturbations.  

For constant sources $z^{aa}$ and $z^{bb}$, the path-generating function
for non-equilibrium, asymptotically-steady transition numbers may
simply be written in terms of a transfer matrix that has the same
modification in every timestep, as 
\begin{eqnarray}
\lefteqn{
  {\psi}^{\mbox{\scriptsize path}} \! 
  \left( z^{aa}, z^{bb} \right) \equiv 
  e^{
    - {\Gamma}^{\mbox{\scriptsize path}}
    \left( {\zeta}^{aa}, {\zeta}^{bb} \right)
  } 
} & & 
\nonumber \\
& = & 
  \sum_{\sigma} \sum_{j \mid \sigma}
  {\left( z^{aa} \right)}^{M^{aa}_j}
  {\left( z^{bb} \right)}^{M^{bb}_j}
  p_{j \mid \sigma}
  {\pi}_{\sigma} 
\nonumber \\
& = & 
  \frac{1}{Q}
  \begin{array}{c}
    \left[ 
      \begin{array}{cc}
        1 & 1 
      \end{array}
    \right] \\
    \phantom{
      \left[ 
        \begin{array}{cc}
          1 & 1 
        \end{array}
      \right]
    }
  \end{array}
  {
    \left( 
      \left[
        \begin{array}{cc}
          {\gamma}^{aa}
          z^{aa}        & {\gamma}^{ab} \\
          {\gamma}^{ba} & {\gamma}^{bb} 
                          z^{bb}
        \end{array}
      \right]
    \right) 
  }^M
  \left[
    \begin{array}{c}
      {\pi}_a \\
      {\pi}_b 
    \end{array}
  \right] . 
\label{eq:path_gen_fun_def}
\end{eqnarray}
The starting probabilities $\left\{ {\pi}_{\sigma} \right\}$ in
Eq.~(\ref{eq:path_gen_fun_def}) will no longer be those preserved by
the transfer matrix at nonzero $\left\{ {\zeta}^{\alpha \beta}
\right\}$, so the function ${\Gamma}^{\mbox{\scriptsize path}} \!
\left( {\zeta}^{aa}, {\zeta}^{bb} \right)$ will generally differ from
the form~(\ref{eq:loq_Q_Gibbs_reln}) given for $-\log Q$.  The
difference will be associated with transient decay from $\left\{
{\pi}_{\sigma} \right\}$ to the perturbed stationary state, and will
scale as ${\left( T \right)}^0$ at large $T$.  If we denote this
constant as ${\Lambda}^0$, the asymptotic expression for the
generating function becomes
\begin{eqnarray}
  {\Gamma}^{\mbox{\scriptsize path}} \!
  \left( 
    {\zeta}^{aa}, {\zeta}^{bb} 
  \right)
& = & 
  \sum_{\alpha \beta}
  { 
    \left( 
      \lambda + \zeta
    \right)
  }^{\alpha \beta}
  {\bar{M}}^{\alpha \beta}_{\left( \lambda + \zeta \right)} - 
    S^{\mbox{\scriptsize path}} \! 
  \left[ 
    \left\{ 
      {\bar{M}}^{\alpha \beta}_{\left( \lambda + \zeta \right)}
    \right\}
  \right] 
\nonumber \\
& & \mbox{} + 
  {\Lambda}^0 + 
  \log Q  . 
\label{eq:psi_path_asympt}
\end{eqnarray}

The asymptotic stationary values ${\bar{p}}^a$, ${\bar{p}}^b$ may be
computed as functions of $\lambda + \zeta$ as before, and the
functional relation may then be inverted to assign values to the
${\zeta}^{\alpha \beta}$.  It is possible to check that only the
combination ${\zeta}^{aa} - {\zeta}^{bb}$ appears in any of the
transition-number expressions, so we could originally have set $z^{aa}
= 1/z^{bb}$ as a single argument to ${\psi}^{\mbox{\scriptsize
    path}}$.  This construction exactly follows the construction of
the two-argument and the one-argument moment-generating functions in
Sec.~\ref{sec:eq_gen_fn_SEA}.  

An important secondary relation that is not merely a definition but
results from the choice ${\zeta}^{ba} = {\zeta}^{ab} \equiv 0$, is
that the off-equilibrium transition rates satisfy
\begin{equation}
  \frac{
    {\bar{M}}^{ba}_{\left( \lambda + \zeta \right)}
  }{
    T 
  } = 
  \frac{
    {\bar{M}}^{ba}_{\left( \lambda + \zeta \right)}
  }{
    T 
  } = 
  \sqrt{{\bar{p}}_a k_{+}}
  \sqrt{{\bar{p}}_b k_{-}} .
\label{eq:transition_forms_zs}
\end{equation}
Eq.~(\ref{eq:transition_forms_zs}) is the counterpart to
Eq.~(\ref{eq:transition_forms_eq}) for the equilibrium distribution.
This property will be responsible for the cancellation of terms
between the path entropy and external weight factors which could
diverge as $\Delta t \rightarrow 0$, and recovery of the hydrodynamic
limit. 

The transition rates~(\ref{eq:transition_forms_zs}) equal the rate at
which the Langevin field correlation
function~(\ref{eq:lambda_corr_fn}) injects noise into the number-field
correlation function~(\ref{eq:nu_pr_corr_fn}).  Thus we confirm that,
for this simple example of non-interacting particles, the Langevin
equation~(\ref{eq:Langevin}) describes the fluctuating trajectories of
single-particle states.  The Langevin description is indirect, as
${\nu}^{\prime}$ is formally the expectation value of a Poisson
distribution, and not a discrete single-particle trajectory.
Following the injection of noise by the correlation function, it is
then the classical relaxation time of these fluctuations, given by
Eq.~(\ref{eq:rate_decay_eval}), that causes the single-time
fluctuations~(\ref{eq:nu_pr_corr_fn}) to agree with those in the
equilibrium generating-function distribution from
Eq.~(\ref{eq:rho_eq_binomial_source}).  We thus establish the mutual
consistency of all of the approaches used.

\subsection{The effective action from caliber and its interpretation}

From here we compute the effective action by Legendre transform, just
as in Sec.~\ref{sec:EQ_large_dev} for the single-time equilibrium
distribution.  The result is
\begin{eqnarray}
\lefteqn{
  S^{\mbox{\scriptsize eff}} \! 
  \left[ 
    \left\{ 
      {\bar{M}}^{\alpha \beta}
    \right\}
  \right] = 
  {\Gamma}^{\mbox{\scriptsize path}} \!
  \left( 
    {\zeta}^{aa}_{\bar{M}} , {\zeta}^{bb}_{\bar{M}}  
  \right) - 
  {\zeta}^{aa}_{\bar{M}} 
  {\bar{M}}^{aa} - 
  {\zeta}^{bb}_{\bar{M}}  
  {\bar{M}}^{bb}
} & & 
\nonumber \\
& = & 
  \sum_{\alpha \beta}
  {\lambda}^{\alpha \beta}
  {\bar{M}}^{\alpha \beta} - 
  S^{\mbox{\scriptsize path}} \! 
  \left[ 
    \left\{ 
      {\bar{M}}^{\alpha \beta}
    \right\}
  \right] + 
  {\Lambda}^0 + \log Q 
\nonumber \\
& = & 
  \sum_{\alpha}
  {\lambda}^{\alpha a}
  {\bar{M}}^{\alpha a} - 
  \log 
  \left(
    \begin{array}{c}
      M {\bar{p}}_a \\
      {\bar{M}}^{ba}
    \end{array}
  \right) 
\nonumber \\
& & 
  \mbox{} + 
  \sum_{\alpha}
  {\lambda}^{\alpha b}
  {\bar{M}}^{\alpha b} - 
  \log 
  \left(
    \begin{array}{c}
      M {\bar{p}}_b \\
      {\bar{M}}^{ab} 
    \end{array}
  \right) 
\nonumber \\
& &
  \mbox{} +  
  {\Lambda}^0 + \log Q . 
\label{eq:S_eff_paths}
\end{eqnarray}
In the last expression we recall the
decomposition~(\ref{eq:path_ent_multinomials}) of the stationary-state
path entropy into the logarithms of two independent binomials.  For
the equilibrium transition rates ${\Lambda}^0 \rightarrow S^0 \!
\left( \left\{ {\pi}^{\alpha} \right\} \right)$ and the expression
vanishes as required.  Recalling that $S^0 \!  \left( \left\{
{\pi}^{\alpha} \right\} \right) + \log Q$ in
Eq.~(\ref{eq:loq_Q_Gibbs_reln}) is simply the maximizer of the terms
written explicitly in Eq.~(\ref{eq:S_eff_paths}), we recover exactly
the difference of Gibbs free energies which was the stochastic
effective action in Eq.~(\ref{eq:S_eff_G_diffs}).

It now remains only to check that the
expression~(\ref{eq:S_eff_paths}) recovers the non-divergent entropy
rate computed from the master equation, and to assign physical
meanings to the terms that appear. 

Consider the limit in which $M^{ba} = M^{ab}$ remains finite as
$\Delta t \rightarrow 0$ and $M \rightarrow \infty$, and suppose all
$k_{\pm} \Delta t \ll 1$.  The external source term that appears in
the effective action~(\ref{eq:S_eff_paths}) evaluates to 
\begin{widetext}
\begin{eqnarray}
  \frac{
    \sum_{\alpha \beta}
    {\lambda}^{\alpha \beta}
    {\bar{M}}^{\alpha \beta} 
  }{
    T 
  } 
& = & 
  - \frac{M {\bar{p}}^a - M^{ba}}{M \Delta t}
  \log 
  \left( 
    1 - 
    k_{+} \Delta t 
  \right) -   
  \frac{M {\bar{p}}^b - M^{ab}}{M \Delta t}
  \log 
  \left( 
    1 - 
    k_{-} \Delta t 
  \right) -   
  \frac{M^{ba}}{T}
  \log 
  \left( k_{+} \Delta t \right) + 
  \frac{M^{ab}}{T}
  \log 
  \left( k_{-} \Delta t \right) 
\nonumber \\
& = & 
  {\bar{p}}_a
  k_{+} + 
  {\bar{p}}_b
  k_{-} - 
  \sqrt{
    {\bar{p}}_a k_{+}
    {\bar{p}}_b k_{-}
  }
  \log 
  \left( 
    k_{+} k_{-} {\Delta t}^2 
  \right) . 
\label{eq:ext_rates_path_Gibbs}
\end{eqnarray}
The first two terms in the second line, although proportional to the
expected escape rates from the two wells with occupancies
${\bar{p}}_a$, ${\bar{p}}_b$, actually come from the accumulation of
time intervals in which the particle \emph{does not escape} and whose
likelihoods differ from unity only by $k_{\pm} \Delta t$.  This
accumulation of terms would not be affected even if we made finite
changes in $M^{ba}$ or $M^{ab}$, as long as total $M$ was sufficiently
large.  The other term, containing the logarithm, comes from the
unlikely events in which sufficient energy accumulates in a
fluctuation to induce a barrier crossing.  It is simply the product of
Boltzmann factors on a set of independent intervals of length $\Delta
t$, whose total number is $M^{ba} + M^{ab}$.  

The terms in the purely combinatorial path entropy evaluate to 
\begin{eqnarray}
  \frac{
    S^{\mbox{\scriptsize path}} \! 
    \left[ 
      \left\{ 
        {\bar{M}}^{\alpha \beta}
      \right\}
    \right]
  }{
    M 
  } 
& = &
  - \frac{M {\bar{p}}^a - M^{ba}}{M}
  \log 
  \left( 
    1 - 
    \frac{M^{ba}}{M {\bar{p}}^a}
  \right) - 
  \frac{M {\bar{p}}^b - M^{ab}}{M}
  \log 
  \left( 
    1 - 
    \frac{M^{ab}}{M {\bar{p}}^b}
  \right) - 
  \frac{M^{ba}}{M}
  \log 
  \frac{M^{ba}}{M {\bar{p}}^a} - 
  \frac{M^{ab}}{M}
  \log 
  \frac{M^{ab}}{M {\bar{p}}^b}
\nonumber \\
& \approx & 
  \frac{M^{ba}}{M} 
  \left[ 
    2 - 
    \log 
    \left( 
      \frac{
        M^{ba} M^{ab}
      }{
        M^2 {\bar{p}}_a {\bar{p}}_b
      }
    \right)
  \right]
\nonumber \\
& = & 
  \Delta t 
  \sqrt{
    {\bar{p}}_a k_{+}
    {\bar{p}}_b k_{-}
  }
  \left[ 
    2 - 
    \log 
    \left( 
      k_{+} k_{-}
      {\Delta t}^2 
    \right)
  \right] . 
\label{eq:path_ent_scaling}
\end{eqnarray}
\end{widetext}
We have placed the terms in the same order as those in the external
linear function~(\ref{eq:ext_rates_path_Gibbs}), and the term with the
logarithm -- again arising from the events where the state changes --
is in fact identical in the entropy and in
Eq.~(\ref{eq:ext_rates_path_Gibbs}).  The two log terms exactly cancel
in the subtraction that defines the effective
action~(\ref{eq:S_eff_paths}), leaving a finite remainder, which we
will see is the hydrodynamic entropy rate. 

It is the first two, non-divergent term, which differ between the path
entropy and the external probabilities.  The total numbers $M^{ba}$
and $M^{ab}$ are frequencies entering the binomial factors in the path
entropy, and both terms contribute identical factors proportional to
$M^{ba} = M^{ab}$.  The difference of
Eq.~(\ref{eq:ext_rates_path_Gibbs}) and
Eq.~(\ref{eq:path_ent_scaling}) gives the non-divergent effective
action by Eq.~(\ref{eq:S_eff_paths}), which goes at large $T$
(ignoring the contribution of terms $\sim {\left( T \right)}^0$) to
\begin{eqnarray}
  \frac{
    S^{\mbox{\scriptsize eff}} \! 
    \left[ 
      \left\{ 
        {\bar{M}}^{\alpha \beta}
      \right\}
    \right]
  }{
    T 
  } 
& \rightarrow & 
  {\bar{p}}_a
  k_{+} + 
  {\bar{p}}_b
  k_{-} - 
  2 \sqrt{
    {\bar{p}}_a k_{+}
    {\bar{p}}_b k_{-}
  }
\nonumber \\
& = & 
  {
    \left( 
      \sqrt{
        {\bar{p}}_a k_{+}
      } - 
      \sqrt{
        {\bar{p}}_b k_{-}
      }
    \right) 
  }^2 . 
\label{eq:S_eff_eval}
\end{eqnarray}
Eq.~(\ref{eq:S_eff_eval}) recovers precisely the 1-particle
coefficient of $t_C$ in Eq.~(\ref{eq:S_eff_holder_eval}), when
converted to descaled time ${\tau}_C$ by the factor $k_{+} + k_{-}$.

Thus we again find that $T$ is the scaling variable in the
large-deviations principle associated with independent path
fluctuations for a single particle.  This scaling applies only to the
extended-time partition function.  It acts independently of the
scaling with $N$, which is the same in the extended-time and
single-time levels of thermodynamic description. 

\subsection{Discretization divergences and the natural scale}

As was explained in Sec.~\ref{sec:natural_scale}, the discrete time
interval $\Delta t$ may be taken only down to a certain scale, before
the approximation by a discrete two-state theory becomes
inappropriate.  The lower limit to $\Delta t$ from the continuum
double-well model will come from the diffusive relaxation time, for a
trajectory starting at the saddle point and ending in whichever well
is most-quickly reached.  On timescales shorter than this, it becomes
invalid to characterize escapes as elementary events counted by
$M^{ba}_j$ or $M^{ab}_j$.

We expect that when the discretization scale is set to this value --
termed the \emph{natural scale} of the problem following
Ref.~\cite{Polchinski:RGEL:84} -- the arguments of the logarithms in
Equations~(\ref{eq:ext_rates_path_Gibbs},\ref{eq:path_ent_scaling})
become
\begin{equation}
  \Delta t k_{\pm} \approx
  2 \pi 
  e^{
    - \beta 
    \left( {\mu}_{\ddagger} - {\mu}_{a/b}^1 \right)
  } . 
\label{eq:exps_nat_scale}
\end{equation}
The exponential waiting times in $k_{\pm}$ are not affected, but the
factor $\Delta t$ cancels the dimensional prefactors in $k_{\pm}$,
which have dimensions $1 / \mbox{time}$, and have a characteristic
scale set by the diffusive relaxation rate. 

If we plug the approximation~(\ref{eq:exps_nat_scale}) into the
formula~(\ref{eq:ext_rates_path_Gibbs}) for the environment terms in
the path-space Gibbs free energy, this term evaluates approximately to 
\begin{eqnarray}
  \frac{
    \sum_{\alpha \beta}
    {\lambda}^{\alpha \beta}
    {\bar{M}}^{\alpha \beta} 
  }{
    T 
  } 
& \approx & 
  {\bar{p}}_a
  k_{+} + 
  {\bar{p}}_b
  k_{-} 
\nonumber \\
& & 
  \mbox {} + 
  \frac{{\bar{M}}^{ba}}{T}
  \beta 
  \left( {\mu}_{\ddagger} - {\mu}_a^1 \right) + 
  \frac{{\bar{M}}^{ab}}{T}
  \left( {\mu}_{\ddagger} - {\mu}_b^1 \right) . 
\nonumber \\
\label{eq:ext_rates_path_nat_scale}
\end{eqnarray}

We have thus arrived at a mathematical expression of the graphical
heuristic for thinking about dynamical ensembles given in
Fig.~\ref{fig:fixed_saddle} of Sec.~\ref{sec:fp_skeleton}.  The
attracting fixed points of the underlying free-energy landscape carry
charge-valued, extensive state variables such as the occupation
frequencies ${\bar{p}}_a$, ${\bar{p}}_b$ for states.  They are related
under Legendre transform to dual, intensive entropy gradients from the
environment -- here $k_{+}$ and $k_{-}$ respectively -- which
determine the probabilities for each state.  The extensive,
current-valued state variables -- here the numbers ${\bar{M}}^{ba}$
and ${\bar{M}}^{ab}$ of transition events -- live on the saddle points
of the landscape.  Under Legendre transform, they should be coupled to
the dual chemical potentials that determine the probability of
transitions: here ${\mu}_{\ddagger} - {\mu}_a^1$ and ${\mu}_{\ddagger}
- {\mu}_b^1$.  The asymmetry between states and transitions arises
because chemical potentials appear exponentially in the leaving rates
$k_{\pm}$, and only linearly in the transition-state terms involving
${\mu}_{\ddagger}$.  

\subsection{Connection to energy dissipation}

The path-space Gibbs free energy~(\ref{eq:S_eff_paths}) has separated
events into categories, which we can associate with different kinds of
energy transfers.  The events in which the particle exchanges wells
could correctly be associated with either absorption of energy from
the bath, or dissipation to it, depending on the sign of the
free-energy difference in the starting and ending well.  However,
these terms entirely cancel between the path entropy and the
environment-mediated probability, and do not contribute to the entropy
rate expression in Eq.~(\ref{eq:S_eff_holder_eval}) in the
hydrodynamic limit.  This cancelling term may be ascribed to ``entropy
production'' in the path ensemble which is exported to the
environment, but it does not appear in the large-deviations formula.
Moreover, we are lucky that we do not need it to compute
path-probabilities, because it appears to depend on the model
resolution $\Delta t$, which is not physically meaningful.

The terms that survive in the entropy rate, and contribute to its
distinctive form~(\ref{eq:holding_ent_rate}), come from all the
moments when thermal fluctuations \emph{do not} lead to escapes.  For
all these, energy is neither absorbed nor dissipated.  In summary, we
acknowledge that there are cases~\cite{Jarzynski:fluctuations:08} in
which a collection of non-equilibrium transitions is constrained by
consistency with the equilibrium distribution that is their starting
or ending point.  Since the equilibrium distribution is dictated by
energy, these constrained transformations must also have
characterizations in terms of energy.  However, for a large collection
of non-equilibrium fluctuation theorems, energy alone is not serving
as the primary constraint on kinetics, and we must look for the
explanation of path probabilities in terms of other parameters of the
stochastic process. 

\section{Discussion and conclusions}
\label{sec:conclusions}

\subsection{Why generating functions for large-deviations formulae?}

We remarked in Sec.~\ref{sec:liouville_forms} that generating
functions provide a more natural representation for large-deviation
formulae, and better approximation methods, than the discrete
probability distribution, because generating functions expand in
moments of collective many-particle motion, rather than in
single-particle hops.  We have also seen, comparing
Eq.~(\ref{eq:rho_eq_binomial_from_Gibbs}) to
Eq.~(\ref{eq:S_eff_G_diffs}), that the cumulant-generating function
extracts the leading exponential term in from the probability
distribution, bypassing sub-leading terms from normalization.  But how
does it perform this separation?  We may understand that the isolation
of leading exponentials is in fact a direct reflection of the way the
generating function expands in moments.

In elementary statistical mechanics~\cite{Kittel:TP:80}, we argue that
the microcanonical and canonical ensembles produce the same entropy,
because the probability of the most-likely element in a sharply peaked
distribution is equal at leading exponential order to the weight of
all typical elements in that distribution.  The two probabilities
differ only by the width of the distribution, which is sub-leading as
we saw in Eq.~(\ref{eq:resid_fluct_ents}).

The moment-generating function may be understood as extracting, not
only the probability to observe a single sample-fluctuation from the
equilibrium distribution, but in fact the projection of the original
distribution onto \emph{the entire sub-distribution} within which that
sample fluctuation is typical.  This is the shifted binomial
distribution appearing in Eq.~(\ref{eq:gen_func_reduced_def}).  The
constant of proportionality between the two distributions, whose
logarithm is the cumulant-generating function, is then precisely the
leading exponential difference of probabilities, because normalization
terms in the original distribution and the projected distribution have
canceled.

\subsection{Operator algebras in the Doi formalism; two-field methods
  in general, and the difference between classical and quantum systems}

The Doi~\cite{Doi:SecQuant:76,Doi:RDQFT:76} construction of
time-dependent ensembles uses a further property of generating
functions which can lead to confusion on first exposure, but which is
important to understand in order to assign the correct meaning to the
operators and states used in the construction.

Any probability distribution may be written as a sum of modes, and the
generating function associated with that probability distribution may
likewise be written as a sum of polynomials.  Therefore, generating
functions satisfy a property of \emph{superposition}, similar in many
respects to the superposition satisfied by states in quantum
mechanics.  Time evolution acts on superpositions, either of quantum
states, or of the modes of a generating function, independently.  Much
of the time-dependence in either quantum mechanics or stochastic
processes can therefore be understood from the sequencing of the
time-evolution operators, without reference to the particular
function-spaces in which they act.  Doi's contribution was to distill
the dependence in the order of time-evolution operators into an
\emph{operator algebra}, which has the same form as the algebra of the
simple harmonic oscillators of quantum mechanics or quantum field
theory.

For this reason, Doi-Peliti methods are
sometimes~\cite{Mattis:RDQFT:98} termed ``Quantum Field Theory''
methods for stochastic processes.  However, the operator algebra
itself carries no implication of a \emph{physical} property of
quantum-superposition of distinct eigenstates of observables, and
derives only from the linearity of time evolution.  The physical
commitment to the classical or quantum nature of a system comes from
specifying the space of functions representing states and their
observable properties, and on the inner product in that space.  For
further discussion of the many similarities implied, solely by
superposition and causality, between classical and quantum systems,
see Ref.~\cite{Kamenev:DP:01}.  It is under the correspondence of
operator algebras that the stochastic effective action and the quantum
effective action are equivalent objects.

The use of time-dependent sources to probe complex histories with the
stochastic effective action does not seem common in reaction-diffusion
applications of Doi-Peliti theory.  We hope that such methods can come
to be used to ask a wider range of structurally explicit questions
about non-equilibrium paths.

As the simplest non-trivial examples, this paper has demonstrated new
entropy-rate formulae for steady non-equilibrium paths, both for
independent-particle random walks, and for stoichiometrically
non-trivial chemical reactions.  An area for future work is to extend
such formulae to study the large-deviations behavior of random walks
on graphs or the fluctuations of realistic chemical-reaction networks.

\subsection{The information conditions that embed single-time
  ensembles within ensembles of histories}

The equilibrium (canonical) thermodynamic ensemble was recovered in
Sec.~\ref{sec:DP_gen_fn_srcs} as the projection of the larger ensemble
over histories onto a single time slice, in a region where the
histories had settled into their stationary distribution over both
states and transitions.  ``Fluctuations'' in the equilibrium ensemble
have no dynamical connotation, and are simply sample values.
Histories have both dynamical correlation -- a property of a path
itself -- and stochastic fluctuation that represents sampling from the
ensemble over paths.

Fluctuation values in the equilibrium ensemble, when embedded
within the larger ensemble of histories, have the interpretation of
instantaneous conditions on histories accompanied by \emph{no other
  information}.  These information conditions imply distributions over
histories conditioned on the single-time fluctuation, which appears as
an independent variable.  Freidlin-Wentzell theory reproduces the
probabilities for such single-time fluctuations from the integrals, over
the most-probable histories, of conditional probabilities for each
successive increment of change.

The equilibrium distribution is not the only ensemble that can be
recovered from an ensemble of histories by projection onto single
instants.  Projections onto slices where paths have not settled into
their stationary distribution result in non-equilibrium entropies at
single times.  These entropies require both charge and current state
variables as arguments~\cite{Smith:dual:05,Smith:DP:08}, and they can
be written in terms of the \emph{Wigner distribution}, considered in
Ref.~\cite{Smith:evo_games:}.  We may say, then that there is a
hierarchy of ensembles, from the most-complete ensemble over
histories, to single-time non-equilibrium ensembles, which lose the
time-correlation of histories but preserve current-valued state
variables, finally reaching the equilibrium ensemble, which projects
onto only the charge-valued state variables.  By connecting state
variables to fixed points of free-energy landscapes in
Fig.~\ref{fig:fixed_saddle} , and to Legendre duality in
Eq.~(\ref{eq:ext_rates_path_nat_scale}), we have attempted to provide
several perspectives to understand this duality.

\subsection{Thermodynamics beyond Zeno}

We may understand the dynamical system of Freidlin-Wentzell theory as
doing, for thermodynamics, what Hamiltonian dynamics does to free
mechanics from Zeno's paradoxes.  The problem in both cases is to
represent motion as an inherent property, independent of position,
within single instants of time.  In quantum mechanics this
independence is manifest -- traveling waves are distinct
superpositions from standing waves -- but it comes at the price that
position and momentum are non-commuting
observables~\cite{Smith:dual:05,Smith:DP:08}.  The correspondence
between two-field methods for stochastic processes and dissipative
quantum systems (annotated bibliography in App.~\ref{sec:literature}),
shows that the Doi-Peliti construction is revealing essentially the
same form of inherent duality as the standing/traveling wave duality
in quantum mechanics.  The remarkable insight from Freidlin-Wentzell
theory, not obvious beforehand, is that this dual momentum variable
for classical stochastic processes should be an ``inference field'',
and that in reversible random walks it takes the form of a chemical
potential.

\subsection{The master equation versus maximum caliber}

Ref's.~\cite{Ghosh:caliber:06,Stock:caliber:08,Wu:caliber:09}
emphasize the potential for master equations to give only a
coarse-grained description compared to maximum-caliber methods that
track distinguishable particle trajectories.  For interacting
particles, this difference may be important.  However, for
non-interacting particles, the $N$-particle moment-generating function
is simply the $N^{\mbox{\scriptsize th}}$ power of the one-particle
moment-generating function~\cite{Wilf:gen_fun:06}.  Indeed, this
relation is what causes the field
normalizations~(\ref{eq:descale_fields},\ref{eq:action_angle_transform})
to lead to an exact (and not merely asymptotic) factorization of
scale-factor $N$ from the remaining large-deviation rate functions for
relative numbers, not only for the two-state system, but for any
non-interacting particle system.  We also see that the Doi-Peliti
construction ``contains'' a description of single-particle
trajectories, from the agreement of the transition
rates~(\ref{eq:transition_forms_zs}) in the discrete-time model with
the correlation function~(\ref{eq:lambda_corr_fn}) of the Langevin
stochastic differential equation, whose solution is the Gaussian
approximation to discrete random walk.  

A useful further exercise would be to map the generating functions for
the master equation of Sec.~\ref{sec:NEQ_FW_large_dev} and the
discrete-time transfer matrix of Sec.~\ref{sec:caliber} term-by-term,
to understand why the latter readily yields a decomposition with the
structure of the Gibbs free energy, while in the former this is
hidden.  In this way it might be possible to extend the flexibility of
Doi-Peliti effective action methods directly to maximum caliber.

\subsection{On the necessity of thermodynamic limits}

Our rather conservative extension of large-deviations calculations,
from equilibrium to non-equilibrium statistical mechanics, has the
advantage of starting in well-known territory, and taking a small step
across which we can construct all relations between the different
ensembles explicitly.  It does not do justice to the generality of
large-deviations formulae, emphasized in
Ref.~\cite{Touchette:large_dev:08}, and up to now we have not
adequately expressed the dependence throughout science on the
existence of thermodynamic descriptions.  

The ability to coarse-grain has been our most important tool to bring
an infinitely complicated world within reach of analysis and
comprehension.  In the early years of statistical mechanics, when the
formulations of
Boltzmann~\cite{Boltzmann:complexions:77,Boltzmann:pop_schrift:05} and
Gibbs~\cite{Gibbs:papers_1:93} were seen as ``replacing'' classical
thermodynamics, Fermi~\cite{Fermi:TD:56} was quick to emphasize that
thermodynamic descriptions have their own internal consistency and, at
the appropriate level of scale, are self-contained with respect to
empirical validation.  We have learned over the last half-century that
this point is true and important to a degree that Fermi could not have
foreseen.  As renormalization methods have become well-understood
conceptually~\cite{Wilson:RG:74}, we have learned that \emph{all}
descriptions are -- or may as well be -- thermodynamic
descriptions~\cite{Weinberg:QTF_I:95,Polchinski:RGEL:84}.\footnote{Properly
  they are called \emph{effective theories}.  The class of equilibrium
  mechanical theories generally recognized within the conventional
  term ``thermodynamics'' are a class of effective theories.  The
  defining properties used in this review are, however, properties of
  the whole class, and so the terms ``thermodynamic description'' and
  ``effective theory'' could be used interchangeably to refer to them.
  This concept of ``effective'' theory is also the source of the name
  for the quantum and stochastic effective actions.}

Our view of the meaning of fundamental theory has changed accordingly.
All theories once regarded as fundamental have been effective
theories, and at the same time, from criteria of scale-dependent
statistical reduction, each has been fundamental.  Whether the series
terminates is now understood not to be critical; any level at which
one can form a thermodynamic description is a valid starting point for
deduction at that scale and larger scales of aggregation.

Hierarchies of nested phase
transitions~\cite{Weinberg:TFTM:93,Weinberg:DoFT:92}, connected by
regions of large-deviations scaling, govern the behavior of
equilibrium matter at all scales now known, within theories that are
already understood.  Recognizing that neither large-deviations scaling
nor phase transition are phenomena linked inherently to equilibrium,
we may ask how many non-equilibrium cases of symmetry breaking and
large-deviations scaling are responsible for complex dynamics in
nature that are not yet well-understood in terms of collective
phenomena~\cite{Goldenfeld:evol_collect:10}.

\subsection*{Acknowledgments}

I am grateful to acknowledge Insight Venture Partners for support of
this work, and Cosma Shalizi for many helpful conversations and
references over several years. 

\appendix

\section{Related or equivalent approaches to large-deviations theory
  for non-equilibrium processes, and pointers to literature}
\label{sec:literature}

The main text brings together Doi-Peliti construction methods, the
larger framework of Freidlin-Wentzell theory, and Jaynes's methods of
defining combinatorial entropies for paths.  Mathematical differences
among the methods are small; Doi-Peliti theory may be seen simply as a
construction method for the Freidlin-Wentzell quasipotential, and the
only important addition from path entropies is the finer resolution it
gives to distinguishable-particle trajectories.

What will not be apparent in the main text is that several distinct
literatures exist for these nearly-identical mathematical methods,
which have been repeatedly discovered for quantum mechanics, and for
discrete- and continuous-variable classical stochastic processes.
From the collection of different special cases, it has become
clear~\cite{Kamenev:DP:01} that a single framework of two-field
functional integrals subsumes all the different approaches.  Moreover,
all integrals of this type share particular structural features, which
reflect superposition or causality, independently of the physical
substrate to which they apply.

We summarize in order the key ideas behind two-field methods,
functional Legendre transforms, and entropy-rate methods, and finally
mention a parallel treatment within the mathematics literature that
emphasizes observables rather than states.

\subsection{Two-field functional integral methods} 

The most widely-used (formally exact)\footnote{With this condition we
exclude very widely used but intentionally approximate methods such as
Langevin equations or the van Kampen
expansion~\cite{vanKampen:Stoch_Proc:07}.  Any of the full methods
cited here may be reduced to Langevin or van Kampen approximations by
standard methods of Gaussian integration.} methods in the physics
literature to compute both deterministic and fluctuation properties of
non-equilibrium systems are a class of generating-functional methods,
all of which have expressions as two-field, coherent-state functional
integrals.  The earliest of these were the density-matrix methods of
Schwinger~\cite{Schwinger:MBQO:61} and
Keldysh~\cite{Keldysh::65,Lifshitz:LandL:80} to combine dissipation
and dynamics in quantum systems.\footnote{These methods, in turn,
build on the earlier work of
Matsubara~\cite{Matsubara:quant_stat_mech:55,Mahan:MPP:90} using
field-theoretic representations of the finite-temperature partition
function to compute time-dependent correlation functions for
fluctuations about thermal equilibrium.}  The Schwinger-Keldysh form
was later proposed as a general framework to handle dissipative
dynamics in classical or quantum systems by Martin, Siggia, and Rose
(MSR)~\cite{Martin:MSR:73}.  It was observed by
Doi~\cite{Doi:SecQuant:76,Doi:RDQFT:76} that the superposition
principle for generating functions for multi-particle classical
stochastic processes is similar enough to that for quantum density
matrices that the same operator algebra may be used to describe both,
and only their Hilbert spaces differ.  The Doi operator algebra for
discrete Markov processes was later expanded as a coherent-state
functional integral of MSR form by
Peliti~\cite{Peliti:PIBD:85,Peliti:AAZero:86}.  Alex Kamenev has
recently emphasized~\cite{Kamenev:DP:01} that the causal structure
originally recognized by Keldysh, and reflected in a characteristic
``tri-diagonal'' form of the MSR action functional, is the unifying
physical commitment that all field-theoretic methods of this kind
share.

Explanations of the common elements of superposition and operator
algebras are given in the main text in Sec.~\ref{sec:DP_gen_fn_srcs},
and the tri-diagonal form of Gaussian kernels and Green's functions is
illustrated in App.~\ref{sec:exact_gen_fun}.

Among the two-field methods, the one that most explicitly emphasizes
the large-deviations structure of the macroscopic approximation is the
ray-theoretic, or eikonal-based, approach of Freidlin and
Wentzell~\cite{Freidlin:RPDS:98,Graham:path_int:77,Graham:potential:84,%
Graham:non_diff:85,Eyink:action:96}.  This approach, introduced in
basic form by Eyring~\cite{Eyring:ToRP:41} to understand chemical
reaction rates, has been most extensively developed for escape and
first-passage
problems~\cite{Maier:escape:93,Maier:non_grad:92,Maier:exit_dist:97,%
Maier:caustics:93,Maier:scaling:96,Maier:bifurc:00,Maier:oscill:96,%
Maier:sloshing:01,Maier:droplets:01}.  It refines and generalizes the
equilibrium barrier-penetration formulae of Wentzel, Kramers, and
Brillouin (WKB)~\cite{Sakurai:ModQM:85}.  More generally the
Freidlin-Wentzell method is of interest to mathematicians for the
solution of boundary-value problems for diffusion in potentials.
Freidlin-Wentzell theory introduces a quantity known as the
\emph{quasipotential}, so-named because it is thought of as a
generalization of the equilibrium thermodynamic
potentials\footnote{``Thermodynamic potential'' as used here is often
a somewhat indefinite reference to the classical thermodynamic entropy
or any of its Legendre transforms.  It will be shown quite precisely
below, exactly what combination of such potentials the
Freidlin-Wentzell quasipotential generalizes.}  The quasipotential is
a large-deviations rate function, appropriately scaled for system
size.  It may be evaluated as an action-functional integral along a
ray or ``eikonal'' of the diffusion
operator~\cite{Graham:path_int:77,Graham:potential:84}, which has the
interpretation of the most probable trajectory leading to an escape or
boundary value.

\subsection{Functional Legendre transform and stochastic effective
  actions} 

Generating-functional methods for dynamical systems -- especially
emphasizing efficient approximation -- have been extensively refined
in vacuum quantum field theory (QFT) and to some extent in
condensed-matter physics.  They are used in these domains to extract
the leading deterministic approximation to fluctuating
quantum-mechanical observables, a use very close to the extraction of
leading exponential dependence of probabilities in large-deviations
theory.  The functional Legendre transform of the cumulant-generating
functional is known in QFT as the \emph{quantum effective
  action}~\cite{Weinberg:QTF_II:96}.  It is the basis of
background-field methods used to identify the ground states or vacua
of field theories, and it is the starting point for a variety of
renormalization schemes.  Efficient and elegant graphical methods have
been derived to approximate it within the moment expansions of
Gaussian functional integrals.

In the Doi-Peliti construction for discrete stochastic processes, the
counterpart to the quantum effective action is the deficit in entropy
for any configuration from the maximum entropy attainable under the
same constraints.  This entropy deficit is precisely the
large-deviations rate function for fluctuations in
macrostates~\cite{Ellis:ELDSM:85}.  For closed equilibrium systems
this rate function is simply a difference between two entropies.  For
open equilibrium systems the contribution from entropies of the
environment leads to the expression of the rate function as a
difference of Gibbs free energies.  The latter case leads to the forms
developed in the main text.

In the large-deviations limit for non-equilibrium stochastic
processes, where probabilities are defined on histories rather than on
instantaneous states, the rate function becomes the difference between a
combinatorial entropy defined on a path, and an entropy rate from the
Markov process that generates the path in a stationary environment, as
we have illustrated in Sec.~\ref{sec:caliber}.  

We have termed this large-deviations rate function, counterpart to the
quantum effective action in QFT, the \emph{stochastic effective
  action}.  The quasipotential used in Freidlin-Wentzell theory to
compute boundary values of diffusion equations in potentials is an
instance of the stochastic effective action.  For a one-dimensional
system it is the form produced by the moment-generating function for
observations at a single time.  The general stochastic effective
action, for arbitrary path fluctuations, is computed as the
self-consistent expectation value of the Freidlin-Wentzell field
action, in which the Liouville operator acts as Hamiltonian.

\subsection{Path entropies and maximum caliber} 

The approach to non-equilibrium thermodynamics taken by
E.~T.~Jaynes~\cite{Jaynes:caliber:80} is a direct outgrowth of methods
based on the entropy rate of a stochastic process.  (Related methods
were developed for chaotic deterministic dynamical systems, where the
entropy rate was replaced by the Kolmogorov-Sinai, or \emph{metric},
entropy.)  Probabilities are assigned to histories, and the measure on
the space of histories, together with any constraints, then determines
the path-space entropy.  Jaynes coined the term \emph{caliber} for the
path entropy, to suggest the fluctuation width of a tube of
micro-histories about any coarse-grained macro-history.

In the text, we have used a formulation of the maximum-caliber
principle developed in
Ref's.~\cite{Ghosh:caliber:06,Stock:caliber:08,Wu:caliber:09},
suitable for computing the probability of steady non-equilibrium
distributions.  In cases where the transition probabilities (including
those altered by insertions from generating functions) are stationary,
the calculation of the entropy rate simplifies using the chain rule
for conditional entropies~\cite{Cover:EIT:91}.  The chain rule
separates the path entropy into an equilibrium entropy on the
stationary distribution, and a conditional entropy of transitions from
that distribution.  This compact representation appears in
Sec.~\ref{sec:var_approach}.  

\subsection{Large-deviations theory within the probability
  literature}  

The methods used in this paper are those common in the physics
literature, particularly the literature from reaction-diffusion
theory, of which Ref's.~\cite{Mattis:RDQFT:98,Cardy:FTNEqSM:99} are
representative.  Parallel to this physics literature is a very large
literature on stochastic processes and large-deviations theory, framed
within the language of modern probability
theory~\cite{Kurtz:SIAM_largedev:81,Ethier_Kurtz:Markov:86,%
  Fristedt_Gray:Prob_Th:97,Feng_Kurtz:large_dev:06}.  The approaches
omitted in this review have in common a starting point in the
\emph{reverse Kolmogorov equation}, which emphasizes the
time-evolution of operators corresponding to observables, rather than
the forward-Kolmogorov or master equation used here, which models the
time-evolution of probability distributions.  Rigorous proofs of
convergence, which have not been the main concern in this article, are
sometimes made easier by using the observable representation.

\section{A fluctuation-oriented summary of the entropy, the roles of
  state variables, and the structure expressed by constraints, in
  classical equilibrium thermodynamics}
\label{sec:free_energies}

The entropy is the state function maximized subject to the constraints
on the system configurations.  The values of the constraints are
given by the state variables that are the arguments of the
entropy. 
\begin{displaymath}
  S \! \left( U, V, \left\{ {\rm n}_i \right\} \right) . 
\end{displaymath}
Common sources of constraint are internal energy $U$, volume $V$, or
the numbers $\left\{ {\rm n}_i \right\}$ of different kinds of
particles.

The entropy arises as the leading exponential term in the
\emph{relative} probabilities of different states of the system,
grouped according to the values of the constraints $U$, $V$, for which
that entropy would be maximum, 
\begin{displaymath}
  e^{
    S \left( U, V, \left\{ {\rm n}_i \right\} \right)
  } .  
\end{displaymath}
In this respect, the relative probability is a more fundamental
quantity than the normalized probability.  The normalization is the
sum or integral of terms $e^S$ over whichever states the system can
take, which can depend on the setting.  For independent systems, the
relative probabilities multiply, so that the sum of the entropies is
maximized at the overall-most-likely state.  It is this relation
between entropy and relative probability that defines the macroscopic
fluctuation properties of ensembles~\cite{Ellis:ELDSM:85}.

The way entropy governs the structure of interacting systems comes
from the properties of its gradients.  The manipulation of these
gradients is the main enterprise of classical
thermodynamics~\cite{Fermi:TD:56}. 

The equation of state is merely a definition of the gradients of the
entropy in terms of observable quantities: 
\begin{eqnarray}
  \delta S 
& = &
  \frac{\partial S}{\partial U} 
  \delta U + 
  \frac{\partial S}{\partial V} 
  \delta V + 
  \sum_i 
  \frac{\partial S}{\partial {\rm n}_i } 
  \delta {\rm n}_i 
\nonumber \\
& \equiv & 
  \beta \, \delta U + 
  \beta p \, \delta V - 
  \sum_i \beta {\mu}_i \, \delta {\rm n}_i . 
\label{eq:EOS_gradient}
\end{eqnarray}
In chemical thermodynamics $1 / \beta = k_B {\tt T}$ corresponds to
the temperature in energy units, $p$ is the pressure, and $\left\{
{\mu}_i \right\}$ are the chemical potentials.  These gradients of $S$
with respect to its extensive arguments are the \emph{intensive state
  variables} in the thermodynamic description.

Eq.~(\ref{eq:EOS_gradient}) may be re-arranged to express entropy
change in the \emph{units} of any of the constraining state variables
that are shared between system sub-components, or between the system
and its environment.  When energy is chosen as the unit of measure --
a natural choice since it is conserved among nearly all forms of
systems in contact -- the result is the usual thermodynamic statement
of ``conservation of energy'',
\begin{equation}
  \delta U = 
  k_B {\tt T} \, \delta S - 
  p \, \delta V + 
  \sum_i {\mu}_i \, \delta {\rm n}_i , 
\label{eq:EOS_as_cons_of_U}
\end{equation}
In this representation $p \, \delta V$ and $\left\{ {\mu}_i \, \delta
n_i \right\}$ are interpreted respectively as mechanical and chemical
increments of work, and $k_B {\tt T} \, \delta S$ labeled ``heat''.
Eq.~(\ref{eq:EOS_as_cons_of_U}) is only a ``conservation law'' in the
sense that $\delta U$ is a \emph{lower bound} on the energy change
required to permit a change $\delta S$, in the context where volume
also changes by $\delta V$ and particle numbers by $\left\{ \delta
{\rm n}_i \right\}$.  

The expression of competition of entropy terms in the relative
probability, when a small system is coupled to a much larger
environment, is that we may often expand the entropy in the
environment to linear order in the exchanged amounts of any conserved
quantity that the system and environment dynamically apportion between
them.  The result for the total relative probability of an
apportionment that leaves $U$ and $V$ in the system is
\begin{equation}
  e^{
    S \left( U, V, \left\{ {\rm n}_i \right\} \right)
    - {\beta}_E U - {\beta}_E p_E V  
  } . 
\label{eq:fluct_form_env}
\end{equation}
We have supposed in the expression~(\ref{eq:fluct_form_env}) that
energy and volume can be exchanged, but not particle number, and that
intra-system equilibration leads to the relative probability $e^S$
that would be maximum with $U$ and $V$ as constraints.  Here
${\beta}_E$ and $p_E$ are regarded as fixed descriptors of the
environment, by supposing that it is large compared to the system, so
that only the first derivative of its entropy need be known.
${\beta}_E$ and $p_E$ therefore are not \emph{functions} of $U$ or $V$.

The expected or stable state of the system is the one that maximizes
the joint probability~(\ref{eq:fluct_form_env}) over $U$ and $V$.
Ordinarily this maximization is regarded to take place within the
fluctuations of the ensemble, and values other than the maximizers are
never used in the combination $S \left( U, V, \left\{ {\rm n}_i
\right\} \right) - {\beta}_E U - {\beta}_E p_E V$ in classical
thermodynamics.

If we wish to use the classical theory to understand the effects of
constraints on system states, we think of using ${\beta}_E$ and $p_E$
as control variables, imposed through the interaction between the
system and the environment.  The log-probability as a function of
these control variables then becomes $\max_{U,V} \left[ S \!  \left( U
  , V , \left\{ {\rm n}_i \right\} \right) - {\beta}_E U - {\beta}_E
  p_E V \right]$.  At the maximizers, $\beta = {\beta}_E$ and $p =
p_E$.  If we continue to denominate entropy changes in units of
energy, and if we choose signs so that the $U$, $V$ extremum is a
minimum,\footnote{This convention is arbitrary, of course, but it
  helps to realize that the choice to minimize the free energy is a
  relic of the notion from ${19}^{\mbox{\scriptsize{th}}}$-century
  mechanics that energy minimization identifies stable states.  A
  convention that keeps maximization of entropy foremost expresses
  directly the origin of stability in statistical degeneracy, and we
  will use such a convention throughout the paper.} then the
log-probability is proportional to the \emph{Gibbs Free Energy}
\begin{equation}
  G \! 
  \left( 
    k_B {\tt T} , p , \left\{ {\rm n}_i \right\} 
  \right) \equiv 
  \min_{U,V}
  \left[ 
    U + pV - k_B {\tt T} 
    S \! 
    \left( 
      U , V , \left\{ {\rm n}_i \right\} 
    \right) 
  \right] .
\label{eq:Gibbs_def}
\end{equation}
Now $p$ and $\beta$ have been set equal in the system and
environment.  The subscripts $p_E$, ${\beta}_E$ have therefore been
dropped, and the minimizing $U$ and $V$ are now functions of $p$ and
$\beta$.  

If we wish to express classical thermodynamic relations in the form
that relates most directly to the fluctuation-origins of the entropy,
we may simply work directly with the logarithm of the relative
probability, which is
\begin{equation}
  - \beta G \! 
  \left( 
     k_B {\tt T} , p , \left\{ {\rm n}_i \right\} 
  \right) \equiv 
  \max_{U,V}
  \left[ 
    S \! 
    \left( 
      U , V , \left\{ {\rm n}_i \right\} 
    \right) - 
    \beta 
    \left( U + p V \right)
  \right] . 
\label{eq:beta_Gibbs_S}
\end{equation}
We will take the sign and normalization from
Eq.~(\ref{eq:beta_Gibbs_S}) to define a standard \emph{Legendre
  transform} of $S$.

The fluctuation theorems in Sec.~\ref{sec:EQ_large_dev}, for particle
exchange, are similar in structure to the intermediate
expression~(\ref{eq:fluct_form_env}) for exchanges of energy or
volume.  $-\beta G$ takes the place of $S$, because it is assumed that
energy and volume equilibrate on times much shorter than the
equilibration times for particle exchange.  In addition, the two
states of the system exchange particles only with each other and not
with the environment, so neither Gibbs free energy is evaluated only
to linear order.  

\section{Exact solution for continuous sources in the time-dependent
  generating functional}
\label{sec:exact_gen_fun}

The non-local relation between current sources and the non-equilibrium
paths they produce makes computation of the functional Legendre
transform technically challenging for all but simple cases.  However,
the simplicity of the Gaussian integral in coherent-state variables
makes it possible to understand this non-locality with exact
solutions.  The coherent-state variables in this appendix will follow
the diagonalization of Sec.~\ref{sec:field_quasipotential}

The coherent-state field action~(\ref{eq:action_twofields_descaled_j})
may be recast in the matrix form~\cite{Kamenev:DP:01} characteristic
of all two-field action functionals,\footnote{For relations among
  these, see the literature review in App.~\ref{sec:literature}.} as
\begin{widetext}
\begin{eqnarray}
  S_j 
& = & 
  N \int d\tau
  \begin{array}{c}
    \left[ 
      \begin{array}{cc}
        {\hat{\phi}}_b & 
        {\hat{\phi}}_a
      \end{array}
    \right] \\
    \phantom{{\phi}_a}
  \end{array}
  \left[ 
    \begin{array}{cc}
      - {\partial}_{\tau} + {\bar{\nu}}_a - j / 2 & 
      - {\bar{\nu}}_a \\
      - {\bar{\nu}}_b & 
      - {\partial}_{\tau} + {\bar{\nu}}_b - j / 2 
    \end{array}
  \right]
  \left[ 
    \begin{array}{c}
      {\phi}^{\dagger}_b \\
      {\phi}^{\dagger}_a
    \end{array}
  \right]
\nonumber \\
& = & 
  N \int d\tau
  \begin{array}{c}
    \left[ 
      \begin{array}{cc}
        \sqrt{2} \hat{\phi} & 
        \sqrt{1/2} \hat{\Phi}
      \end{array}
    \right] \\
    \phantom{{\phi}_a}
  \end{array}
  \left\{
    - {\partial}_{\tau} + 
    \frac{1}{2} 
    {\sigma}_0 + 
    \frac{1}{2} 
    {\sigma}_3 - 
    \left( \bar{\nu} + \frac{j}{2} \right)
    {\sigma}_1 + 
    i \bar{\nu} 
    {\sigma}_2 
  \right\}
  \left[ 
    \begin{array}{r}
      \sqrt{1/2} {\phi}^{\dagger} \\
      \sqrt{2} {\Phi}^{\dagger}
    \end{array}
  \right] . 
\label{eq:S_j_coh_state_matrix}
\end{eqnarray}
\end{widetext}
In the second line, ${\sigma}_0$ designates the $2 \times 2$ identity
matrix and the other ${\sigma}_i$ are the Pauli
matrices~\cite{Sakurai:ModQM:85} 
\begin{eqnarray}
  {\sigma}_1 
& = & 
  \left[ 
    \begin{array}{cc}
        & 1 \\
      1 &   \\
    \end{array}
  \right] 
\nonumber \\
  {\sigma}_2 
& = & 
  \left[ 
    \begin{array}{cc}
        & -i \\
      i &    \\
    \end{array}
  \right] 
\nonumber \\
  {\sigma}_3 
& = & 
  \left[ 
    \begin{array}{cc}
      1 &    \\
        & -1 \\
    \end{array}
  \right] . 
\label{eq:Pauli_matrices}
\end{eqnarray}

The solution to the stationary point conditions, and for the effective
action more generally, follows closely what has already been done for
the single-time generating function.  The major difference is that we
must replace the solution of scalar differential equations with the
time-ordered matrix solution of a linear differential equation
expressed in terms of the matrix in
Eq.~(\ref{eq:S_j_coh_state_matrix}).  The $2 \times 2$ matrix kernel,
which depends on parameters of the potential and on $j$, will be
denoted by 
\begin{equation}
  \sigma \! \left[ \bar{\nu} , j \right] \equiv 
    \frac{1}{2} 
    {\sigma}_0 + 
    \frac{1}{2} 
    {\sigma}_3 - 
    \left( \bar{\nu} + \frac{j}{2} \right)
    {\sigma}_1 + 
    i \bar{\nu} 
    {\sigma}_2 . 
\label{eq:sigma_sources_def}
\end{equation}

The expression of the boundary conditions at time $T$ (here again $T$
will be assumed finite but large) may be written
\begin{equation}
  {
    \left[ 
      \begin{array}{r}
        \sqrt{1/2} {\phi}^{\dagger} \\
        \sqrt{2} {\Phi}^{\dagger}
      \end{array}
    \right] 
  }_T = 
  \sqrt{2}
  \left[ 
    \begin{array}{r}
      0 \\ 
      1 
    \end{array}
  \right] , 
\label{eq:upper_phi_dag_bc}
\end{equation}
and in terms of these, the general solution for the ${\phi}^{\dagger}$
fields propagates this constraint backward in time, as 
\begin{equation}
  {
    \left[ 
      \begin{array}{r}
        \sqrt{1/2} {\phi}^{\dagger} \\
        \sqrt{2} {\Phi}^{\dagger}
      \end{array}
    \right] 
  }_{\tau} = 
  \sqrt{2}
  {\mathcal{T}}^{-1}
  e^{
    - \int_{\tau}^{\hat{T}} d\tau 
    \sigma \! \left[ \bar{\nu} , j_{\tau} \right]
  }
  \left[ 
    \begin{array}{r}
      0 \\ 
      1 
    \end{array}
  \right] . 
\label{eq:general_phi_dag_soln}
\end{equation}
Here ${\mathcal{T}}^{-1}$ denotes the inverse time-ordering operator
which arranges terms in the exponential from right to left in order of
decreasing time.  The solution~(\ref{eq:general_phi_dag_soln}) is the
direct generalization of the
solutions~(\ref{eq:phi_dag_soln_genfunc},\ref{eq:Phi_dag_soln_gen_func}). 

In the same manner, for continuous sources (no $\delta$-functions, so
we do not have to worry about exponential tails) $j_{\tau < 0} \equiv
0$, and the equilibrium state as the initial distribution, the lower
boundary condition for the $\hat{\phi}$ fields is
\begin{equation}
  {
    \left[ 
      \begin{array}{cc}
        \sqrt{2} \hat{\phi} & 
        \sqrt{1/2} \hat{\Phi}
      \end{array}
    \right] 
  }_0 = 
  \frac{{\hat{\Phi}}_0}{ \sqrt{2}}
  \left[ 
    \begin{array}{cc}
      2 \bar{\nu} & 1 
    \end{array}
  \right] , 
\label{eq:lower_phi_hat_bc}
\end{equation}
and the general time-dependent solution is 
\begin{equation}
  {
    \left[ 
      \begin{array}{cc}
        \sqrt{2} \hat{\phi} & 
        \sqrt{1/2} \hat{\Phi}
      \end{array}
    \right] 
  }_{\tau} = 
  \frac{{\hat{\Phi}}_0}{ \sqrt{2}}
  \left[ 
    \begin{array}{cc}
      2 \bar{\nu} & 1 
    \end{array}
  \right] 
  {\mathcal{T}}^{-1}
  e^{
    - \int_0^{\tau} d\tau 
    \sigma \! \left[ \bar{\nu} , j_{\tau} \right]
  } . 
\label{eq:general_phi_hat_soln}
\end{equation}
Eq.~(\ref{eq:general_phi_hat_soln}) generalizes the constant
solution~(\ref{eq:phi_diag_const}) due to nonzero $j_{\tau}$.  

The constraint of total number fixes the normalization of
${\hat{\Phi}}_0$ as before, through the relation 
\begin{eqnarray}
  1 
& = & 
  \begin{array}{c}
    {
      \left[ 
        \begin{array}{cc}
          \sqrt{2} \hat{\phi} & 
          \sqrt{1/2} \hat{\Phi}
        \end{array}
      \right] 
    }_{\tau} \\
    \phantom{{\phi}_s}
  \end{array}
  {
    \left[ 
      \begin{array}{r}
        \sqrt{1/2} {\phi}^{\dagger} \\
        \sqrt{2} {\Phi}^{\dagger}
      \end{array}
    \right] 
  }_{\tau} \mbox{ (at any $\tau$)} 
\nonumber \\
& = & 
  {\hat{\Phi}}_0
  \left[ 
    \begin{array}{cc}
      2 \bar{\nu} & 1 
    \end{array}
  \right] 
  {\mathcal{T}}^{-1}
  e^{
    - \int_0^{\hat{T}} d\tau 
    \sigma \! \left[ \bar{\nu} , j_{\tau} \right]
  } 
  \left[ 
    \begin{array}{r}
      0 \\ 
      1 
    \end{array}
  \right] , 
\label{eq:phi_hat_zero_from_norm}
\end{eqnarray}
corresponding to Eq.~(\ref{eq:Phi_from_N_solve}).  The general
time-dependent number expectation $n_{\tau}$ is immediately seen to
satisfy
\begin{widetext}
\begin{eqnarray}
  2 n_{\tau} 
& = & 
  N 
  \begin{array}{c}
    {
      \left[ 
        \begin{array}{cc}
          \sqrt{2} \hat{\phi} & 
          \sqrt{1/2} \hat{\Phi}
        \end{array}
      \right] 
    }_{\tau} \\
    \phantom{{\phi}_s}
  \end{array}
  \left[ 
    \begin{array}{cc}
        & 1 \\
      1 &   \\
    \end{array}
  \right] 
  {
    \left[ 
      \begin{array}{r}
        \sqrt{1/2} {\phi}^{\dagger} \\
        \sqrt{2} {\Phi}^{\dagger}
      \end{array}
    \right] 
  }_{\tau}
\nonumber \\
& = & 
  N 
  {\hat{\Phi}}_0
  \left[ 
    \begin{array}{cc}
      2 \bar{\nu} & 1 
    \end{array}
  \right] 
  {\mathcal{T}}^{-1}
  e^{
    - \int_0^{\tau} d\tau 
    \sigma \! \left[ \bar{\nu} , j_{\tau} \right]
  } 
  {\sigma}_1
  {\mathcal{T}}^{-1}
  e^{
    - \int_{\tau}^{\hat{T}} d\tau 
    \sigma \! \left[ \bar{\nu} , j_{\tau} \right]
  } 
  \left[ 
    \begin{array}{r}
      0 \\ 
      1 
    \end{array}
  \right] 
\nonumber \\
& = & 
  2 \frac{\delta}{\delta j_{\tau}}
  N \log 
  \left\{ 
  \left[ 
    \begin{array}{cc}
      2 \bar{\nu} & 1 
    \end{array}
  \right] 
  {\mathcal{T}}^{-1}
  e^{
    - \int_0^{\hat{T}} d\tau 
    \sigma \! \left[ \bar{\nu} , j_{\tau} \right]
  } 
  \left[ 
    \begin{array}{r}
      0 \\ 
      1 
    \end{array}
  \right] 
  \right\} 
\nonumber \\
& = & 
  - 2 \frac{\delta}{\delta j_{\tau}}
  \Gamma \! \left[ j \right] . 
\label{eq:n_gen_fields_solve}
\end{eqnarray}
\end{widetext}
The functional form in the first two lines of
Eq.~(\ref{eq:n_gen_fields_solve}) directly generalizes
Eq.~(\ref{eq:nu_tau_time_genfunctal}) for the point source.  The last
two lines are the functional equivalent, with a variational derivative
replacing the partial derivative, of the single-time
relation~(\ref{eq:grad_Gamma_number}).  And indeed, by exactly the
evaluation leading to Eq.~(\ref{eq:gen_func_fields_Phis}), we find
that the cumulant-generating functional is given by
\begin{equation}
  \Gamma \! \left[ j \right] = 
  - N \log 
  \left\{ 
  \left[ 
    \begin{array}{cc}
      2 \bar{\nu} & 1 
    \end{array}
  \right] 
  {\mathcal{T}}^{-1}
  e^{
    - \int_0^{\hat{T}} d\tau 
    \sigma \! \left[ \bar{\nu} , j_{\tau} \right]
  } 
  \left[ 
    \begin{array}{r}
      0 \\ 
      1 
    \end{array}
  \right] 
  \right\} .   
\label{eq:gen_functal_exact}
\end{equation}

We now have an exact expression for the effective action which
generalizes the single-time expression~(\ref{eq:S_eff_n_compute}).
Drawing ${\nu}_{\tau} = n_{\tau} / N$ from
Eq.~(\ref{eq:n_gen_fields_solve}), the stochastic effective action is
\begin{equation}
  S^{\mbox{\scriptsize eff}} \! 
  \left[ n \right] = 
  N \int_0^{\hat{T}} d\tau
  j_{\tau} {\nu}_{\tau} + 
  \Gamma \! \left[ j \right] .   
\label{eq:S_eff_exact_fields}
\end{equation}

Eq.~(\ref{eq:S_eff_exact_fields}) is the promised functional which
correctly generalizes the notion of an entropy difference from
Eq.~(\ref{eq:S_eff_G_diffs}) to a space of histories.  It represents a
large-deviations principle for the stochastic process from
Eq.~(\ref{eq:master_equation}), with a rate function $\int_0^{\hat{T}}
d\tau j_{\tau} {\nu}_{\tau} + \Gamma \! \left[ j \right] / N$ which is
strictly $N$-independent for Gaussian fluctuations.  In this sense the
linear reaction is the analog of the Gaussian distribution, and its
large-deviations principle is simply the central limit theorem for a
system with finite variance.  The residual functional determinant for
fluctuations about the mean value -- though it will not be computed
here -- would introduce sub-extensive corrections in $N$ just like
those of the residual entropy in Eq.~(\ref{eq:resid_fluct_ents}).
(See Ref's~\cite{Weinberg:QTF_II:96}~Ch.~16, \cite{Kamenev:DP:01}, and
\cite{Coleman:AoS:85}~Ch.~7 for good treatments of such calculations).

\section{Formulae for multi-particle reactions; equivalently, the
  extension from random walks from graphs to directed hypergraphs}
\label{sec:multi_part_react}

Independent random walks between wells, by a particles of a single
type, may be modeled in the discrete-state approximation by diffusion
on an ordinary graph.  If the the mass-action rates satisfy detailed
balance in the steady state, the graph may be considered undirected;
otherwise it is directed.  Chemical reactions in which multiple
particles jointly move through a transition state from reactants to
products generalize this relation from graphs to hypergraphs.  A graph
need not be directed in the case of detailed balance, because the
nodes at either end of a link are naturally complementary.
Hypergraphs, in contrast, are defined to permit arbitrary collections
of nodes as the ``boundary'' of a hyper-edge.  We therefore require
directed hyper-edges to model chemistry, which distinguish a subset of
nodes as inputs and a second subset as outputs.

The surprising formula~(\ref{eq:holding_ent_rate}) for the probability
rate of persistent non-equilibrium paths may immediately be extended
to arbitrary networks of chemical reactions, hence, from graphs to
directed hypergraphs.  Global network topology will not be considered
here, but the rate formulae for arbitrary multiparticle reactions at
steady-state dis-equilibria will be given.

Consider a general collection of bi-directional reactions indexed
by their reactant and product sides as $r$ and $p$.  Individual
species participating in the reaction generalize the wells $a$ and $b$
of the two-state system, to a many-well problem where indices $a_i$
and $b_j$ are drawn from a common set of species.  The reaction
formula is 
\begin{equation}
  p_{b_1} + \cdots + p_{b_n}  
  \stackrel{ba}{\rightleftharpoons}
  r_{a_1} + \cdots + r_{a_m} . 
\label{eq:general_reaction_stoich}
\end{equation}
To simplify the notation, rather than writing stoichiometric
coefficients other than unity, it will be assumed that the
coefficients $a_i$ or $b_j$ can repeatedly take the same values as
needed.  The order in which the indices $ab$ are written defines the
relation between the sign of the current and the changes of
concentrations, so that $ab$ and $ba$ denote the same reaction but
with opposite sign conventions for the current.  When only the pair
matters, without respect to sign, the pair-index will be denoted by
$\left< ab \right>$.

The complexes of reactants and products behave in the transition state
as a single ``particle'', for which the one-particle free energy (or
chemical potential) will be denoted ${\mu}_{\ddagger}^{\left< ab
  \right>}$.  The half-reaction rate constants are then defined from
the transition-state and single-species one-particle chemical
potentials as
\begin{eqnarray}
  k_{ba}
& = & 
  e^{
    - \beta 
    \left( 
      {\mu}_{\ddagger}^{\left< ab \right>} - 
      \sum_{i=1}^m {\mu}^1_{a_i}
    \right)
  }
\nonumber \\
  k_{ab}
& = & 
  e^{
    - \beta 
    \left( 
      {\mu}_{\ddagger}^{\left< ab \right>} - 
      \sum_{j=1}^n {\mu}^1_{b_j}
    \right)
  } . 
\label{eq:general_half_react_rate_consts}
\end{eqnarray}
The forward and backward half-reaction rates associated with the
reaction~(\ref{eq:general_reaction_stoich}), with the simplified
dilute-solution assumption that activities are proportional to
concentrations, are then given by
\begin{eqnarray}
  r_{ba}
& = & 
  k_{ba}
  \prod_{i=1}^m {\rm n}_{a_i}
\nonumber \\
& = & 
  e^{
    - \beta 
    {\mu}_{\ddagger}^{\left< ab \right>} 
  }
  e^{
    \sum_{i=1}^m 
    \left( 
      \beta {\mu}^1_{a_i} + \log {\rm n}_{a_i}
    \right) 
  }  
\nonumber \\
& \equiv & 
  e^{
    - \beta 
    \left( 
      {\mu}_{\ddagger}^{\left< ab \right>} - 
      \sum_{i=1}^m {\mu}_{a_i}
    \right)
  }
\nonumber \\
  r_{ab}
& = & 
  k_{ab}
  \prod_{j=1}^n {\rm n}_{b_j} . 
\nonumber \\
& = & 
  e^{
    - \beta 
    {\mu}_{\ddagger}^{\left< ab \right>} 
  }
  e^{
    \sum_{j=1}^n 
    \left( 
      \beta {\mu}^1_{b_j} + \log {\rm n}_{b_j}
    \right) 
  }  
\nonumber \\
& \equiv & 
  e^{
    - \beta 
    \left( 
      {\mu}_{\ddagger}^{\left< ab \right>} - 
      \sum_{j=1}^n {\mu}_{b_j}
    \right)
  } . 
\label{eq:general_half_react_rates}
\end{eqnarray}
In the succeeding lines of each rate law, dropping the superscript $1$
on the species chemical potentials reflects the absorption of number
factors to form the $n$-particle chemical potentials, as in
Eq.~(\ref{eq:n_part_chem_pots}).  

\subsubsection{Master equation, Liouville operator, and
  Freidlin-Wentzell action}

The master equation for the stochastic reaction network immediately
generalizes Eq.~(\ref{eq:general_exchange_ME}) to 
\begin{equation}
  \frac{\partial {\rho}_{\vec{{\rm n}}}}{\partial t} = 
  \sum_{ab} 
  \left( 
    e^{
      \sum_{i=1}^m \partial / \partial {\rm n}_{a_i} - 
      \sum_{j=1}^n \partial / \partial {\rm n}_{b_j}
    } - 1 
  \right)
  r_{ba}
  {\rho}_{\vec{{\rm n}}} . 
\label{eq:general_exchange_ME_network}
\end{equation}
Note that the reaction order matters in this expression, so that the
same exponential of shift operators occurs in two terms but with
opposing sign, acting on different products of the species numbers
${\rm n}_{a_i}$ or ${\rm n}_{b_j}$.  

The construction of the field functional integral proceeds just as for
the linear reaction.  Skipping directly to action-angle variables, the
Liouville operator generalizing Eq.~(\ref{eq:Liouville_actionangle})
becomes
\begin{equation}
  \mathcal{L} = 
  \sum_{ba}
  k_{ba}
  \prod_{i=1}^m
  n_{a_i}
  \left(
    1 - 
    e^{
      \sum_{j=1}^n {\eta}_{b_j} - 
      \sum_{i=1}^m {\eta}_{a_i} 
    } 
  \right) . 
\label{eq:Liouville_actionangle_general}
\end{equation}
The corresponding action functional generalizing
Eq.~(\ref{eq:S_ass_diss_AA_red}) (and dropping surface terms
associated with total number) is then
\begin{equation}
  S = 
  \int dt 
  \left\{ 
    - \sum_{k=1}^I
    {\partial}_t {\eta}_k
    n_k + 
    \mathcal{L}
  \right\} . 
\label{eq:S_ass_diss_AA_red_general}
\end{equation}
For the general network, which may have multiple characteristic
timescales in different reactions, it is less convenient to find an
overall non-dimensionalization of time than to simply work in
real-time coordinates, so we will write all integrals with measure
$dt$.   

If, as for the double-well, sources are coupled only to (arbitrary
polynomials in) the $n_k$, the $n_k$ equation of motion will continue
to depend explicitly only on the $\eta$ variables.  Hence, the
response to the sources will be through these variables.  The $n_k$
stationary-point equations therefore continue to determine quantities
such as the value of $\mathcal{L}$ appearing in the steady
non-equilibrium effective action.  From these, the forms of sources
needed to produce such backgrounds can then be inferred.  The equation
for $n_k$ is 
\begin{equation}
  {\partial}_t n_k = 
  \sum_{ab}
  e^{
    - \beta 
    \left( 
      {\mu}_{\ddagger}^{\left< ab \right>} - 
      \sum_{i=1}^m {\mu}_{a_i}
    \right)    
  }
  \frac{\partial}{\partial {\eta}_k}
  \left(
    e^{
      \sum_{j=1}^n {\eta}_{b_j} - 
      \sum_{i=1}^m {\eta}_{a_i} 
    } 
  \right) . 
\label{eq:n_k_EOM_general}
\end{equation}
Again, the sum is symmetric under exchange of $a$ and $b$, so wherever
$\partial / \partial {\eta}_k$ acts, it acts equally and with opposite
sign in two terms.  Cancellation of all such pairs of terms
corresponds to detailed balance, except that the $\eta$ variables in
the exponential can take nonzero values.

Eq.~(\ref{eq:n_k_EOM_general}) can be solved with ${\partial}_t n_k
\equiv 0$ for any configuration of the $\left\{ n_k \right\}$ by
setting 
\begin{equation}
  {\eta}_k = 
  \frac{1}{2}
  \beta 
  {\mu}_k = 
  \frac{1}{2}
  \left( 
    \beta {\mu}^1_k + 
    \log n_k 
  \right) . 
\label{eq:SS_cond_etas_general}
\end{equation}

A sequence of evaluations identical to those leading to
Eq.~(\ref{eq:S_eff_holder_eval}) then produces an equivalent form for
the set of many-particle reactions, now expressed as a sum over
\emph{reactions} only (so unordered pairs $\left< ab \right>$)
\begin{eqnarray}
  \mathcal{L} 
& = & 
  \sum_{\left< ab \right>}
  e^{
    - \beta 
    {\mu}_{\ddagger}^{\left< ab \right>} 
  } 
  {
    \left(
      e^{
        \frac{1}{2} \sum_{i=1}^m
        \beta {\mu}_{a_i}
      } - 
      e^{
        \frac{1}{2} \sum_{j=1}^n
        \beta {\mu}_{b_j}
      }
    \right)
  }^2
\nonumber \\
& = & 
  \sum_{\left< ab \right>}
  {
    \left(
      \sqrt{
        k_{ba} \prod_{i=1}^m n_{a_i}
      } - 
      \sqrt{
        k_{ab} \prod_{j=1}^n n_{b_j}
      }
    \right)
  }^2 . 
\label{eq:SS_cond_L_eval}
\end{eqnarray}
Eq.~(\ref{eq:SS_cond_L_eval}) generalizes
Eq.~(\ref{eq:holding_ent_rate}) for the single linear reaction.  Other
terms in the effective action vanish on a constant-$\vec{n}$ history,
and so this is the expression from Freidlin-Wentzell theory that
governs fluctuations in the manner of an entropy-rate difference.  A
topic for future work is to extend the Jaynes path-entropy methods to
many-particle reactions, to determine whether combinatorial and
environment entropy rates may be isolated for these as for the
double-well.


\end{document}